\documentclass{article}
\usepackage[utf8]{inputenc}
\usepackage{amsmath,graphicx,placeins,amssymb}
\usepackage{tikz}
\usepackage{hyperref}
\usepackage[left=2cm,right=2cm,top=2cm, bottom=2cm]{geometry}
\usepackage{euscript}
\usepackage[normalem]{ulem}
\usepackage{cite}

\newcommand{\partialt}{\frac{\partial}{\partial \tau}}
\newcommand{\intz}{\int_0^1 \rmd z}

\newcommand{\kqq}{\mathcal{P}_{qq}}
\newcommand{\kqg}{\mathcal{P}_{qg}}
\newcommand{\kgq}{\mathcal{P}_{gq}}
\newcommand{\kgg}{\mathcal{P}_{gg}}

\newcommand{\dq}{D_{q_i}}

\newcommand{\dg}{D_g}
\renewcommand{\exp}[1]{e^{#1}}

\definecolor{KKcolor}{rgb}{0,0.3,0}

\definecolor{MRcolor}{rgb}{1,0,1}

\newcommand{\bs}{\boldsymbol}
\def\k{\boldsymbol{k}}
\def\q{\boldsymbol{q}}
\def\r{\boldsymbol{r}}

\def\rmd{{\rm d}}
\def\rme{{\rm e}}

\usepackage{stmaryrd}

\newcommand{\beq}{\begin{eqnarray}}
\newcommand{\eeq}{\end{eqnarray}}
\def\Kc{\EuScript{K}}
\def\lb{{\boldsymbol l}}

\def\p{{\boldsymbol p}}
\def\r{{\boldsymbol r}}

\def\u{{\boldsymbol u}}
\def\v{{\boldsymbol v}}
\def\Q{{\boldsymbol Q}}
\def\P{{\boldsymbol P}}

\newcommand{\nn}{\nonumber\\ }
\def\rme{{\rm e}}

\newcommand{\mincas}{{\sf MINCAS}}
\newcommand{\tmdice}{{\sf TMDICE}}

\renewcommand\mathbf{\boldsymbol}

\begin{document}

\begin{titlepage}

\begin{center}
\end{center}
%

\vspace{5mm}

\begin{center}
    {\LARGE\bf System of evolution equations 
    for quark and gluon 
    \\jet quenching
    with
    broadening$\,^{\star}$} 
\end{center}

\vskip 5mm
\begin{center}
{\large 
  E.\ Blanco$\,{}^a$,
  K.\ Kutak$\,{}^a$,
  W.\ P{\l}aczek$\,{}^b$,
  M.\ Rohrmoser$\,{}^a$ and  
  K.\ Tywoniuk$\,{}^{c,\dagger}$
}
\\
\vskip 2mm
{\em ${}^a\,$Institute of Nuclear Physics, Polish Academy of Sciences,\\
  ul.\ Radzikowskiego 152, 31-342 Krakow, Poland}
\\
\vspace{1mm}
{\em ${}^b\,$Institute of Applied Computer Science, Jagiellonian University,\\
ul.\ {\L}ojasiewicza 11, 30-348 Krakow, Poland}
\\
\vspace{1mm}
{\em ${}^c\,$Department of Physics and Technology, University of Bergen, 5007 Bergen, Norway}
\end{center}
 
\vspace{30mm}
\begin{abstract}
\noindent
We propose a system of evolution equations that describe in-medium time-evolution of transverse-momentum-dependent quark and gluon fragmentation functions. 
Furthermore, we solve this system of equations using  Monte Carlo methods. 
We then quantify the obtained solutions in terms of a few characteristic features, namely the average transverse momentum $\langle |k|\rangle$ and energy contained in a cone, which
allow us to see different behaviour of quark and gluon initiated final-state radiation. 
In particular, the later allows us  to conclude that in the gluon-initiated processes there is less energy in a cone, so that the quark jet is more collimated.

\end{abstract}

\vspace{70mm}
\footnoterule
\noindent
{\footnotesize
${}^{\star}\,$This work is partly supported by
 the Polish National Science Centre grant no. DEC-2017/27/B/ST2/01985.
}\\
\noindent
{\footnotesize
${}^{\dagger\,}$Supported by a Starting Grant from Trond Mohn Foundation (BFS2018REK01) and the University of Bergen.
}

\end{titlepage}


\newpage

\section{Introduction}

Quantum Chromodynamics (QCD) is the well established theory of strong interactions \cite{Ioffe:2010zz}. In heavy-ion collisions, nuclear matter becomes subject to extreme conditions which leads to the appearance of a quark-gluon plasma \cite{Busza:2018rrf}. This new state of matter leads to profound experimental implications, in particular for hard, or high-$p_T$, observables such as jets \cite{dEnterria:2009xfs,Mehtar-Tani:2013pia,Blaizot:2015lma}, a phenomenon referred to as ``jet quenching.''

In this study we plan to focus on certain aspects of jet quenching predicted in \cite{Gyulassy:1990ye,Wang:1991xy}, and observed experimentally at the hadron colliders RHIC \cite{Adler:2002tq} and LHC \cite{Aad:2010bu}. Jet quenching refers generally to the suppression of jets and high-$p_T$ hadrons in heavy-ion collisions due to interactions between hard partons and the quark-gluon plasma (QGP).
This phenomenon is studied using various frameworks: semi-analytical \cite{Baier:1994bd,Baier:1996vi,Zakharov:1996fv,Zakharov:1997uu,Zakharov:1999zk,Baier:2000mf,Gyulassy:2000fs,Guo:2000nz}, see also \cite{Barata:2020rdn,Barata:2020sav,Mehtar-Tani:2019ygg,Mehtar-Tani:2019tvy}, and numerical methods \cite{Caron-Huot:2010qjx,Feal:2018sml,Andres:2020vxs} to approach parton splitting in the medium, kinetic theory \cite{Baier:2000sb,Jeon:2003gi,Arnold:2002ja,Ghiglieri:2015ala,Kurkela:2018wud}, the AdS/CFT based approaches \cite{Liu:2006ug,Ghiglieri:2020dpq} and, finally, Monte Carlo methods \cite{Salgado:2003gb,Zapp:2008gi,Armesto:2009fj,Schenke:2009gb,Lokhtin:2011qq,Casalderrey-Solana:2014bpa,Kutak:2018dim,Blanco:2020uzy}.
In terms of observables, jet quenching induces a broad range of effects that modify the internal jet substructure, contribute to out-of-cone energy loss, the decorrelation of back-to-back jets, and jet thermalization.
In the recent  years the particular interest was focused on the turbulent process of transferring energy from highly energetic gluon jet to soft gluons \cite{Blaizot:2013hx}. This process happens without accumulation of energy by the gluons with moderate values of longitudinal momentum fraction.
One of the recently actively investigated problems is the simultaneous evolution of quarks and gluons in the QGP  studied in connection to turbulent behaviour introduced above. In \cite{Mehtar-Tani:2018zba} it has been demonstrated that the quark to gluon ratio of the soft fragments tends to a universal constant value that is independent of the initial conditions.\\ 
In this paper we would like  generalise this discussion introducing transverse momentum dependence of quarks and gluons. 
Besides confirming the findings of \cite{Mehtar-Tani:2018zba} our analysis will allow to have more detail information about structure of jets as well as to understand better the broadening phenomenon which is directly linked to transverse momentum dependence of fragmentation functions.
In particular, in the recent study by some of us, we solved \cite{Blanco:2020uzy} the equation that takes into account momentum broadening both during branching and via elastic scattering. The resulting distribution is considerably different then the one which accounts only for broadening during elastic scattering. Furthermore, the distributions have harder spectrum than the usually used Gaussian distributions which are used to generate transverse momentum distribution factorised from distribution in longitudinal momentum. 
To address this problem also in the quark case,  we generalise the discussion in~\cite{Blaizot:2013vha,Blaizot:2014rla}  and obtain a system of equations linking quarks and gluons. 
In this approach, QGP is modelled by static centres and a jet interacts with it weakly. The jet propagating through plasma branches according to BDMPS-Z mechanism \cite{Baier:2000mf,Baier:2000sb,Jeon:2003gi,Zakharov:1996fv,Zakharov:1997uu,Zakharov:1999zk,Baier:1994bd,Baier:1996vi,Arnold:2002ja,Ghiglieri:2015ala} and gets broader due to elastic scattering with plasma.  
Furthermore, we solve the equations in full generality, i.e.\ accounting for broadening during branching and due to elastic collisions.

The paper is organised as follows. In the Section~2, we derive branching kernels for quarks and gluons. The kernels allow for splitting of quarks and gluons and also account for transverse momentum dependence. 
In Section~3, we write a system of evolution equations for quarks and gluons and present its formal solution. 
In Section~4, we present distributions resulting from numerical solutions of the equations as well as results for an average transverse momentum and jet energy in a cone as a function of the cone opening angle. 
Then, Section~5 concludes the paper. 
Some further details on the evolution equations and their solutions are
collected in three appendices. In Appendix~\ref{sec:splitting-func}, we give explicit formulae 
for the splitting functions we use in our equations. 
In Appendix~\ref{sec:evol-eq}, we provide evolution equations derived from the general ones after partial integration over some variables.For reference, in Appendix~\ref{sec:evol-first} we provide the first-order perturbative estimate of the full distributions. Finally, Appendix~\ref{sec:numerics} contains brief descriptions of numerical methods used to solve the above equations, 
i.e.\ two Monte Carlo algorithms 
and a method based on the Chebyshev polynomials, 
as well as results of their numerical cross-checks.

\section{Transverse-momentum-dependent splitting kernels}
\label{sec:kernels}

\begin{figure}[!htbp]
    \centering
    \includegraphics[width=0.35\textwidth]{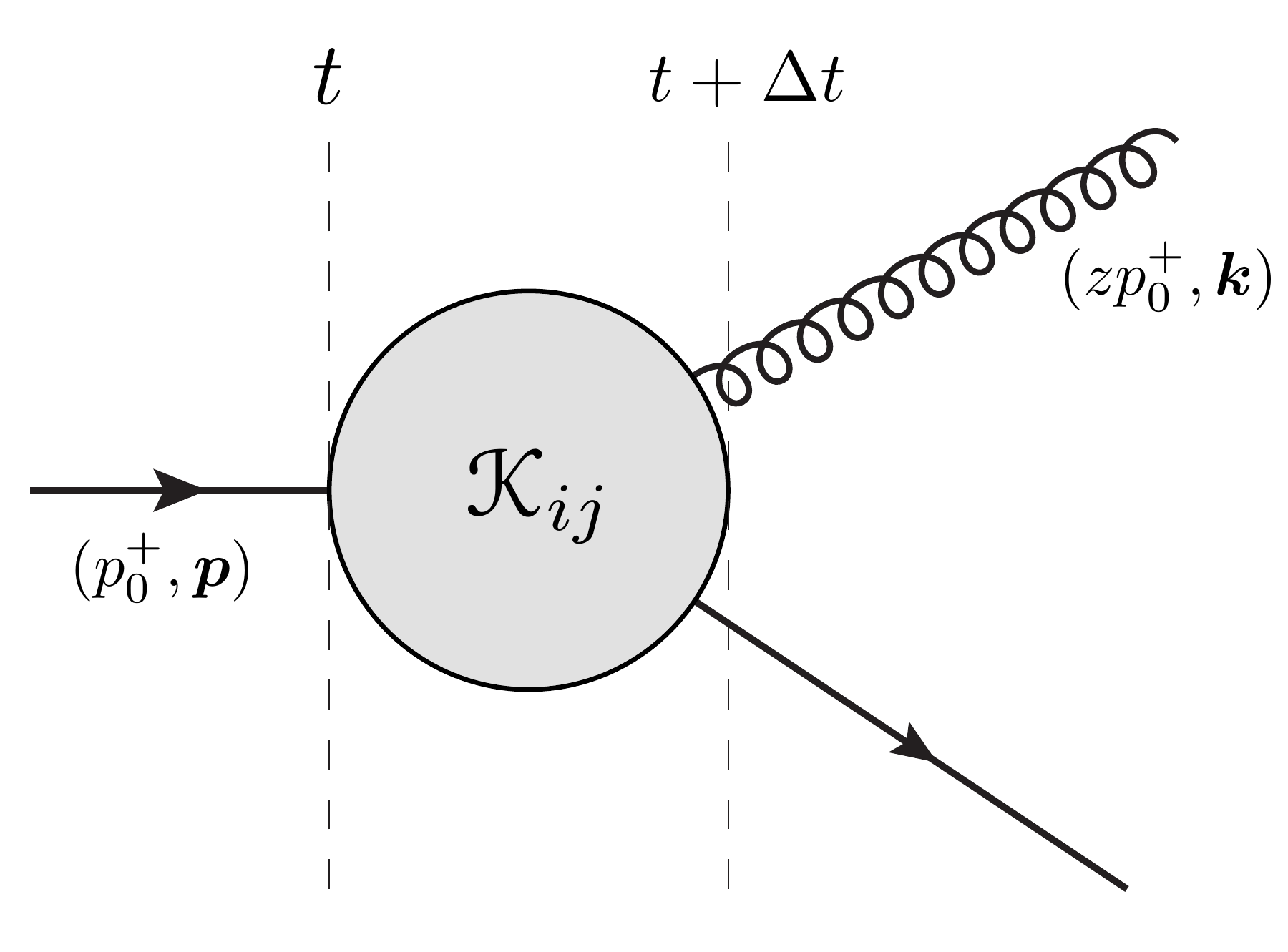}
    \caption{Illustration of the splitting function $\EuScript{K}_{ij}({\bs Q},z,t)$ for the $q \to g+q$ splitting, where $\Q = \k - z\p$.}
    \label{fig:split}
\end{figure}

\begin{figure}[!htbp]
    \centering
    \includegraphics[width=0.2\linewidth]{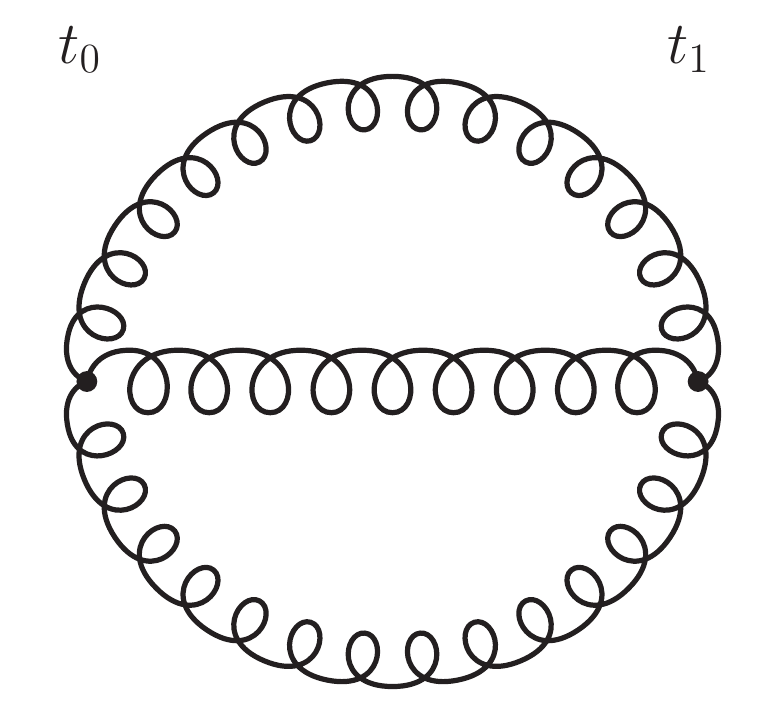}%
    \hspace{2em}
    \includegraphics[width=0.2\linewidth]{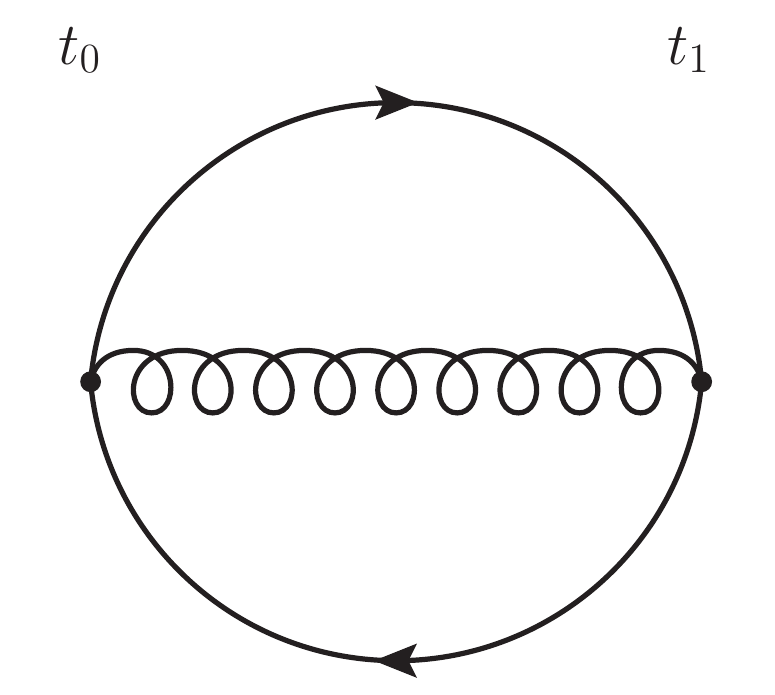}%
    \hspace{2em}
    \includegraphics[width=0.2\linewidth]{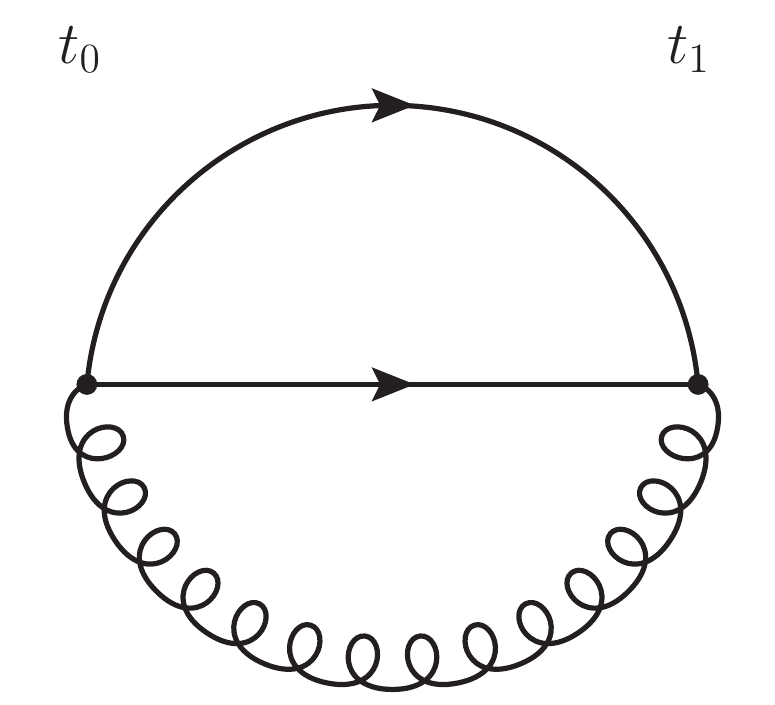}
    \caption{Three of the considered colour-singlet medium correlators contributing to splitting processes in the medium. All lines represent dressed propagators resumming multiple scattering with the medium between time $t_0$, corresponding to the time of splitting in the amplitude, and $t_1$, corresponding to the splitting time in the complex conjugate amplitude. The two upper lines live in the amplitude, and have ``mass'' $\omega_2=(1-z)\omega_0$ and $\omega_1=z\omega_0$, respectively from the top, while the lower line lives in the c.c. amplitude, and carries $\omega_0$.}
    \label{fig:Kernels}
\end{figure}

We shall be interested in computing the $1\to 2$ in-medium splitting kernel, given by \cite{Blaizot:2013vha}
\beq
\Kc_{ij}(\Q,z,p_0^+) = \frac{P^{(k)}_{ij}(z)}{\omega_0^2} {\rm Re} \int_0^\infty \rmd \Delta t \int \frac{\rmd^2 \P}{(2\pi)^2} \frac{\rmd^2 \lb}{(2\pi)^2}
\, ( \P \cdot \Q) \tilde S_{ij}^{(3)}( \P,  \Q,\lb,z,\Delta t,t) \,,
\eeq
where $\omega_0 = z(1-z) p_0^+$, and $P^{(k)}_{ij}(z)$ are the (unregularised) Altarelli--Parisi splitting functions, for different QCD splitting processes.\footnote{As usual, the index ``$j$'' refers to the parton that is splitting, and ``$i$'' refers to the parton that takes the momentum fraction $z$. The parton with an index ``$k$'' takes the momentum fraction $1-z$.}
The three-point correlator $\tilde S_{ij}^{(3)}$ in momentum space, reads
\begin{align}
\label{eq:3b-S}
S_{ij}^{(3)}( \P,  \Q,\lb,z,\Delta t,t) &= \int \rmd^2 \u_1 \rmd^2 \u_2 \rmd^2 \v \, {\rm e}^{i \u_1\cdot  \P - i \u_2 \cdot   \Q - i \v \cdot \lb}\, {\cal I}_{ij}(\u_2,\Delta t+t;\u_1,t) \,,
\end{align}
where ${\cal I}_{ij}$ refers to the path integral,
\begin{equation}
{\cal I}_{ij}(\u_2,t_2;\u_1,t_1) = \int_{\u(t_1) =\u_1}^{\u(t_2) =\u_2} {\cal D} \u \, \rme^{ i\frac{\omega_0}{2} \int_{t_1}^{t_2} \rmd s\, \dot\u^2(s) - \int_{t_1}^{t_2} \rmd s \, n(s) \sigma_{\rm eff} ( \u(s),\v )}\,.
\end{equation}
Here, $n(s)$ is the density of scattering centres in the medium, and 
\label{eq:sigmaeff_2}
\beq
\sigma_{\rm eff}(\u,\v) = \frac{C_i + C_k - C_j}{2}\, \sigma(\u) + \frac{C_i + C_j - C_k}{2}\, \sigma(\v +(1-z) \u) + \frac{C_k + C_j - C_i}{2}\, \sigma(\v -  z \u) \,.
\eeq
This path integral describes the relative motion of the three internal lines of the three-point correlator during the time interval $t_2 - t_1$. This can be made clear if we introduce the variables $\u = \r_i - \r_k$ and $\v = z\r_i + (1-z) \r_k - \r_j$. In this case, the effective potential takes the form 
\beq
\label{eq:sigmaeff_1}
\sigma_{\rm eff}(\r_0,\r_1,\r_2) = \frac{C_i + C_k - C_j}{2}\, \bar\sigma(\r_k - \r_i) + \frac{C_i + C_j - C_k}{2}\, \bar\sigma(\r_i - \r_j) + \frac{C_k + C_j - C_i}{2}\, \bar\sigma(\r_k - \r_j) \,,
\eeq
where 
\begin{equation}
\label{eq:3body-potential}
    \bar\sigma(\r) = \int \frac{\rmd^2 \q}{(2\pi)^2} \left( 1-\rme^{i \q \cdot \r} \right) \bar w(\q)\,,
\end{equation}
and $\bar w(\q) = \rmd^2\bar\sigma_{\rm el}/\rmd^2 \q$ is the elastic scattering potential of the medium stripped of the relevant colour factor (e.g. $w_g(\q) = N_c \bar w(\q)$ and $w_q(\q)= C_F \bar w(\q)$). In this work, we work with the thermal HTL potential, see \eqref{eq:wgql2} below. In Eqs.~\eqref{eq:sigmaeff_1} and \eqref{eq:3body-potential}, the $C_i$'s are the squared Casimir operators of the colour representation of the three correlated particles. Equation~\eqref{eq:3body-potential} can be proven directly by writing down the relevant 3-point functions for each individual splitting in Fig.~\ref{fig:Kernels}. However, a more general argument relies on the properties of colour conservation \cite{Zakharov:1997uu}, and can be extended to higher-order correlators as well \cite{Arnold:2019pxd}. Basically, the subtraction terms in \eqref{eq:sigmaeff_1} correspond to the {\it combined} colour representations of the two interacting particles. For a 3-body problem, and due to colour conservation in the medium, the subtraction terms, e.g.\ $\sim -C_0 \sigma(\r_i-\r_k)$, necessarily have to be related to the Casimir of the particle not involved in the scattering.  

For a medium with constant density, i.e.\ $n(t) = n_0$ for $0<t<L$, and in the harmonic oscillator approximation, $n_0 \bar \sigma(\r) \approx \hat{\bar q} \r^2/4$, we get an effective jet quenching parameter 
\beq
\hat q_{ij}(z) = f_{ij}(z) \hat{\bar q}\,,
\eeq
where
\beq
f_{ij}(z) =\frac{C_i + C_k - C_j}{2} + \frac{C_i + C_j - C_k}{2}\, (1-z)^2 + \frac{C_k + C_j - C_i}{2} \,z^2 \,.
\eeq
Explicitly, we have
\begin{equation}
\begin{aligned}
    f_{gg}(z) &= (1-z) C_A + z^2 C_A \,, \\
    f_{qg}(z) &= C_F - z(1-z) C_A \,,\\
    f_{gq}(z) &= (1-z) C_A + z^2 C_F \,,\\
    f_{qq}(z) &= z C_A + (1-z)^2 C_F \,.
\end{aligned}
\label{eq:fij}
\end{equation}

It is worth keeping in mind that, in the jet quenching literature, the jet transport parameter $\hat q$ often refers to the gluon contribution, i.e.\ $\hat q \equiv \hat q_A = N_c \hat{\bar q}$.
To summarise the results, for processes involving a gluon emission, i.e.\ $R \to g+R$, where the gluon takes away the momentum fraction $z$, we get
\beq
\hat q_{gR}(z) = \frac{N_c}{2} \left[1+ \left(\frac{2 C_R}{N_c} -1 \right) z^2 + (1-z)^2 \right] \hat{\bar q} \,,
\eeq
and in the special case of $g \to q + \bar q$, we get
\beq
\hat q_{qg}(z) = \frac{N_c}{2} \left[ \left(\frac{2 C_F}{N_c} -1 \right)+ z^2 + (1-z)^2 \right] \hat{\bar q} \,.
\eeq
With these approximations, the solution to the path integral ${\cal I}_{ij}(\u_2;\u_1)$ can be written as
\beq
{\cal I}_{ij}(\u_2;\u_1) = \frac{\omega_0 \Omega_{ij}}{2\pi i \sinh \Omega_{ij} \Delta t} \rme^{ i\frac{\omega_0 \Omega_{ij}}{4}\left[ (\u_2 - \u_1)^2 \coth \frac{\Omega_{ij} \Delta t}{2}  \,+ \,(\u_2+\u_1)^2\tanh \frac{\Omega_{ij} \Delta t}{2}  \right] }\,,
\eeq
where 
$\Omega_{ij} = \frac{1+i}{2} \sqrt{f_{ij}(z)\hat{\bar q}/ \omega_0}$. 
After performing the Fourier transforms in Eq.~\eqref{eq:3b-S}, we finally obtain
\beq
S^{(3)}_{ij}( \P,  \Q,z,\Delta t,t) = \frac{2 \pi i}{\omega_0 \Omega_{ij}  \sinh \Omega_{ij} \Delta t} \rme^{- \frac{i}{4\omega_0 \Omega_{ij}} \left[( \P +  \Q)^2\tanh \frac{\Omega_{ij} \Delta t}{2}  \,+\, ( \P -  \Q )^2 \coth \frac{\Omega_{ij} \Delta t}{2}  \right]} \,,
\eeq
which, within these approximations, is time-independent.
Finally, we integrate out $\P$ and $\Delta t$ to obtain the splitting function
\begin{equation}
\Kc_{ij}( \Q,z,p_0^+) = \frac{2 P_{ij}(z)}{z(1-z)p_0^+}\, \sin\left(\frac{\Q^2}{2 k_{\rm br}^2}\right)
\mathrm{exp}\left({-\frac{ \Q^2}{2 k_{\rm br}^2}}\right) \,,
\label{eq:Kij_final}
\end{equation}
where $k^2_{\rm br} = \sqrt{z(1-z)p_0^+ f_{ij}(z) \hat{\bar q}}$ is the typical transverse momentum accumulated during the time it takes to split, also called the formation time. This agrees with the expression derived in \cite{Blaizot:2012fh}.

To make contact with previous works, that did not include the transverse-momentum dependence in the splitting function, we can also rewrite Eq.~\eqref{eq:Kij_final} as
\begin{equation}
\label{eq:Kij-final-2}
\alpha_s \Kc_{ij}(\Q,z,p_0^+) = \frac{1}{2 \bar t_\ast} \, P_{ij}(z) \sqrt{\frac{f_{ij}(z)}{z(1-z)}} \, \mathcal{R}_{ij}(\Q,k_{\rm br}^2) \,,
\end{equation}
where
\begin{equation}
    \mathcal{R}_{ij} (\Q,k_{\rm br}^2) = \frac{4\pi}{ k_{\rm br}^2} \sin \left(\frac{\Q^2}{2 k_{\rm br}^2} \right) \mathrm{exp}\left({-\frac{\Q^2}{2 k_{\rm br}^2}}\right) \,.
\end{equation}
The factor $\mathcal{R}_{ij}(\Q,k_{\rm  br}^2)$ represents the broadening that takes place during the formation time of the splitting which is characterised by $\langle k_\perp^2 \rangle \approx k_{\rm br}^2$, and is normalised
\begin{equation}
    \int \frac{\rmd^2 \Q}{(2\pi)^2} \mathcal{R}_{ij}(\Q,k_{\rm br}^2) = 1 \,.
\end{equation}
Note also that this distribution reduces to a Dirac $\delta$-function when the transverse momentum accumulated during the formation time $k_{\rm br}^2$ tends to zero, $\lim_{k_{\rm br}^2 \to 0} \mathcal{R}_{ij}(\Q,k_{\rm br}^2) = (2\pi)^2 \delta(\Q)$.

Finally, in Eq.~\eqref{eq:Kij-final-2}, we have also defined the analog of the stopping time for a jet with $p_0^+$, namely
\begin{equation}
    \bar t_\ast \equiv \frac{\pi}{\alpha_s} \sqrt{\frac{p_0^+}{\hat{\bar q}}} \,.
\end{equation}
When dealing with both quark and gluon contributions to the splitting processes, this expression is stripped of the relevant colour factors. For purely gluon cascades, the correct stopping time is rather $t_\ast = N_c^{-3/2}\, \bar t_\ast$.


\section{Evolution equations and their formal  Monte Carlo solutions}
\label{sec:EvolEqs}
Using the branching kernels of the previous sections together with scattering kernels it is possible to obtain a system of equations for the 
evolution with time $t$ of fragmentation functions 
 $D_a$ for particles of type $a$ ($a=g$ for gluons or $a=q_i$ for quarks and antiquarks of the flavour $i$) or equivalently of multiplicity distributions $F_a$, which are defined as
\begin{equation}
    D_a(x,\mathbf{k},t):=xF_a(x,\mathbf{k},t)\,,\qquad \textrm{with }\:F_a(x,\mathbf{k},t):=\frac{\rmd^3 N_a}{\rmd x\,\rmd^2 \mathbf{k}}\,.
\end{equation}
%
%

In this section we formulate the evolution equations for the fragmentation functions and describe their Monte Carlo solutions (implemented in the \mincas\ program), while the equivalent equations for multiplicity distributions have analogous Monte Carlo solutions (implemented in the \tmdice\ program).
The evolution equations for the fragmentation functions can be written as
%
%
\begin{equation}
\begin{aligned}
\frac{\partial}{\partial t} D_g(x,\mathbf{k},t) = & \:  \int_0^1 \rmd z\, \int\frac{\rmd^2 \mathbf{q}}{(2\pi)^2} \alpha_s \bigg\{2\Kc_{gg}\left(\mathbf{Q},z,\frac{x}{z}p_0^+\right) D_g\left(\frac{x}{z},\mathbf{q},t\right) 
+ \Kc_{gq}\left(\mathbf{Q},z,\frac{x}{z}p_0^+\right) \sum_i  D_{q_i}\left(\frac{x}{z},\mathbf{q},t\right) 
\\&
-  \Big[ \Kc_{gg}(\mathbf{q},z,xp_0^+) +
\Kc_{qg}(\mathbf{q},z,xp_0^+)\Big]\, D_g(x,\mathbf{k},t) \bigg\} 
+ \int \frac{\rmd^2\mathbf{l}}{(2\pi)^2} \,C_g(\mathbf{l})\, D_g(x,\mathbf{k}-\mathbf{l},t), \\
\frac{\partial}{\partial t} D_{q_i}(x,\mathbf{k},t) = & \:  
 \int_0^1 \rmd z\, \int\frac{\rmd^2\mathbf{q}}{(2\pi)^2} \alpha_s\bigg\{ \Kc_{qq}\left(\mathbf{Q},z,\frac{x}{z}p_0^+\right) D_{q_i}\left(\frac{x}{z},\mathbf{q},t\right) 
+ \frac{1}{N_F} \Kc_{qg}\left(\mathbf{Q},z,\frac{x}{z}p_0^+\right) D_g\left(\frac{x}{z},\mathbf{q},t\right) 
\\&
-  \Kc_{qq}(\mathbf{q},z,xp_0^+)\, D_{q_i}(x,\mathbf{k},t) \bigg\} 
+ \int \frac{\rmd^2\mathbf{l}}{(2\pi)^2} \,C_q(\mathbf{l})\, D_{q_i}(x,\mathbf{k}-\mathbf{l},t)\,,
\end{aligned}
\label{eq:BDIMsys1}
\end{equation}
where $\mathbf{Q} = \mathbf{k} - z \mathbf{q}$, and the index $i$ runs over all active quarks and antiquarks ($i = 1,\ldots, 2N_F$ where $N_F$ is the number of active quark flavours in the cascade). 
The strong coupling constant $\alpha_s$ in Eq.~\eqref{eq:BDIMsys1} is a function of the relative transverse momentum $\Q^2 \approx k_{\rm br}^2$. However, we treat it here as constant: $\alpha_s \approx 0.3$ (for the concrete parameter choices, see below).
The elastic collision kernel $C_{q(g)}(\mathbf{l})$ is given by
\begin{equation}
C_{q(g)}(\mathbf{l}) = w_{q(g)}(\mathbf{l}) - \delta(\mathbf{l}) \int\rmd^2\mathbf{l}'\,w_{q(g)}(\mathbf{l}')\,,
\label{eq:Cl}
\end{equation}
where
\begin{equation}
w_g(\mathbf{l})=\frac{N_c \, g^4 n}{\mathbf{l}^2(\mathbf{l}^2 + m_D^2)},\qquad 
w_q(\mathbf{l})=\frac{C_F\, g^4 n}{\mathbf{l}^2(\mathbf{l}^2 + m_D^2)}\,,
\label{eq:wgql2}
\end{equation}
is the HTL in-medium potential, where $m_D$ is the Debye mass and $n$ the density of scattering centres in a thermal medium. At leading-order, they are given by $m_D^2 = (1+n_f/6)g^2 T^2$ and $n = m_D^2 T/g^2$, where $n_f$ is the number of active flavours in the medium. The coupling to the medium $g$ should be evaluated at the scale of the temperature $\sim 2\pi T$. In this work, we however keep it fixed at the same value as the coupling in the medium-cascade, namely $g = (4\pi \alpha_s)^{1/2} \approx 2$, see the concrete parameter choices below. For the HTL potential, the bare jet transport coefficient is then $\hat{\bar q} = 4\pi\alpha_s^2 n$, see e.g. \cite{Barata:2020sav}.

In the limit of the gluon-dominated cascade, where the quark contributions can be neglected, one obtains the following evolution equation:
\begin{equation}
\begin{aligned}
\frac{\partial}{\partial t} D_g(x,\mathbf{k},t) = & \:   \int_0^1 \rmd z\, \int\frac{\rmd^2 \mathbf{q}}{(2\pi)^2} \alpha_s \left[2\Kc_{gg}\left(\mathbf{Q},z,\frac{x}{z}p_0^+\right) D_g\left(\frac{x}{z},\mathbf{q},t\right) 
- \Kc_{gg}(\mathbf{q},z,xp_0^+)\, D_g(x,\mathbf{k},t) \right] \\
&+ \int \frac{\rmd^2 \mathbf{l}}{(2\pi)^2} \,C_g(\mathbf{l})\, D_g(x,\mathbf{k}-\mathbf{l},t) \,,
\end{aligned}
\label{eq:BDIM1}
\end{equation}
which agrees with previous results \cite{Blaizot:2013vha}. In this work, we study the interplay between quark and gluon degrees of freedom in the cascade.


%

%

In order to solve the evolution equations (\ref{eq:BDIMsys1}) with Markov Chain Monte Carlo (MCMC) methods, 
first we need to transform them into the form of integral equations of the Volterra type, 
similarly as in Ref.~\cite{Kutak:2018dim} for the pure gluon case. 
We start from introducing some useful notation that will facilitate expressing the corresponding equations
in a compact and transparent form.

For transparency in the set of coupled evolution equations, let us now redefine the indices $I,J$ to run over all parton flavours, i.e.\ quarks, antiquarks and gluons, so that
\begin{equation}
I,J \in \{q_1,\ldots,q_{N_F},\bar{q}_1,\ldots,\bar{q}_{N_F},g\}
\label{eq:qgIJ}
\end{equation}
and new $\mathbf{Q}$-dependent evolution kernels $\mathcal{K}_{IJ}(z,y,\mathbf{Q})$, defined as
\begin{align}
\mathcal{K}_{IJ}(z,y,\Q) &= 
 \frac{\alpha_s}{(2\pi)^2}\, (1+\delta_{Ig}\delta_{Jg})\, \bar t_\ast\, z\Kc_{IJ}(\Q,z,y p_0^+) \,,\\
&= \frac{z}{\sqrt{y}} (1+\delta_{Ig}\delta_{Jg})\,
\mathcal{P}_{IJ}(z) \, \mathcal{\tilde R}_{IJ}(\Q,z,y p_0^+) \,,
\label{eqKIJzyQ}
\end{align}
where $\mathcal{P}_{IJ}(z)$ are the $z$-dependent in-medium splitting functions,\footnote{In earlier works, the $z$-dependent splitting functions were denoted by $\mathcal{K}_{IJ}(z)$. Here, we have changed the notation in order to avoid confusion with the $z$ and $\Q$ dependent splitting functions $\mathcal{K}_{IJ}(z,y,\Q)$.} $\mathcal{\tilde R}_{IJ}(\Q,z,p_0^+) = \mathcal{R}_{IJ}(\Q,z,p_0^+)/(2\pi)^2$
and 
$\Kc_{gq_i}\equiv \Kc_{gq}$, 
$\Kc_{q_ig} \equiv \Kc_{qg}/N_F$, 
and $\Kc_{q_iq_j}\equiv \delta_{ij}\Kc_{qq}$.
The factor $(1+\delta_{IG}\delta_{JG})$ accounts for the symmetry factor that appears in the gluon--gluon splitting.
The expressions for the full set of the splitting functions are given in Appendix~\ref{sec:splitting-func}.



Then, let us introduce the Sudakov form-factor $\Psi_I(x)$ that resums all unresolved branchings and scatterings:
\begin{equation}
\Psi_I(x) = \Phi_I(x) + W_I\,,
\label{eq:qgPsix}
\end{equation}
where
\begin{equation}
\Phi_{I}(x) = \sum_{J} \Phi_{JI}(x) \,,\qquad
\Phi_{JI}(x) = \int_0^{1-\epsilon} \rmd z \int \rmd^2\mathbf{Q}\, \mathcal{K}_{JI}(z,x,\mathbf{Q})
\label{eq:qgPhixI}
\end{equation}
and
\begin{equation}
W_I  =  \bar t_\ast \int_{|\mathbf{l} |>l_{\mathrm{min}}} \frac{\rmd^2 \mathbf{l}}{(2\pi)^2} \,w_I(\mathbf{l})\,.
\label{eq:WI}
\end{equation}
The full branching--scattering kernel can be expressed as
\begin{equation}
\EuScript{G}_{IJ}(z,y,\mathbf{Q},\mathbf{l}) =
\mathcal{K}_{IJ}(z,y,\mathbf{Q})\theta(1 - \epsilon - z)\delta(\mathbf{l}) +  
\bar t_\ast \frac{w_I(\mathbf{l})}{(2\pi)^2} \,\theta(|\mathbf{l}| -l_{\mathrm{min}})\,\delta(1-z) \delta(\mathbf{Q})\delta_{IJ}\,.
\label{eq:qgesG}
\end{equation}
The analogous branching--scattering kernel $\tilde{\EuScript{G}}_{IJ}$ of the evolution equations of the multiplicity distributions $F_I$ can be obtained with the sole replacement of $\mathcal{K}_{IJ}(z,y,\mathbf{Q})\mapsto \mathcal{\tilde K}_{IJ}(z,y,\Q) =\mathcal{K}_{IJ}(z,y,\mathbf{Q})/z$. 

With the above notation and after introducing a dimensionless evolution time $\tau = t/\bar t_\ast$, Eq.~(\ref{eq:BDIMsys1}) can be cast in a simple form, namely
\begin{equation}
\begin{aligned}
 \frac{\partial}{\partial \tau} D_I(x,\mathbf{k},\tau)\; +\;  &  \Psi_I(x) D_I(x,\mathbf{k},\tau) 
 =  \int_0^1 \rmd y \int_0^1 \rmd z \int \rmd ^2\mathbf{k'} \int \rmd ^2\mathbf{Q} \int \rmd ^2\mathbf{l} 
\\ & \times \delta(x - zy) \,\delta(\mathbf{k} - \mathbf{l} - \mathbf{Q} - z\mathbf{k'}) \,
 \sum_J  \EuScript{G}_{IJ}(z,y,\mathbf{Q},\mathbf{l})\, D_J(y,\mathbf{k'},\tau)\,.
\end{aligned}
\label{eq:qgkQee}
\end{equation}
Their formal solution in terms of the Volterra-type integral equations reads
\begin{equation}
\begin{aligned}
D_I(x,\mathbf{k},\tau) = &\,
D_I(x,\mathbf{k},\tau_0)\, e^{-\Psi_I(x)(\tau-\tau_0)}
\\&  
+  \sum_J  \int_{\tau_0}^{\tau}\rmd \tau'  \int_0^1 \rmd y\, \int_0^1 \rmd z \int \rmd ^2\mathbf{k'} \int \rmd ^2\mathbf{Q} 
   \int \rmd ^2\mathbf{l} \;
  \EuScript{G}_{IJ}(z,y,\mathbf{Q},\mathbf{l}) \, D_J(y,\mathbf{k'},\tau')
\\& \hspace{10mm}  
  \times e^{-\Psi_I(x)(\tau-\tau')}\,
  \delta(x-zy)\,\delta(\mathbf{k} - \mathbf{l} - \mathbf{Q} - z\mathbf{k'})\,,
\end{aligned}
\label{eq:qgeeSol}
\end{equation}
where $\tau_0 = t_0/\bar t_\ast$ is the initial time for the evolution.
The above integral equations can be solved numerically by iteration,
\begin{equation}
\begin{aligned}
 D_I(x,\mathbf{k},\tau)  = & \sum_{J_0} \int_0^1 \rmd x_0 \int \rmd ^2 \mathbf{k}_0 \,
 \Bigg\{
 e^{-\Psi_{J_0}(x_0)(\tau - \tau_0)}\,\delta_{IJ_0}\, \delta(x-x_0)\,\delta(\mathbf{k}-\mathbf{k}_0)
 \\ &
 + \sum_{n=1}^{\infty}\sum_{J_1,J_2,\ldots,J_n} 
 \prod_{i=1}^n \bigg[ \int_{\tau_{i-1}}^{\tau} \rmd \tau_i \, \int_0^1 \rmd z_i\, \int \rmd ^2\mathbf{Q}_i\, \int \rmd ^2\mathbf{l}_i\,
 \\ & \hspace{35mm}
 \EuScript{G}_{J_iJ_{i-1}}(z_i,x_{i-1},\mathbf{Q}_i,\mathbf{l}_i)\,e^{-\Psi_{J_{i-1}}(x_{i-1})(\tau_i - \tau_{i-1})} \bigg] 
 \\&\hspace{5mm}
\times e^{-\Psi_{J_n}(x_n)(\tau - \tau_n)} \, \delta_{J_nI}\,\delta(x-x_n)\,\delta(\mathbf{k}-\mathbf{k}_n) 
\Bigg\}\,
D_{J_0}(x_0,\mathbf{k}_0,\tau_0) \,,
\end{aligned}
\label{eq:qgkQitsol}
\end{equation}
where 
\begin{equation}
x_i = z_i x_{i-1}, \qquad \mathbf{k}_i = \mathbf{Q}_i + \mathbf{l}_i + z_i\mathbf{k}_{i-1}\,.
\label{eq:kQxiki}
\end{equation}
Similar solutions can be obtained for the evolution equations (\ref{eq:eeintQ}) and (\ref{eq:eeintkqg})
given in Appendix~\ref{sec:evol-eq}.
In the case of Eq.~(\ref{eq:eeintQ}) one only needs to replace in Eq.~(\ref{eq:qgesG}):
$\mathcal{K}_{IJ}(z,y,\mathbf{Q}) \rightarrow (1/\sqrt{y}) z{\mathcal{K}}_{IJ}(z)$,
while for Eq.~(\ref{eq:eeintkqg}) in addition set $w_I(\mathbf{l})=0$.
The most efficient way of numerical evaluations of the above iterative solutions is by employing 
the Markov Chain Monte Carlo (MCMC) methods, similar as in Ref.~\cite{Kutak:2018dim}. 
These methods as well as the appropriate algorithms are described in Appendix~\ref{sec:numerics}.

\section{Numerical results}
\label{sec:NumRes}

In this section we present results that we obtained as solutions of equations (\ref{eq:BDIMsys1}). 
We plot transverse 2D distributions as well as their projections.
Furthermore, we use the solutions to construct 
characteristic features that allow us to 
understand better the physical effects of transverse momentum broadening and differences between quark and gluon jets.
Further numerical results concerning cross-checks of different methods and programs
used for solving of the above equations are presented in Appendix~\ref{sec:numerics}.

Our numerical results have been obtained for the following values of the input parameters:
\begin{eqnarray}
x_{\rm min} &=& 10^{-4}, \qquad \epsilon = 10^{-6} , \qquad l_{\rm min} = 0.1\,{\rm GeV},
\label{eq:qgxpar} \\
N_c &=& 3, \quad N_F = 3, \quad \alpha_s = \pi/10,
\label{eq:qgalbar} \\
E &=& 100\,{\rm GeV}, \quad n = 0.243 \,{\rm GeV}^{3}, \qquad \hat{\bar{q}} = 1 \,{\rm GeV^2/fm}, \qquad m_D = 0.993\,{\rm GeV}.
\label{eq:qgnqhpar}
\end{eqnarray}
%
In the case of the initial gluon the starting distributions are:
\begin{equation}
    D_g(x,\k,t=t_0) \equiv D_g^{(0)}(x,\k) = x \delta(1-x) \delta(\k) \,, \qquad D_S(x,\k,t=t_0) \equiv D_S^{(0)}(x,\k) = 0 \,, 
\label{eq:qginglu}
\end{equation}
while the case of the initial quark:
\begin{equation}
    D^{(0)}_g(x,\k) = 0\,, \qquad D^{(0)}_S(x,\k) = x \delta(1-x)\delta(\k) \,,
\label{eq:qginqua}
\end{equation}
where the index $g$ denotes gluons while $S$ -- the quark-singlet, i.e.\ the sum of active quarks and antiquarks.\\

\subsection{Results for fragmentation functions}

As a first result, we present solutions of system of equations that follows from our equations after one performs the integral over transverse momentum. The goal of this calculation is to obtain by independent calculations results presented in \cite{Mehtar-Tani:2018zba}. The integrated fragmentation functions reads
\begin{equation}
    \label{eq:DI-definition}
    D_I(x,t) = \int \rmd^2\mathbf{k} \,D_I(x,\mathbf{k},t) \,.
\end{equation}
Since we present the results in terms of $k_T\equiv |\k|$ dependence, we introduce the distribution
\begin{equation}
    \Tilde{D}_I(x,k_T,t) = \int_0^{2\pi} 
    \rmd \phi\,k_T\,D_I(x,\mathbf{k},t) \text{, \quad such that\:\:}
    D_I(x,t) = \int \rmd k_T\, \Tilde{D}_I(x,k_T,t)\,.
    \label{eq:Dtilde}
\end{equation}
To solve the system of the evolution equations 
we have used two Monte Carlo programs, {\mincas}  and {\tmdice}, as well as the numerical method 
based on the Chebyshev polynomials. 
The corresponding algorithms are explicitly presented in Appendix~\ref{sec:numerics}. 
The results of Monte Carlo programs agree very well, therefore here we present the results
obtained only by \mincas. 
The plots comparing the Monte Carlo solutions obtained by {\mincas}  and {\tmdice} are shown in the Appendix~\ref{sec:numerics}. Furthermore, we compare the distributions to a perturbative estimate in Appendix~\ref{sec:evol-first}.

\begin{figure}[!t]
    \centering
    \includegraphics[scale=0.4]{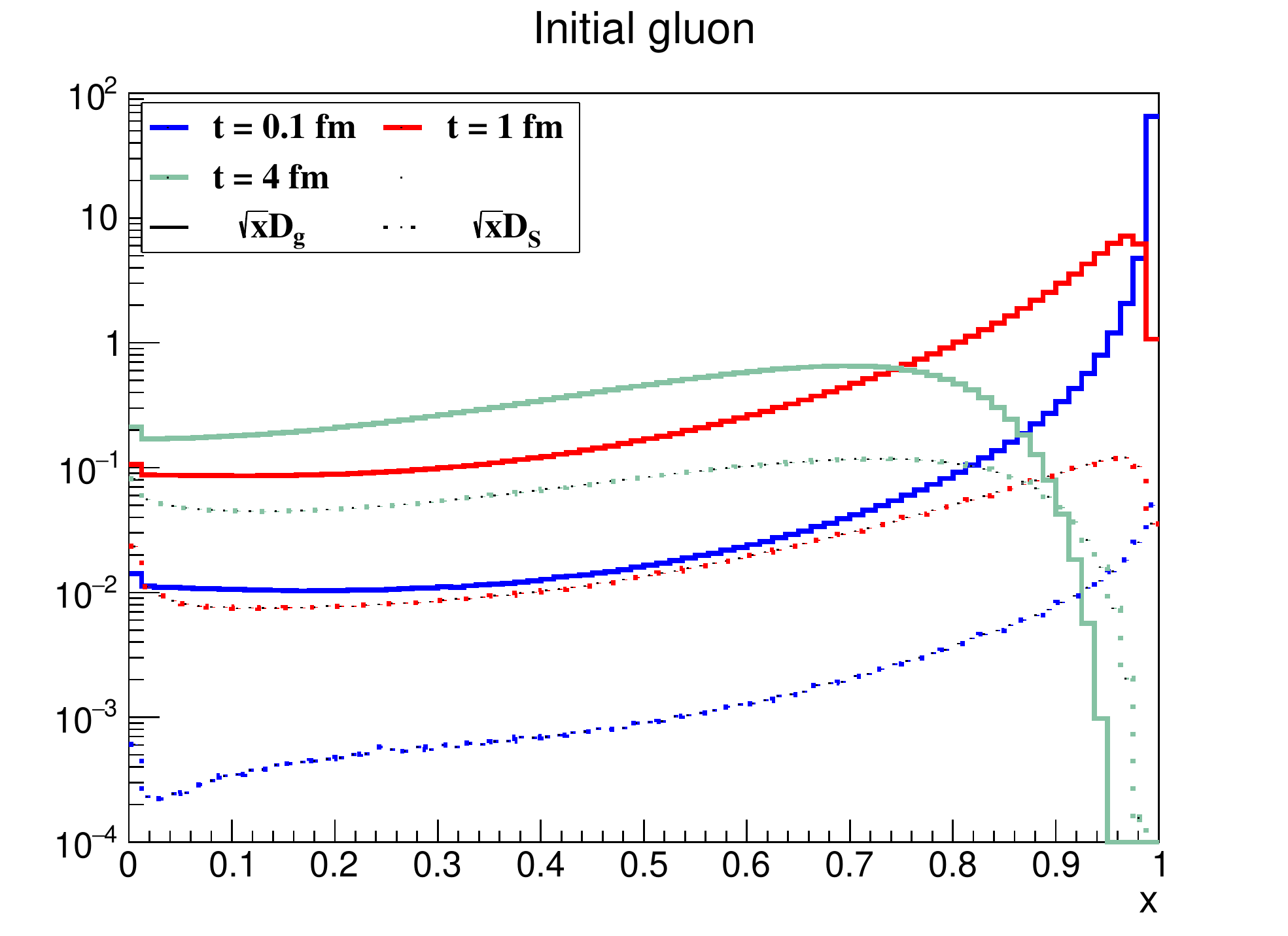}
    \includegraphics[scale=0.4]{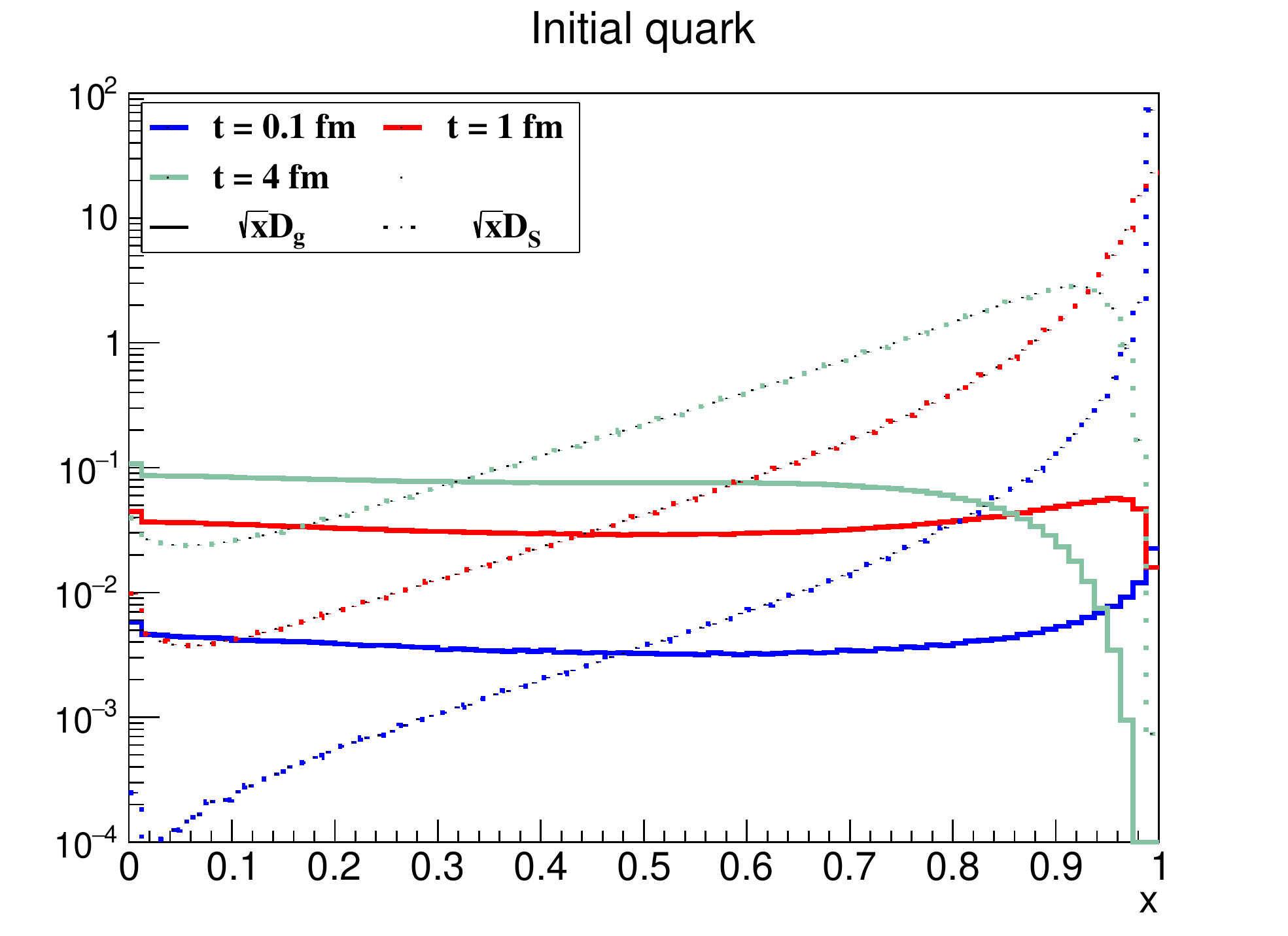}
    \caption{The $\sqrt{x}D(x,t)$ distributions 
    at the time-scales $t=0.1, 1, 4\,$fm:
    cascades initiated by gluon (left) and quark (right). The dashed lines correspond to the quark distributions while the solid lines to the gluon distributions.}
    \label{fig:x_kzq_wq2}
\end{figure}

Furthermore, by visually comparing our results in Fig.~\ref{fig:x_kzq_wq2} 
to the ones presented in \cite{Mehtar-Tani:2018zba}, we see 
that we get the same features of the distributions, i.e.\ energy is not accumulated at the moderate values of $x$. Simultaneously, the distributions increase at small $x$ values, following roughly the $1/\sqrt{x}$ behaviour for the gluons and the quark-singlet for the case of the initial gluon and a similar behaviour for the case of the initial quark. In the case of the initial gluon at late times there is a region, at high $x$, where quarks dominate. 
In the case of the initial quark, gluons tend to dominate if $x$ is low for all time scales, while quarks dominate at $x>0.5$.

\begin{figure}[!htbp]
    \centering
    \includegraphics[scale=0.4]{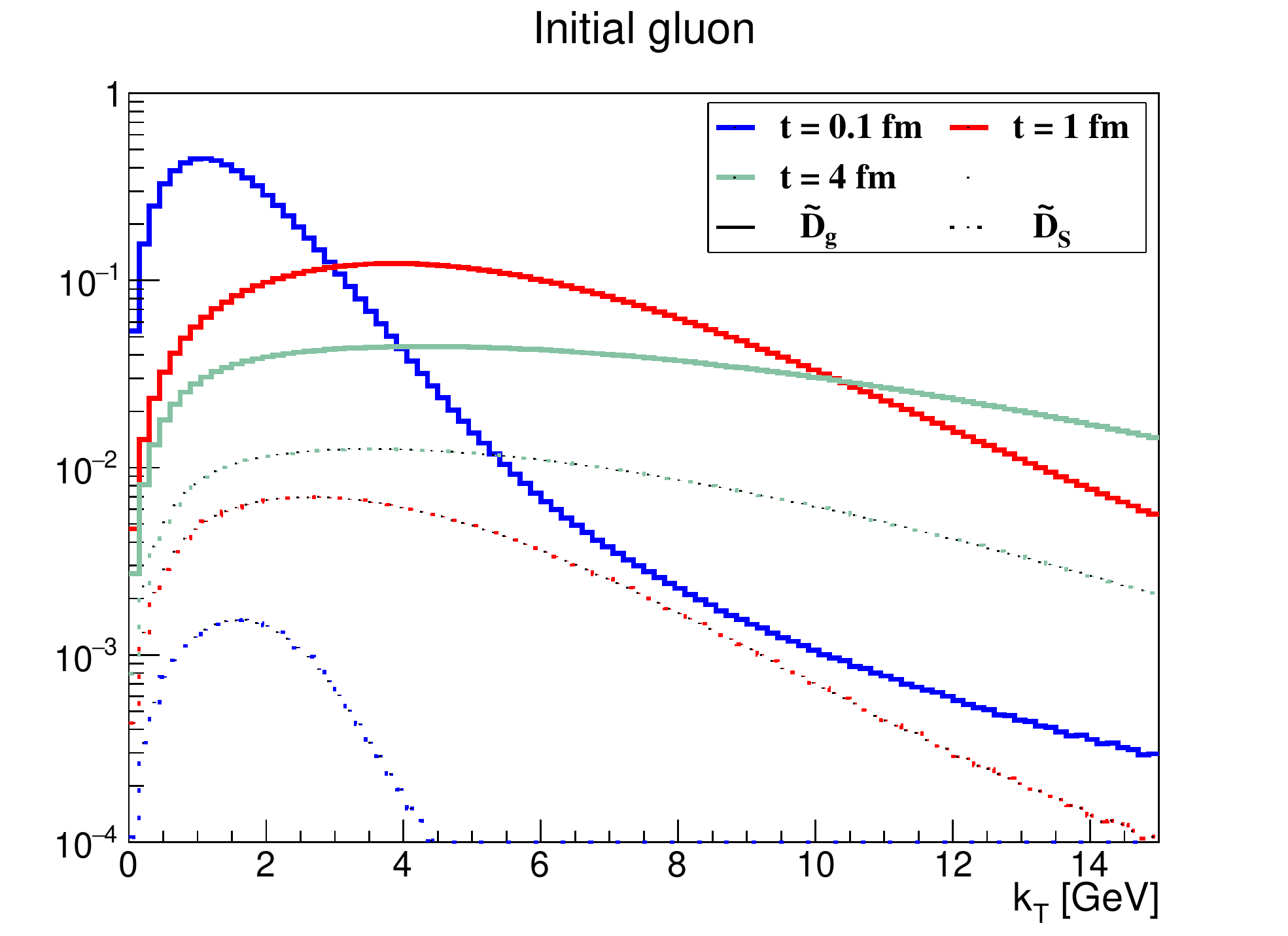}
    \includegraphics[scale=0.4]{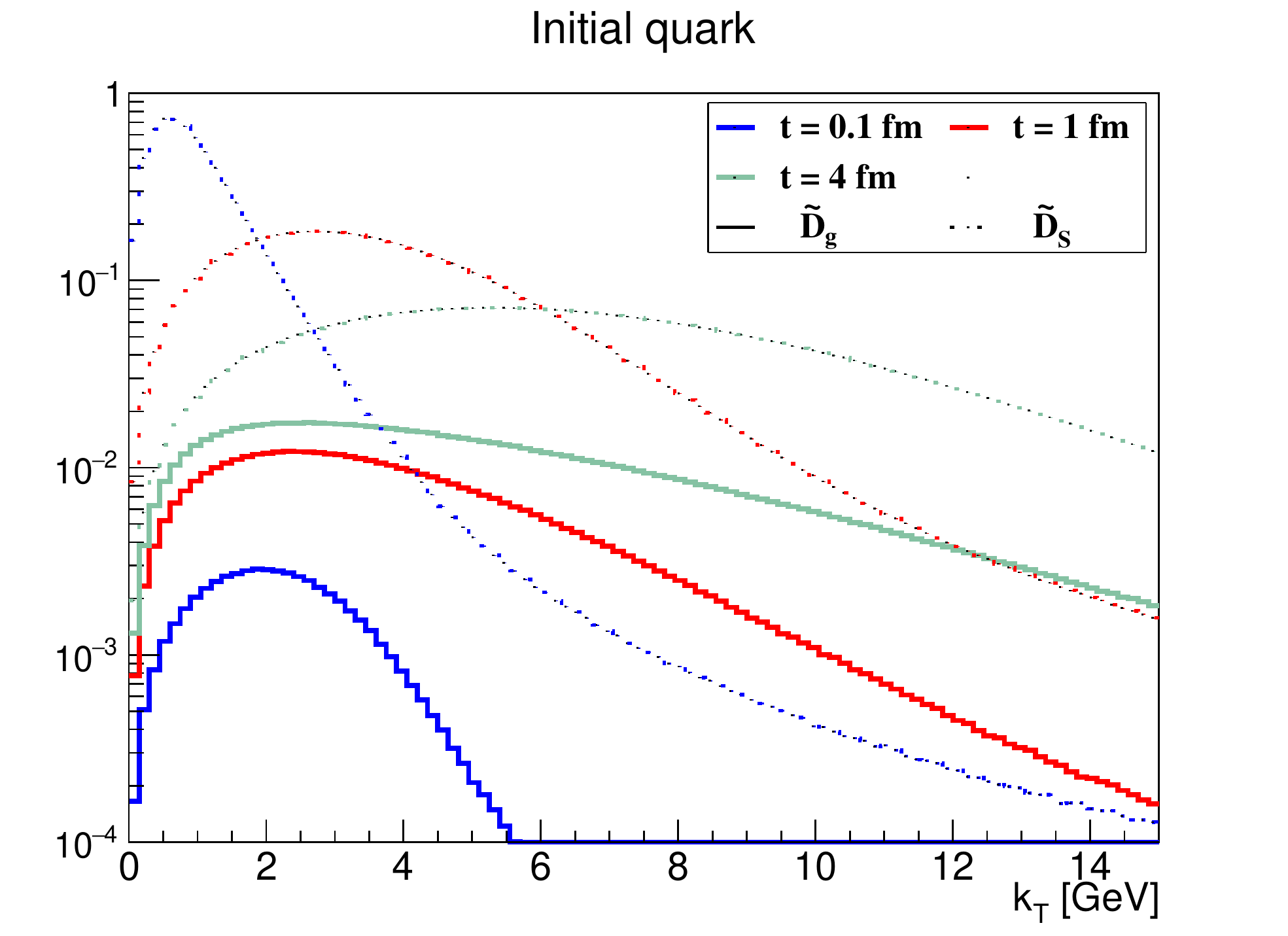}
    \caption{The $\Tilde{D}(k_T,t)$ distributions for 
    $w(\mathbf{l})\propto 1/[\mathbf{l}^2(m_D^2+\mathbf{l}^2)]$
    at the time-scales $t=0.1, 1, 4\,$fm:
    cascades initiated by gluon (left) and quark (right). The dashed lines correspond to the quark distributions while the solid lines to the gluon distributions.}
    \label{fig:kT_kzq_wq2}
\end{figure}

\begin{figure}[!tbp]
    \centering
    \includegraphics[scale=0.4]{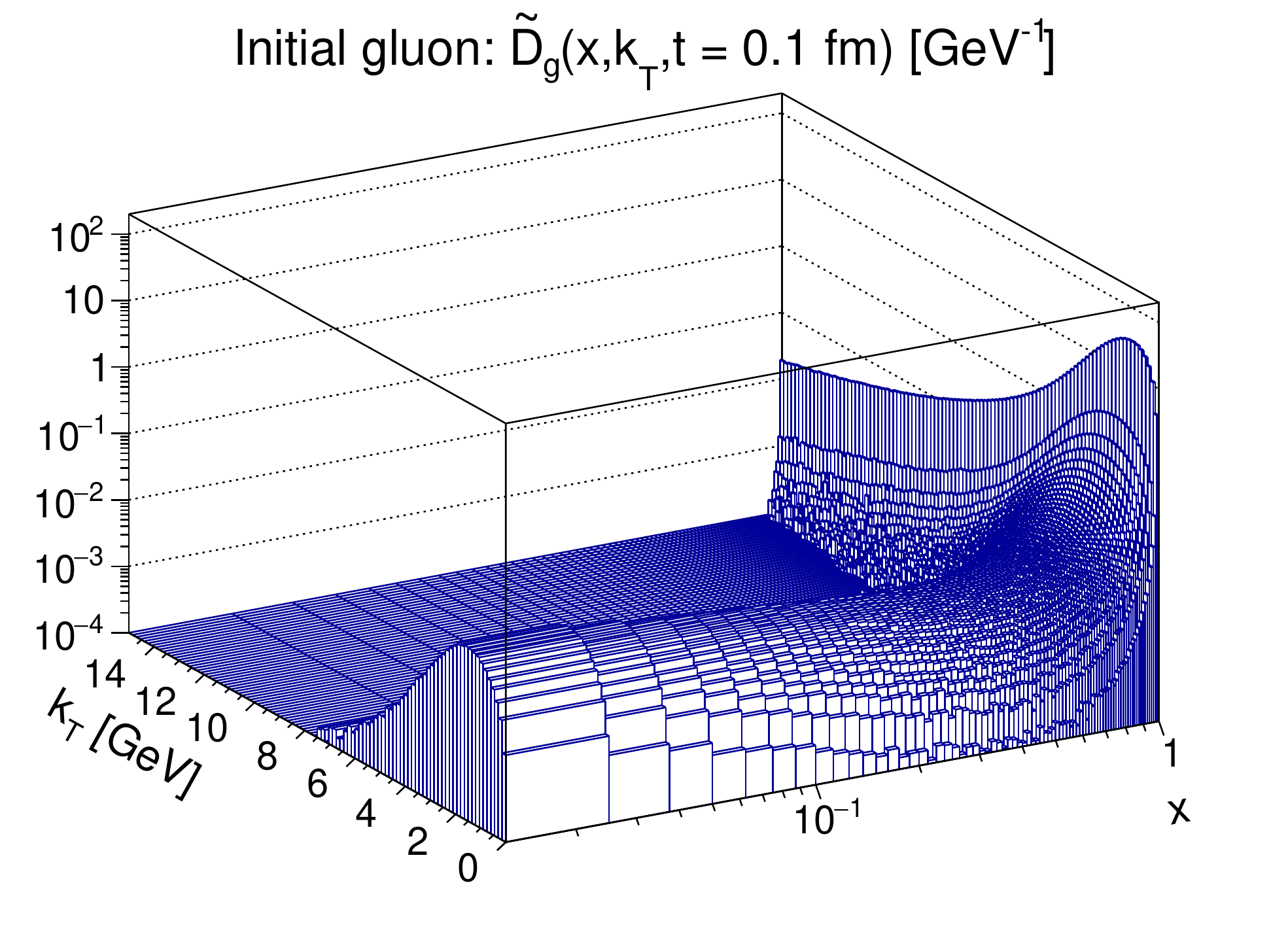}
    \includegraphics[scale=0.4]{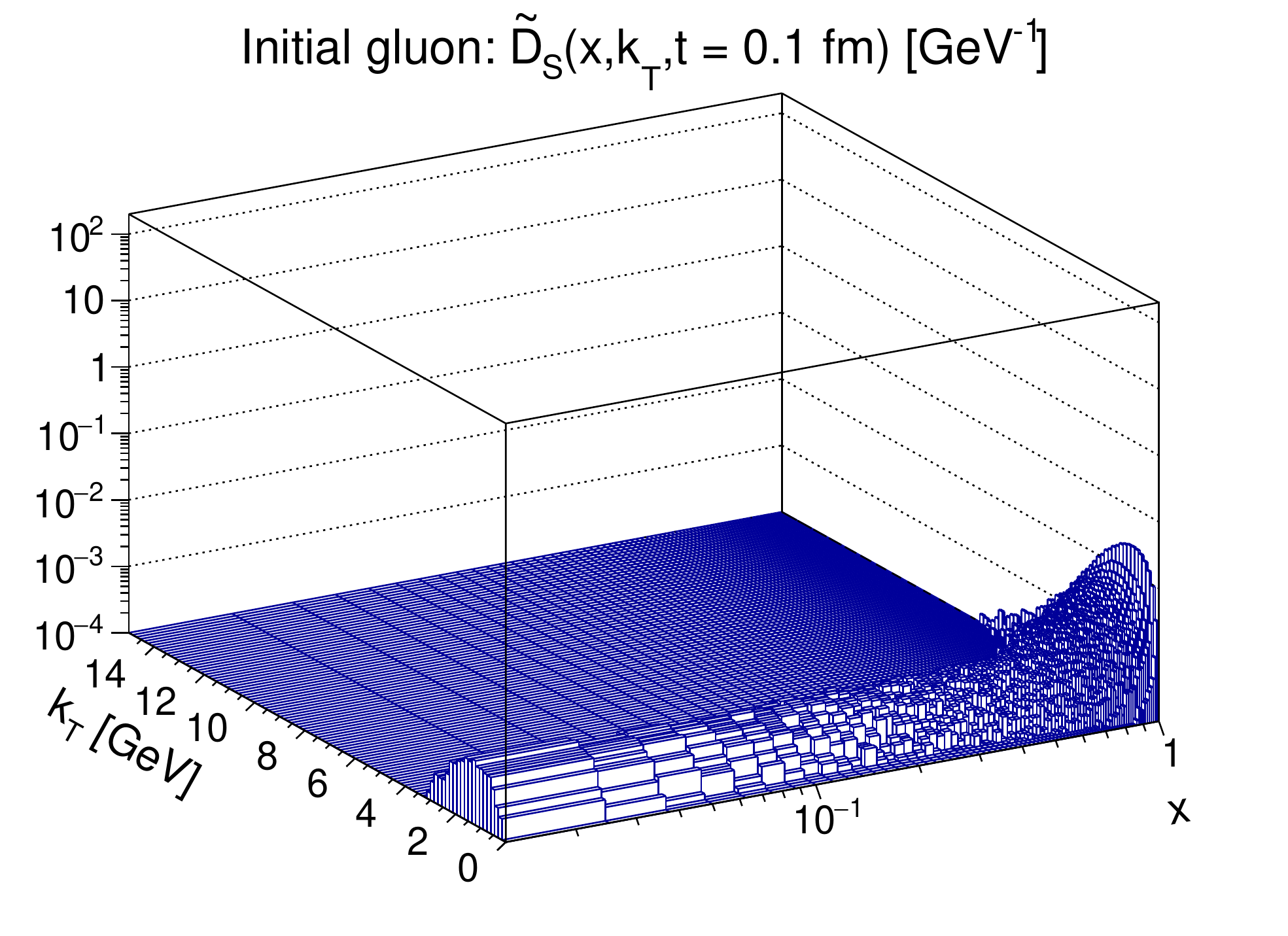}
    \includegraphics[scale=0.4]{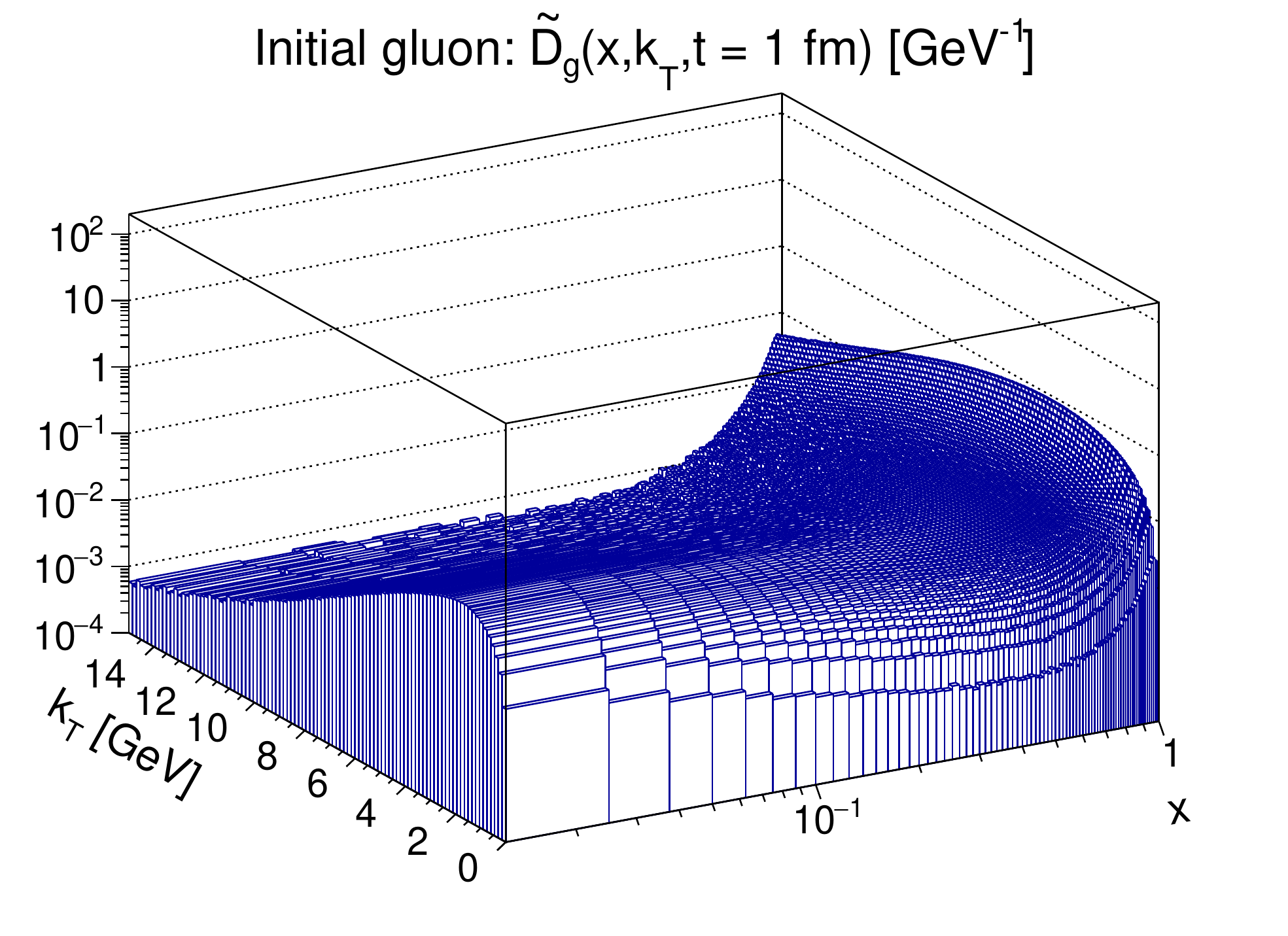}
    \includegraphics[scale=0.4]{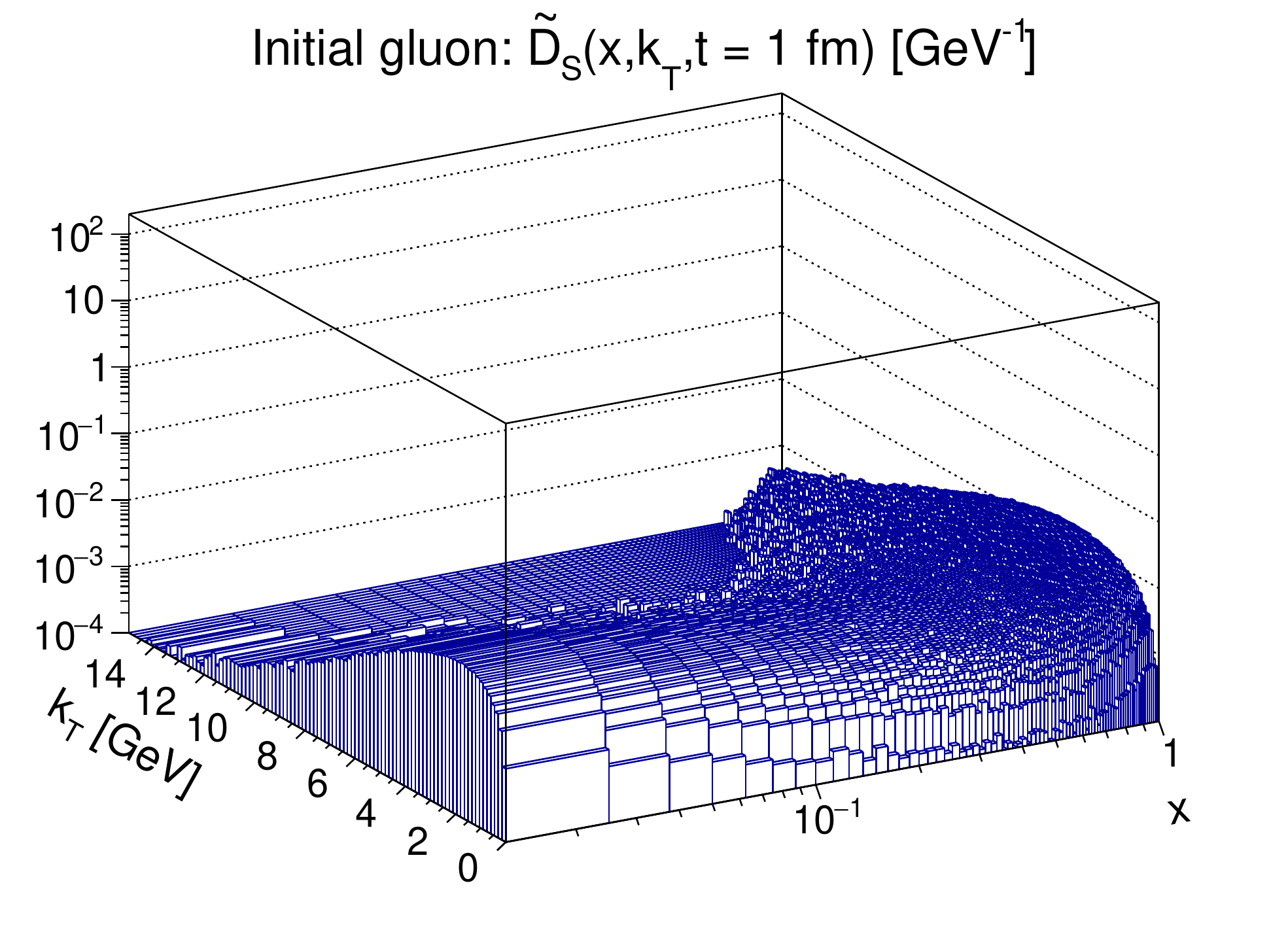}
    \includegraphics[scale=0.4]{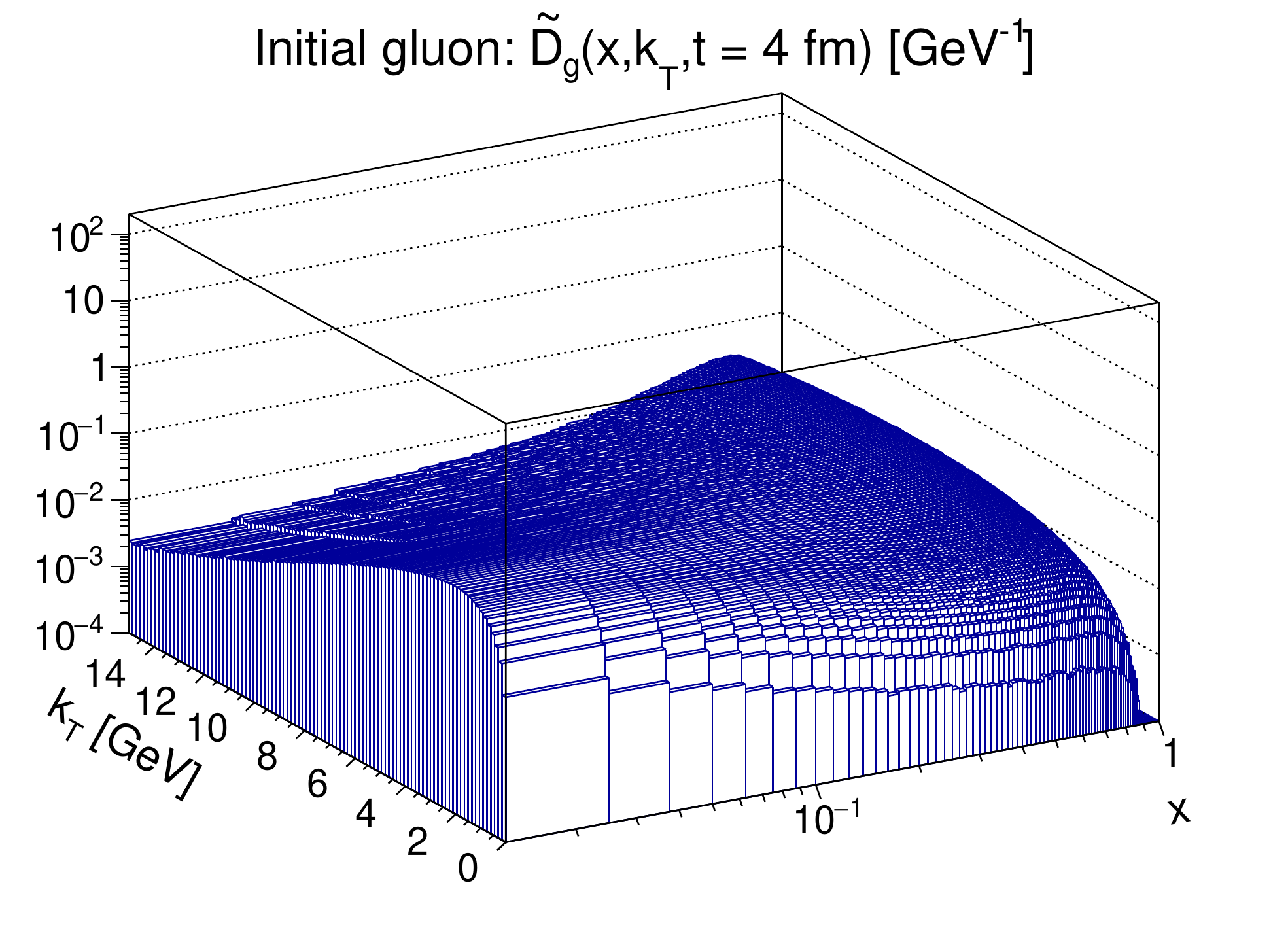}
    \includegraphics[scale=0.4]{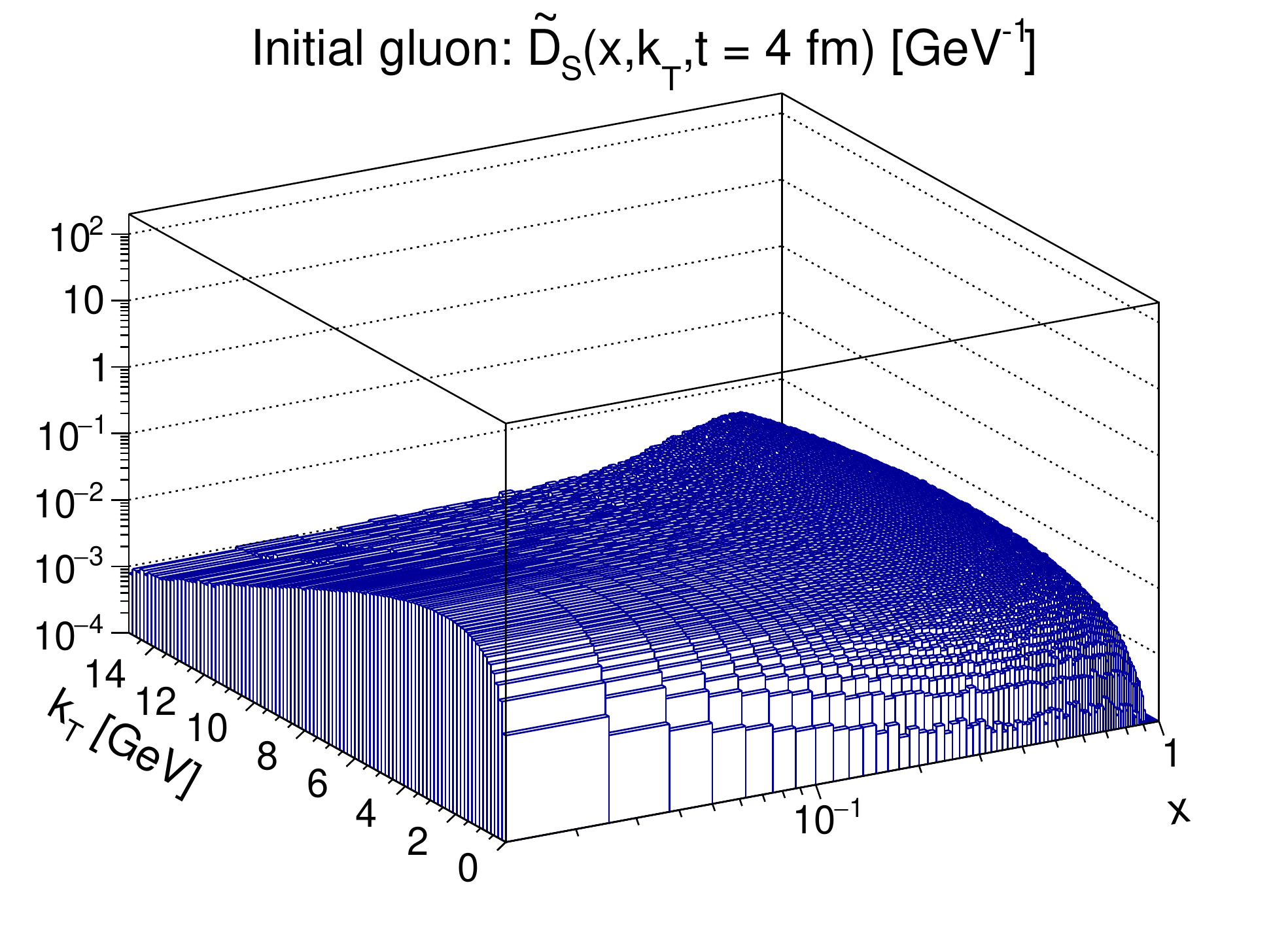}
    \caption{The gluon (left) and quark (right) $k_T$ vs.\ $x$ distributions 
    for cascades initiated by gluons with  
    $w(\mathbf{l})\propto 1/[\mathbf{l}^2(m_D^2+\mathbf{l}^2)]$ 
    at the time-scales $t=0.1, 1, 4\,$fm.}
    \label{fig:ktx_kzq_wq2_inG}
\end{figure}

\begin{figure}[!htbp]
    \centering
    \includegraphics[scale=0.4]{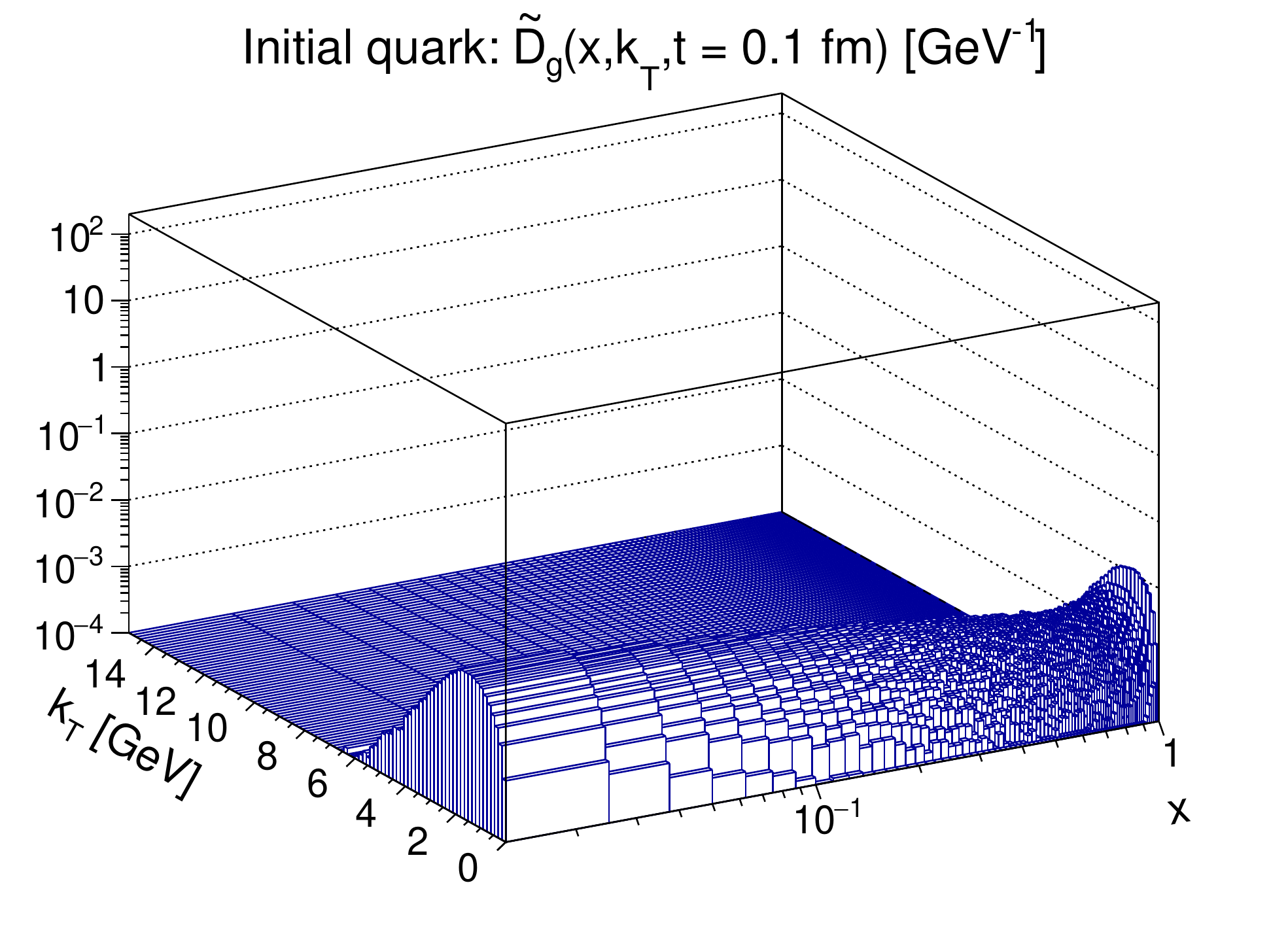}
    \includegraphics[scale=0.4]{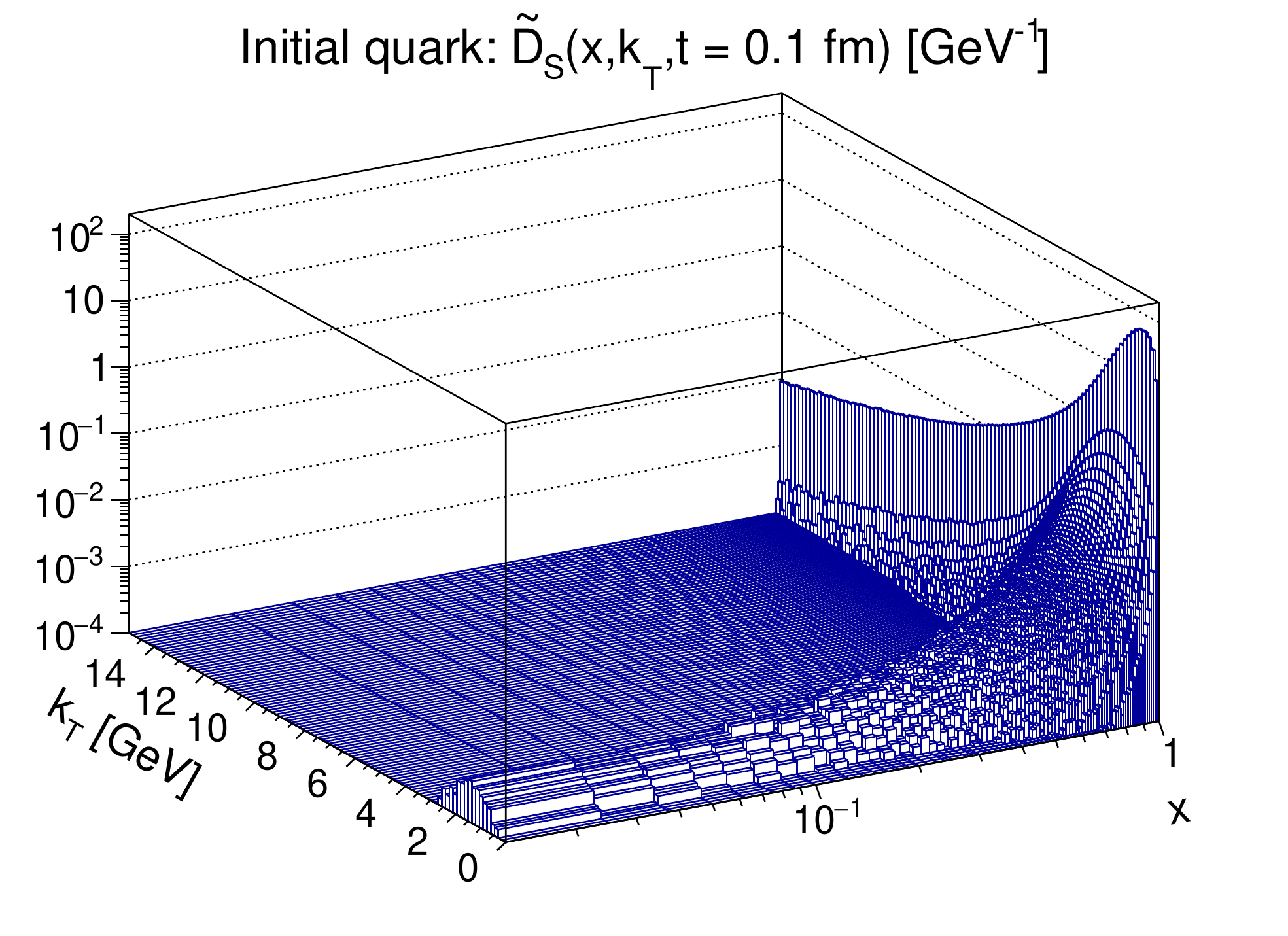}
    \includegraphics[scale=0.4]{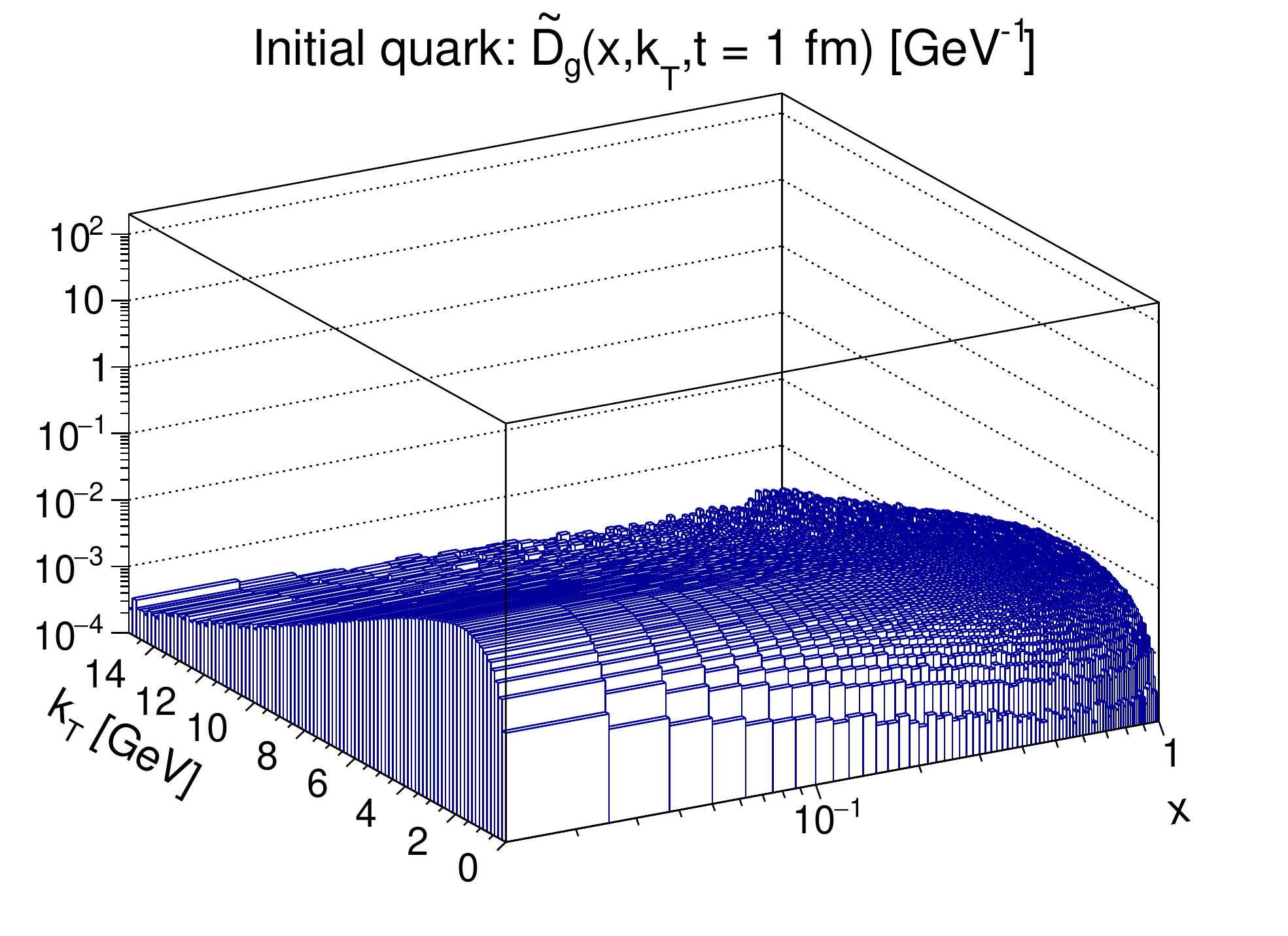}
    \includegraphics[scale=0.4]{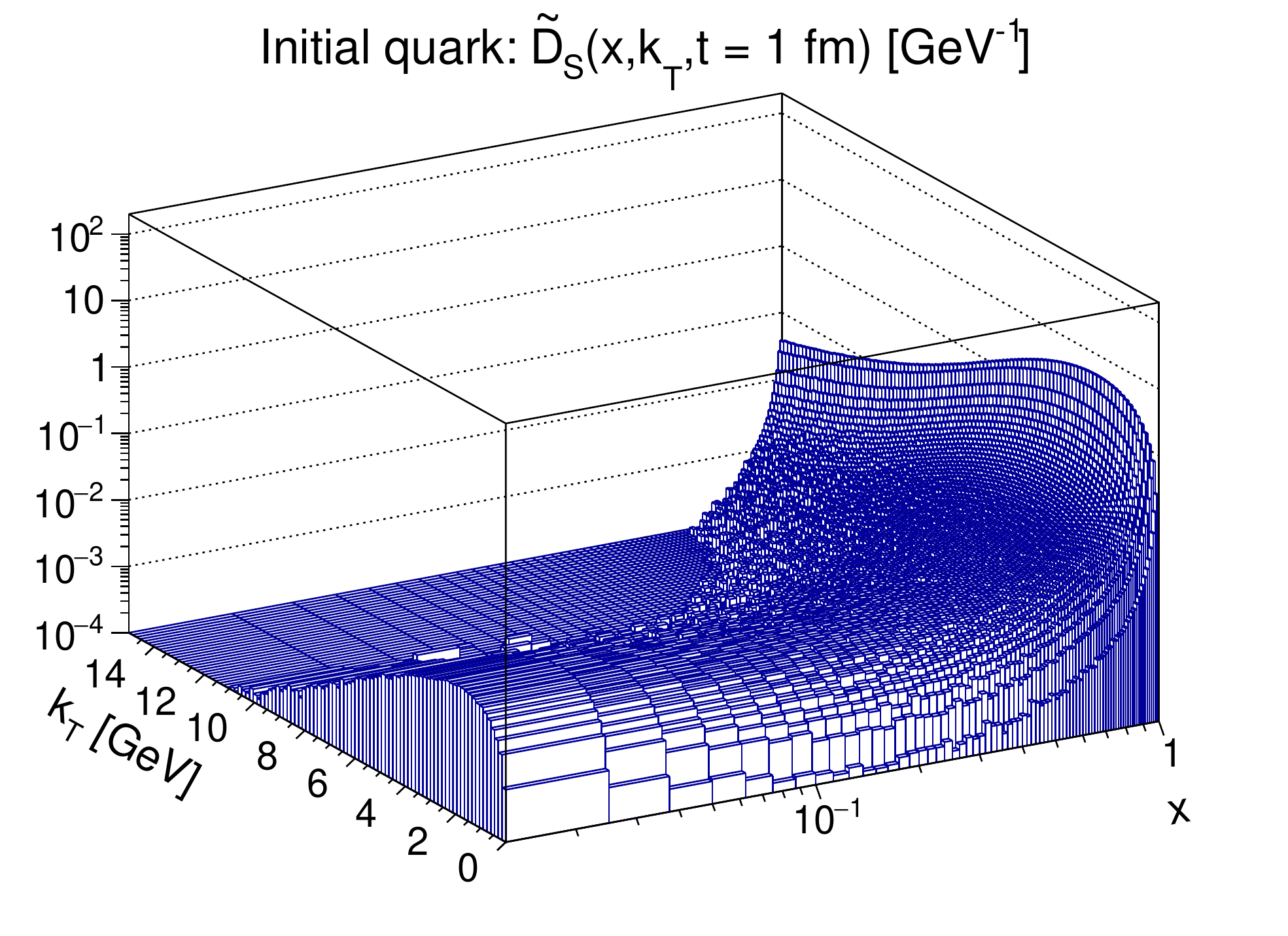}
    \includegraphics[scale=0.4]{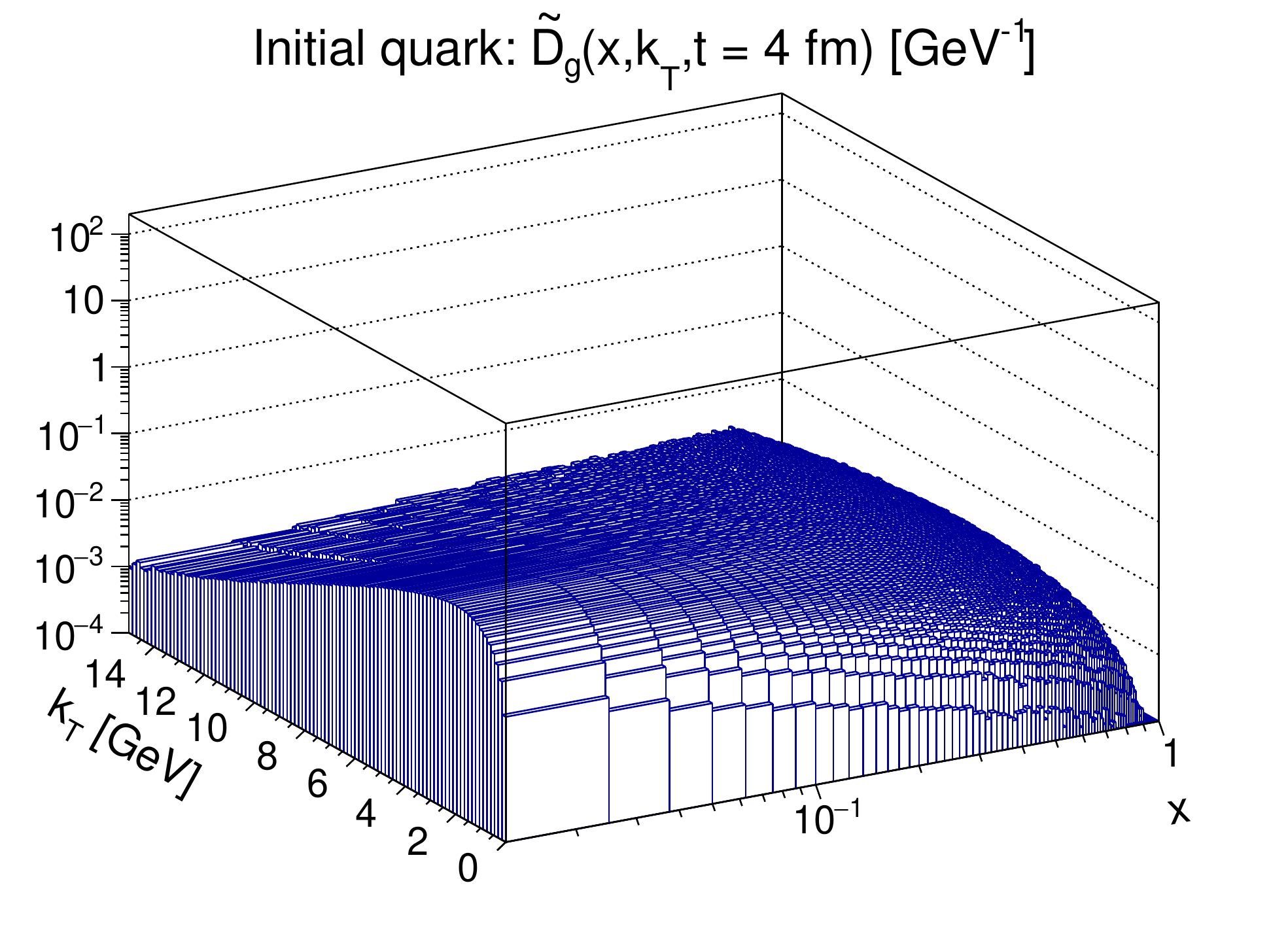}
    \includegraphics[scale=0.4]{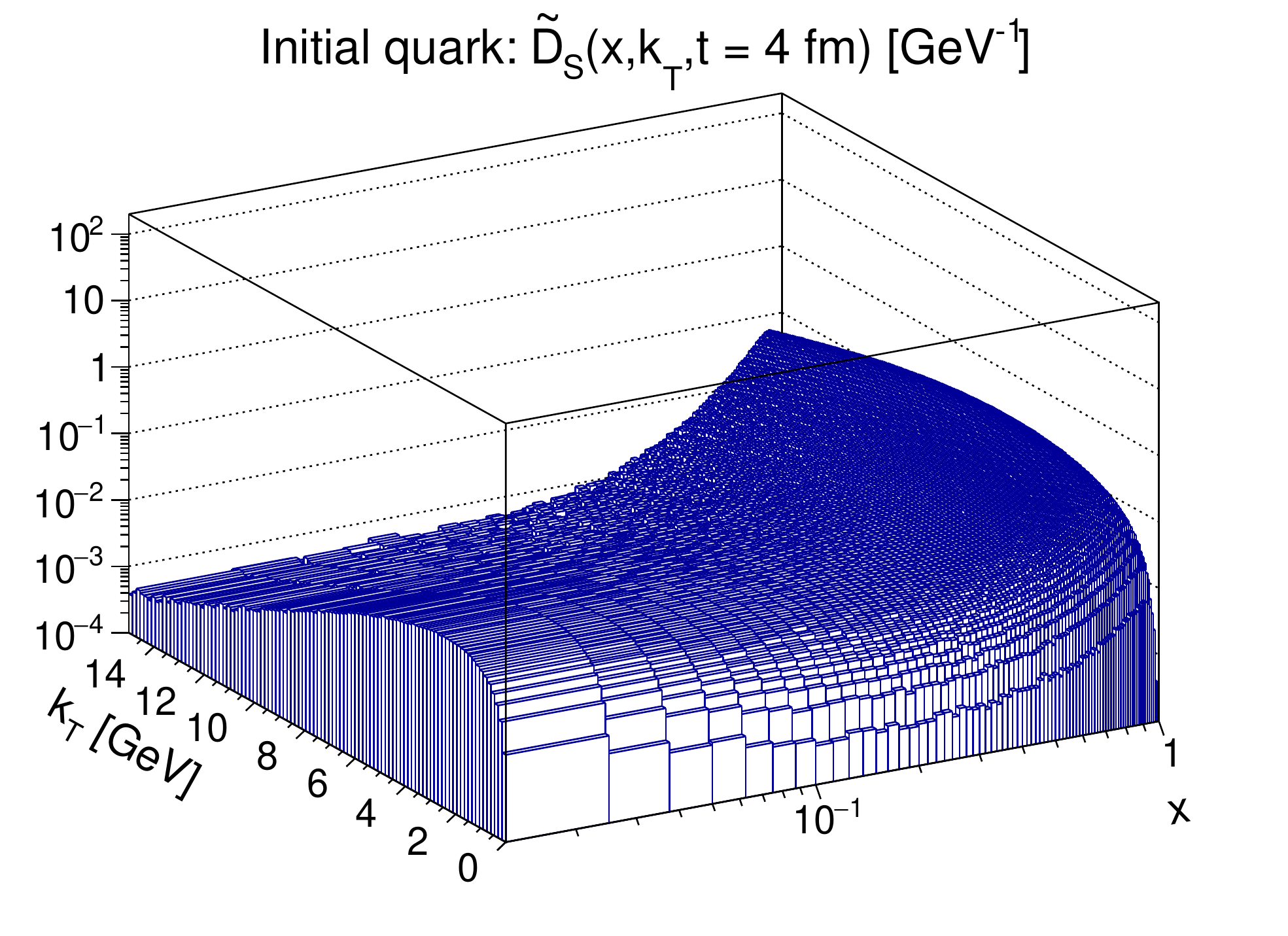}
    \caption{The gluon (left) and quark (right) 
    $k_T$ vs.\ $x$ distributions for cascades initiated by quarks with
    $w(\mathbf{l})\propto 1/[\mathbf{l}^2(m_D^2+\mathbf{l}^2)]$ 
    at the time-scales $t=0.1, 1, 4\,$fm.}
    \label{fig:ktx_kzq_wq2_inQ}
\end{figure}

In Fig.~\ref{fig:kT_kzq_wq2}, we show the $k_T$-dependent distributions integrated over the longitudinal momenta, given in Eq.~\eqref{eq:DI-definition}.
One can clearly see that as the time progresses the distribution for both quarks and gluons become wider. Furthermore, the distributions of gluons are higher than that of quarks if gluons are in the initial state, and similarly, the distributions of quarks are higher than that of gluons if quarks are in the initial state.

The complete 2D distributions, visualising both the $x$ and $k_T$-dependence, are presented in 
Fig.~\ref{fig:ktx_kzq_wq2_inG} and \ref{fig:ktx_kzq_wq2_inQ} for the initial gluon and
quark, respectively. One can can see that the late time behaviour is very similar in processes initiated by quarks or gluons. This behaviour can be linked to diffusive properties of the jet--medium interactions.
Furthermore, while there are some differences in large $x$ part of the spectra, the shape at low $x$ is rather universal.

\subsection{Multiplicity distributions}

\begin{figure}[!htbp]
    \centering
    \includegraphics[scale=0.33]{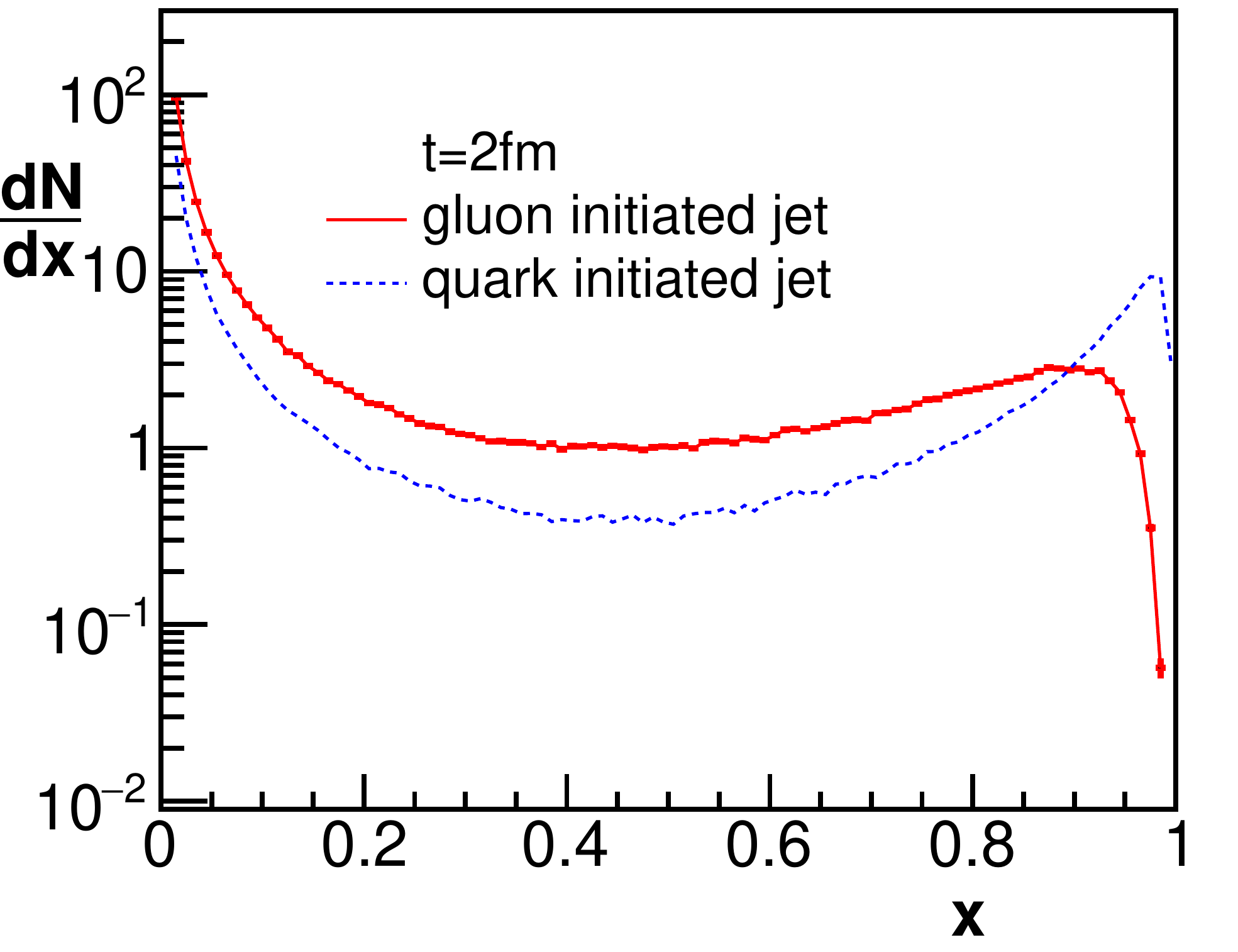}
    \includegraphics[scale=0.33]{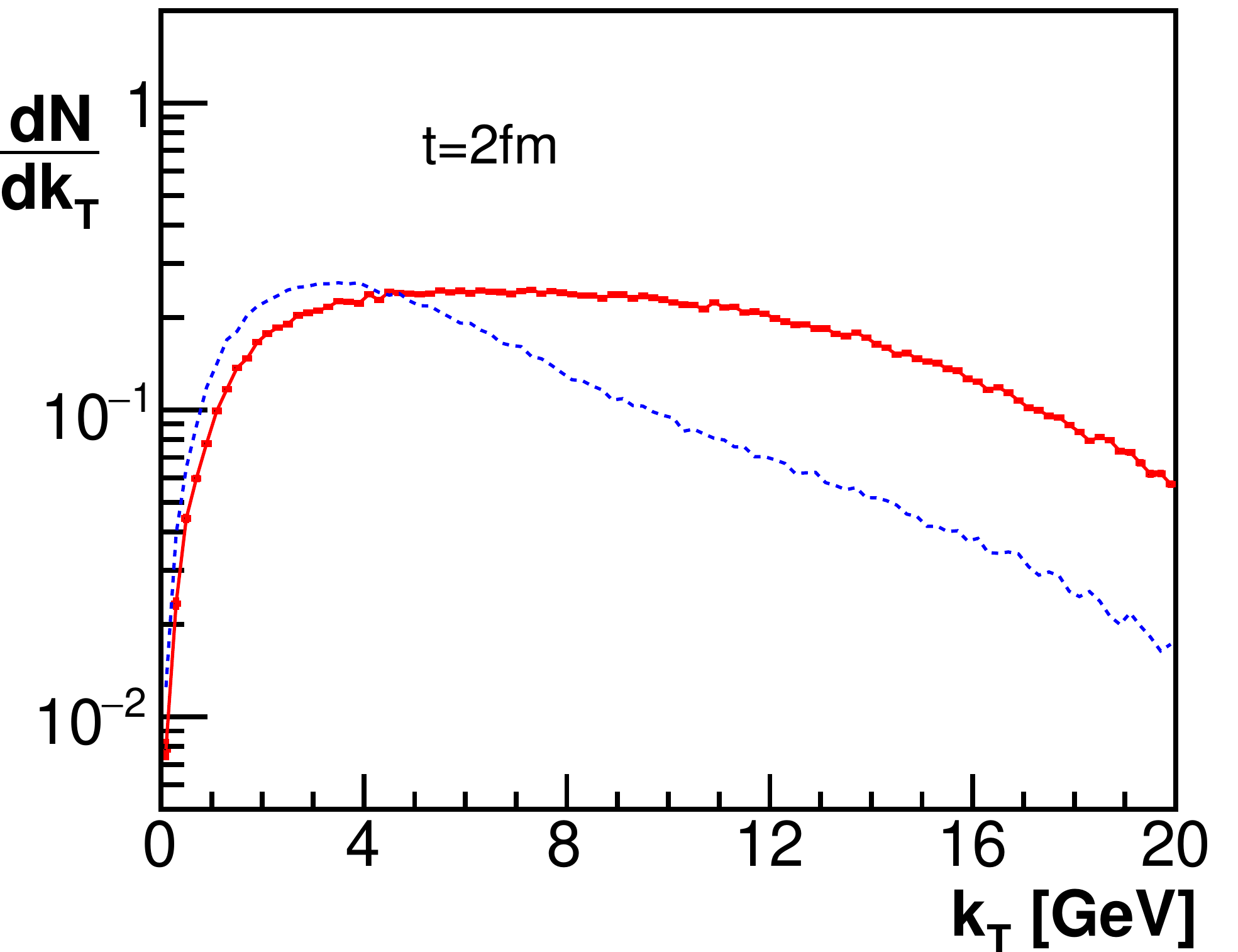}
    \includegraphics[scale=0.33]{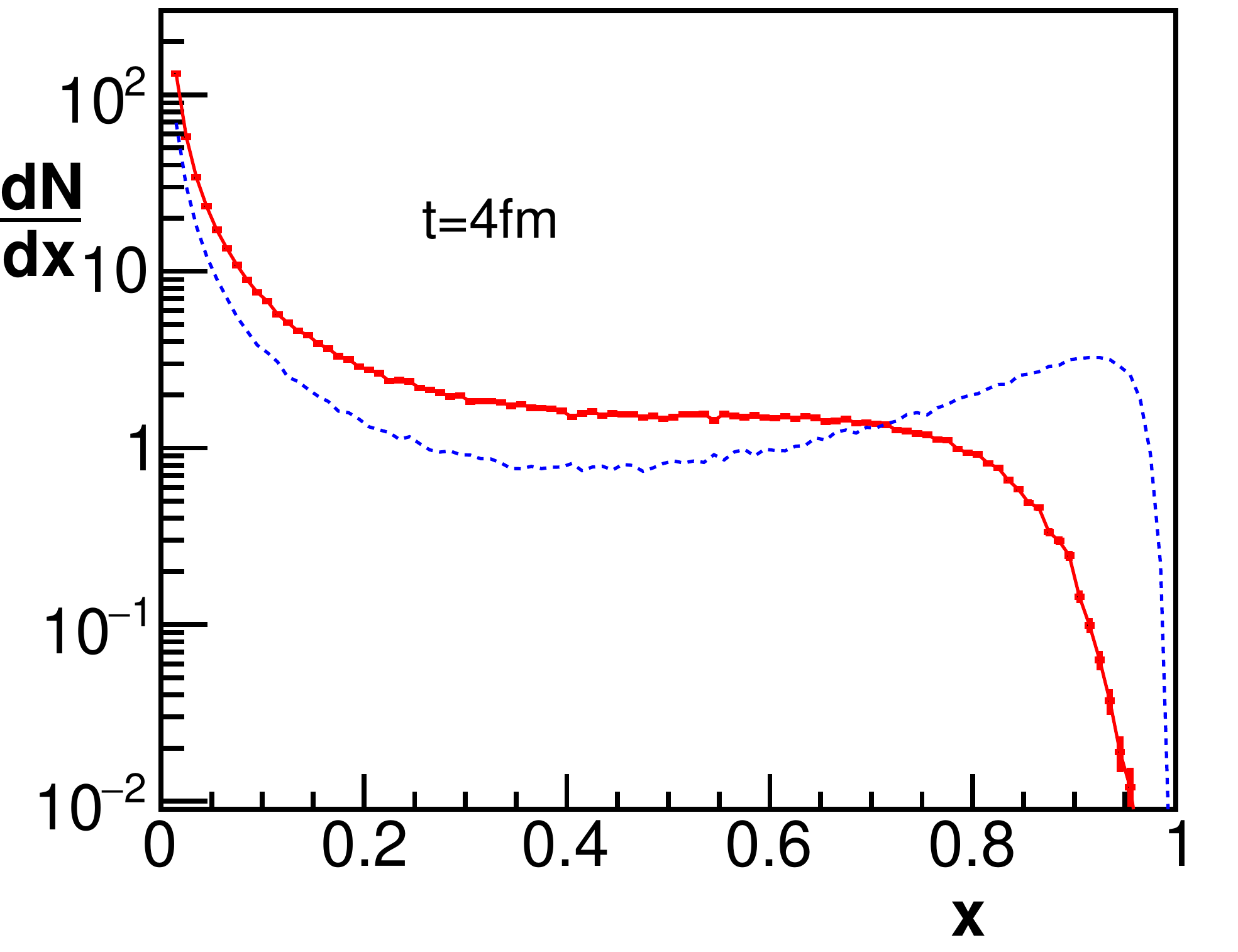}
    \includegraphics[scale=0.33]{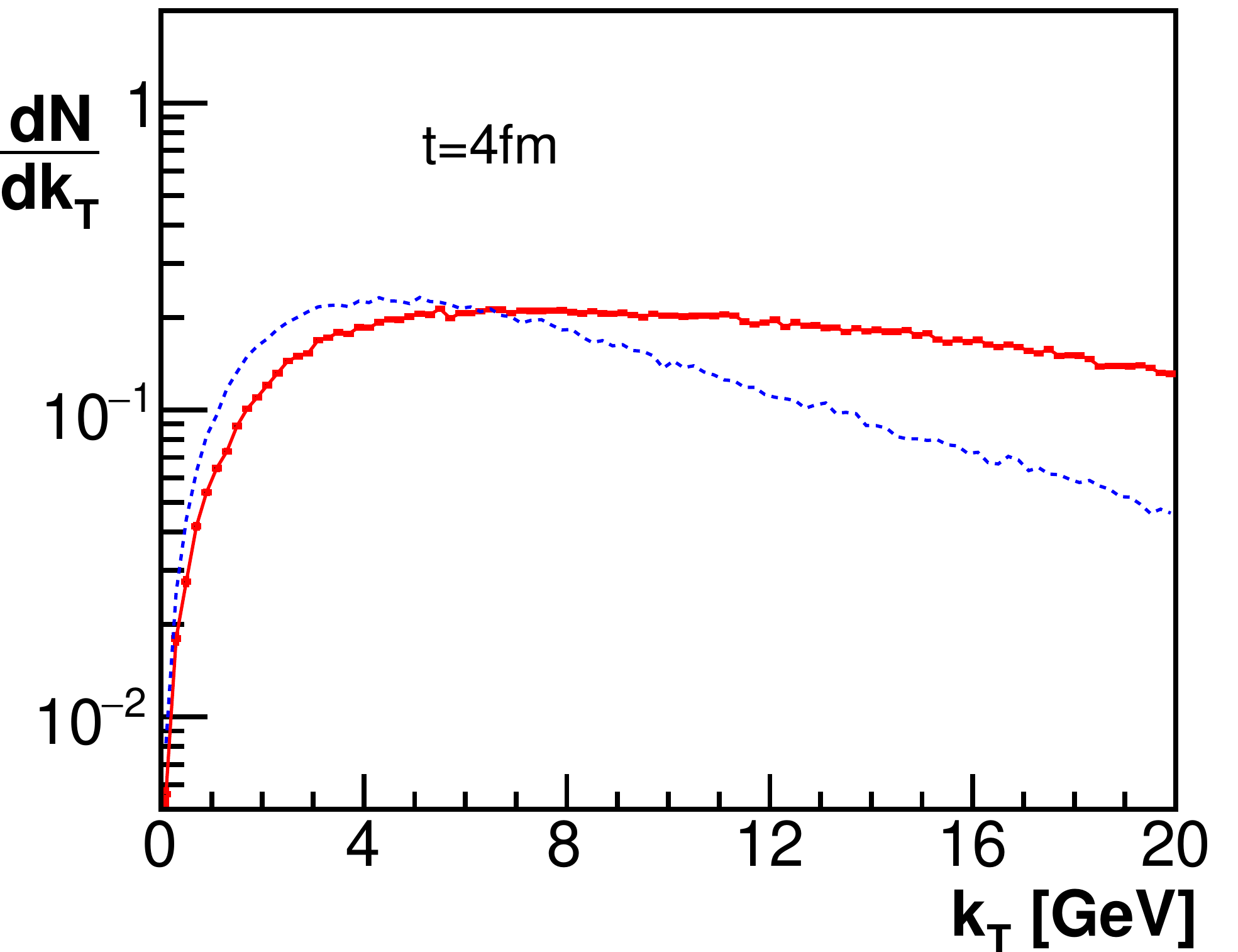}
    \caption{Multiplicity distributions generated by \tmdice\ in $x$ (left) and $k_T$ (right) for cascades initiated by quarks and gluons as indicated  
    for the evolution equations (\ref{eq:BDIMsys1}) with 
    $w(\mathbf{l})\propto 1/[\mathbf{l}^2(m_D^2+\mathbf{l}^2)]$ 
    at the time-scales $t=2\,$fm and $4\,$fm (top and bottom, respectively).}
    \label{fig:fig14}
\end{figure}

We have also obtained the results for specific times $t$ in terms of the multiplicity distributions 
\begin{align}
    \frac{dN}{dx}&=\int_0^\infty k_Tdk_T\int_0^{2\pi}d\phi\left.\left\{F_g(x,\mathbf{k},t)+\sum_iF_{q_i}(x,\mathbf{k},t)\right\}\right.\,,\\
    \frac{dN}{dk_T}&=\int_0^1 k_Tdx\int_0^{2\pi}d\phi\left\{F_g(x,\mathbf{k},t)+\sum_iF_{q_i}(x,\mathbf{k},t)\right\}\,,
\end{align}
for jets in the medium with the \tmdice~algorithm. For the numerical calculations, the same constraints as given in Eqs.~(\ref{eq:qgxpar}), (\ref{eq:qgalbar}), and (\ref{eq:qgnqhpar}) were used, with the exception of $x_{\rm min}$, where $x_{\rm min}=10^{-2}$ was chosen. 
The multiplicity distributions are not infrared and collinearly safe observables, and they considerably depend on the cut-off scale $x_{\rm min}$. In the solution to the evolution equations (\ref{eq:BDIMsys1}), $x_{\rm min}$ corresponds to an energy scale at which the assumption that coherent medium-induced radiation and scatterings occur and dominate breaks down. Thus, we assume for this energy scale an estimate $xE=1$~GeV which is still larger than the medium temperature (where we assume that jet-particles thermalize), but of the same order.
The numerical results are shown in Fig.~\ref{fig:fig14} for the time scales of $t=2\,$fm and $4\,$fm, for jets initiated by either a quark or a gluon. 
As can be seen for the distribution in $x$, the peak in the infrared region that is suppressed by a factor $x$ in the fragmentation functions $D(x)$ in Fig.~\ref{fig:x_kzq_wq2_v2} is much more pronounced in Fig.~\ref{fig:fig14}, showing that the infrared contributions to the jet-multiplicity dominate at large time scales. The distributions in $k_T$ exhibit an increasing broadening with large time scales that is more pronounced for the jets initiated by gluons than those initiated by quarks.

\subsection{Characteristic features of cascades}

In this section, we present our results for some useful characteristic features of the cascades. These should not be confused with observables that can be measured directly in experiment, but rather as projections or ``summary statistics'' containing the most important information of the cascades described in the previous section.

\begin{figure}[!htbp]
    \centering
    \includegraphics[width=0.35\linewidth]{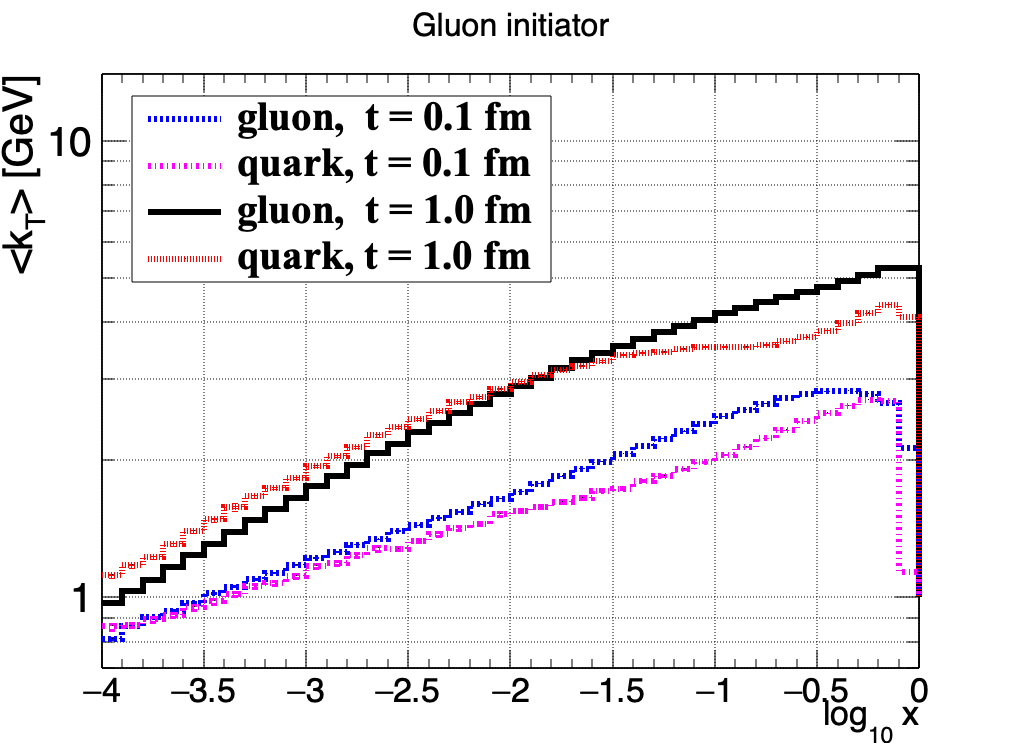}
    \includegraphics[width=0.35\linewidth]{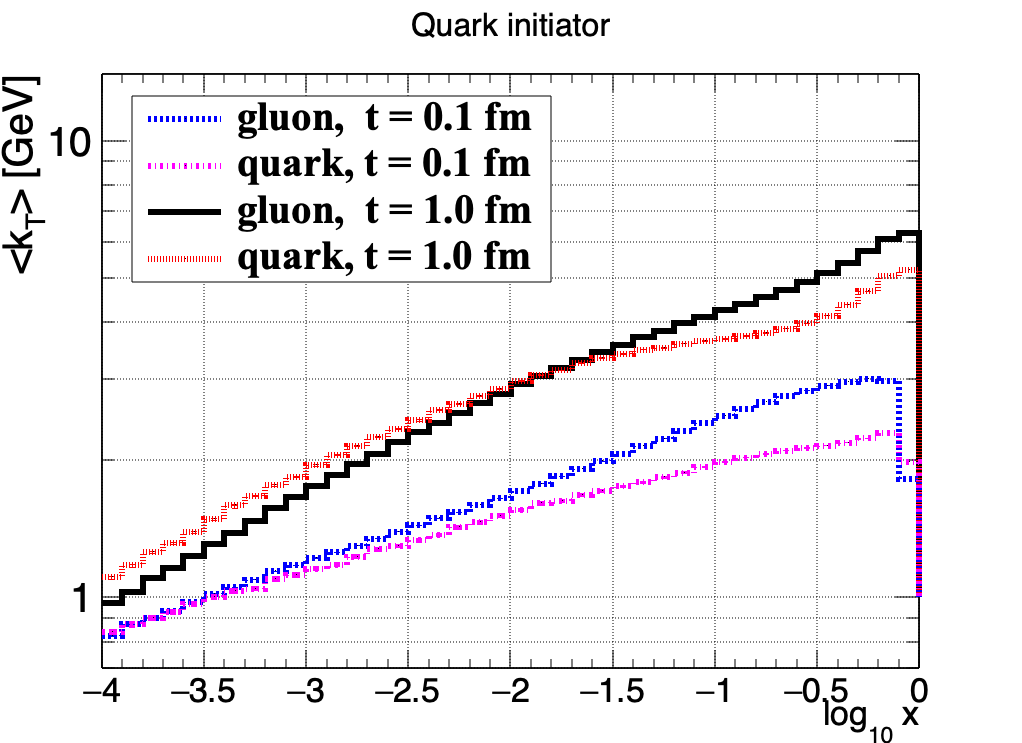}
    \includegraphics[width=0.35\linewidth]{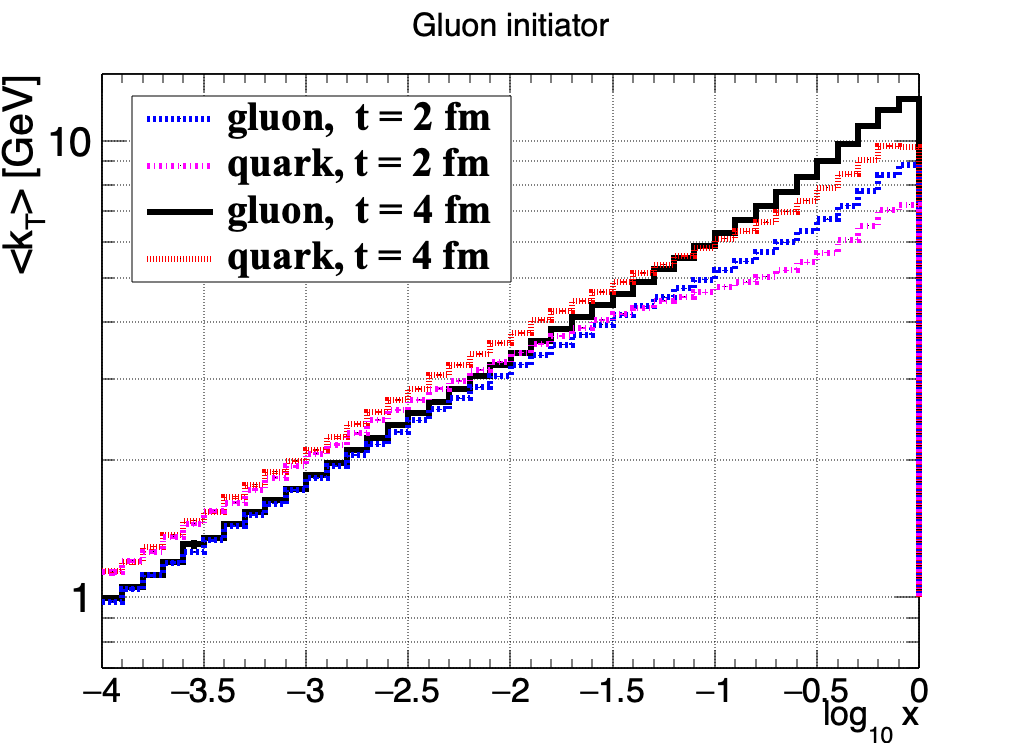}
    \includegraphics[width=0.35\linewidth]{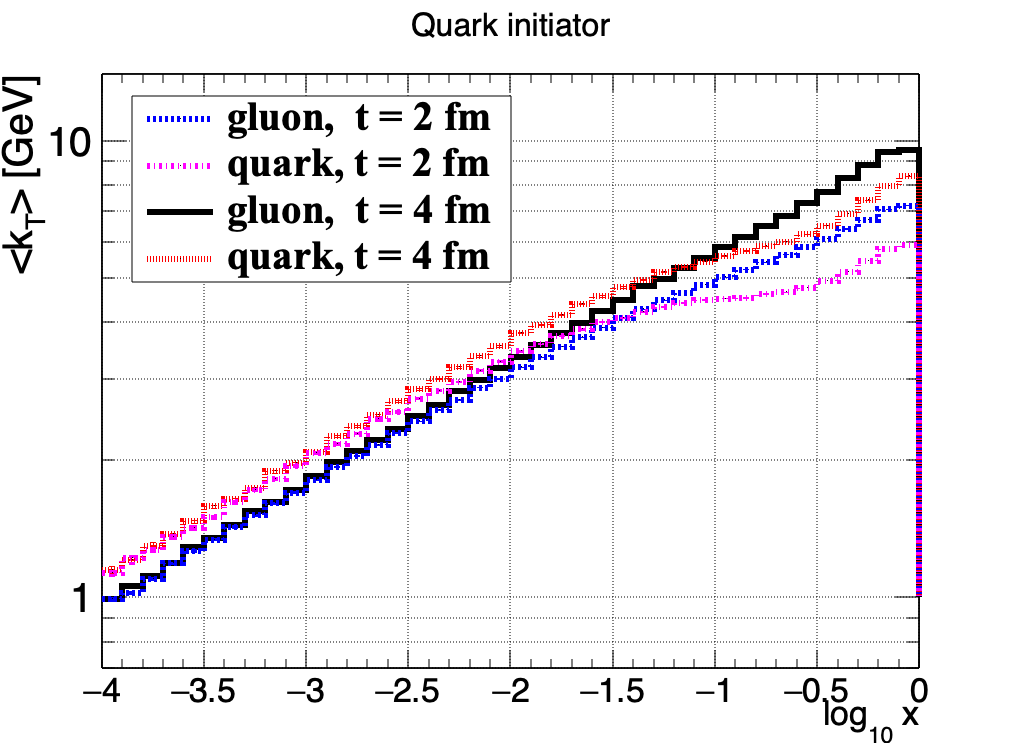}
    \caption{The average transverse momentum $\langle k_T \rangle$ versus $\log_{10}x$ 
    for the evolution equations (\ref{eq:BDIMsys1}) with 
    $w(\mathbf{l})\propto 1/[\mathbf{l}^2(m_D^2+\mathbf{l}^2)]$
    for the time-scales $t=0.1, 1, 2, 4\,$fm.}
    \label{fig:avkt}
\end{figure}

The first feature we discuss, is the average transverse momentum. It is defined as
\begin{equation}
\langle k_T\rangle=
\frac{\int \rmd^2 \k\, |\mathbf{k}| D(x,\mathbf{k},t)}{\int \rmd^2 \k\, D(x,\mathbf{k},t)} = \frac{\int_0^\infty \rmd k_T \, k_T^2 D(x,k_T,t)}{\int_0^\infty \rmd k_T \, k_T D(x,k_T,t)}  \,,
\end{equation}
where $k_T \equiv |\k|$,
as a function of $x$ evaluated for different evolution time values. Overall, we see in Fig.~\ref{fig:avkt} that the distributions both for the initial-state gluons and initial-state quarks are rather similar, i.e.\ as $x$ gets smaller and smaller the average $k_T$ gets smaller, meaning that soft mini-jets become delocalised. The distributions for different times tend to merge as $x$ gets small enough. One can see certain differences if one compares the distributions for the final quark vs.\ the final gluon as time progresses. The slopes of the distributions become different. This is consistent with the obtained earlier results for the fragmentation functions and suggests that gluons have harder momenta than quarks and dominate at larger values of $x$.

\begin{figure}[!htbp]
    \centering
    \includegraphics[width=0.35\linewidth]{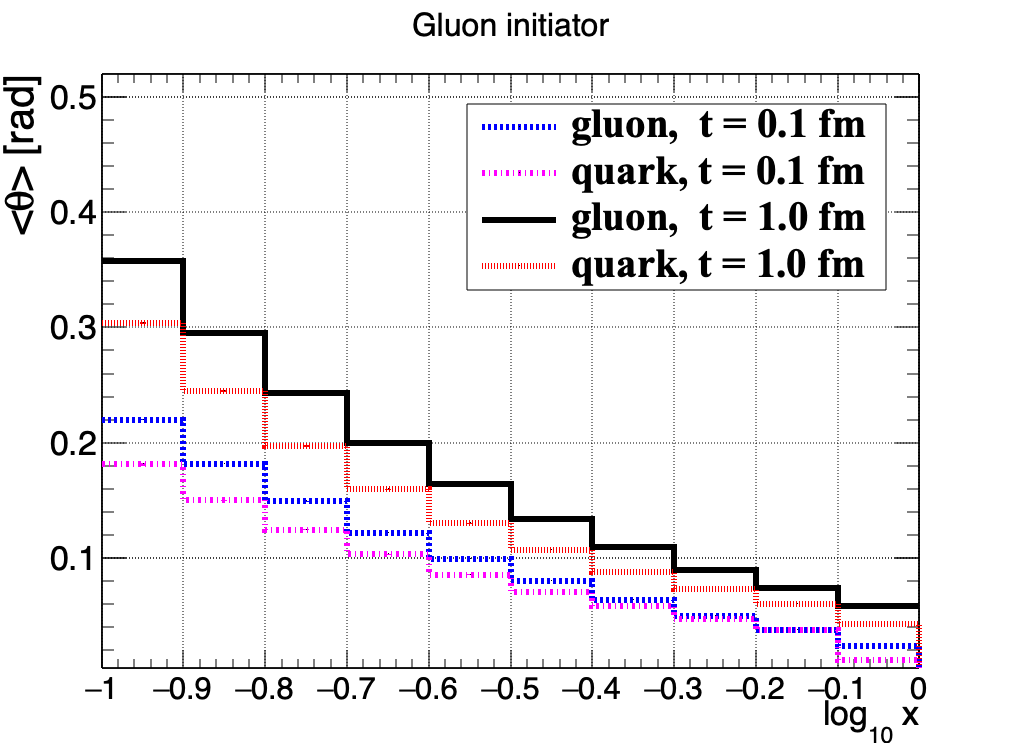}
    \includegraphics[width=0.35\linewidth]{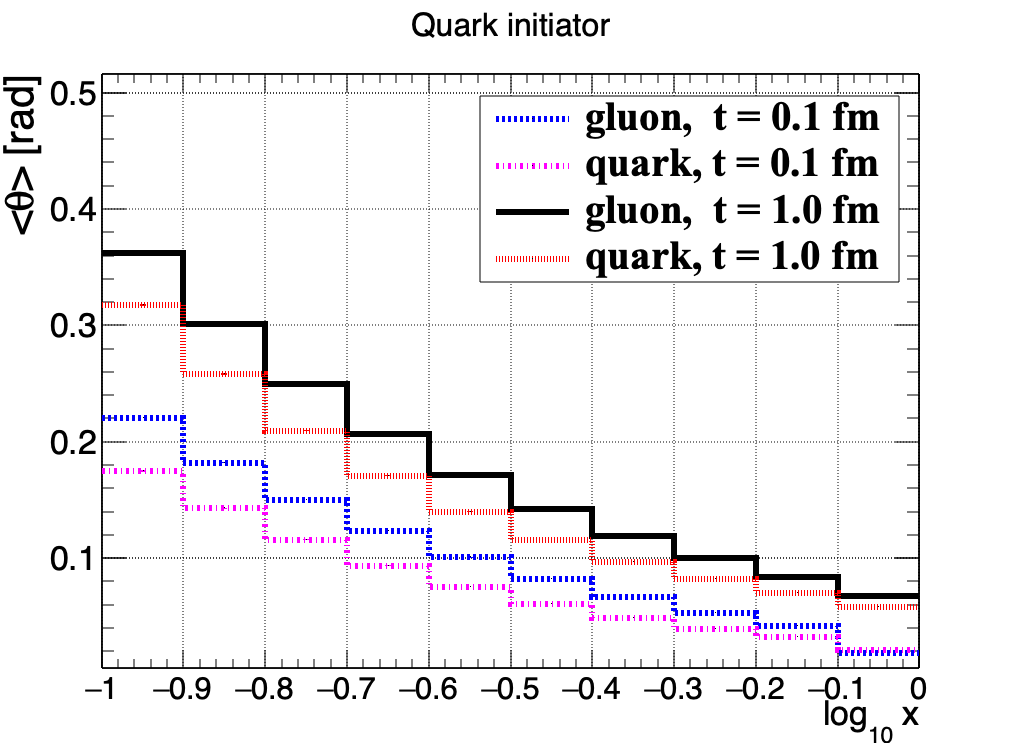}
    \includegraphics[width=0.35\linewidth]{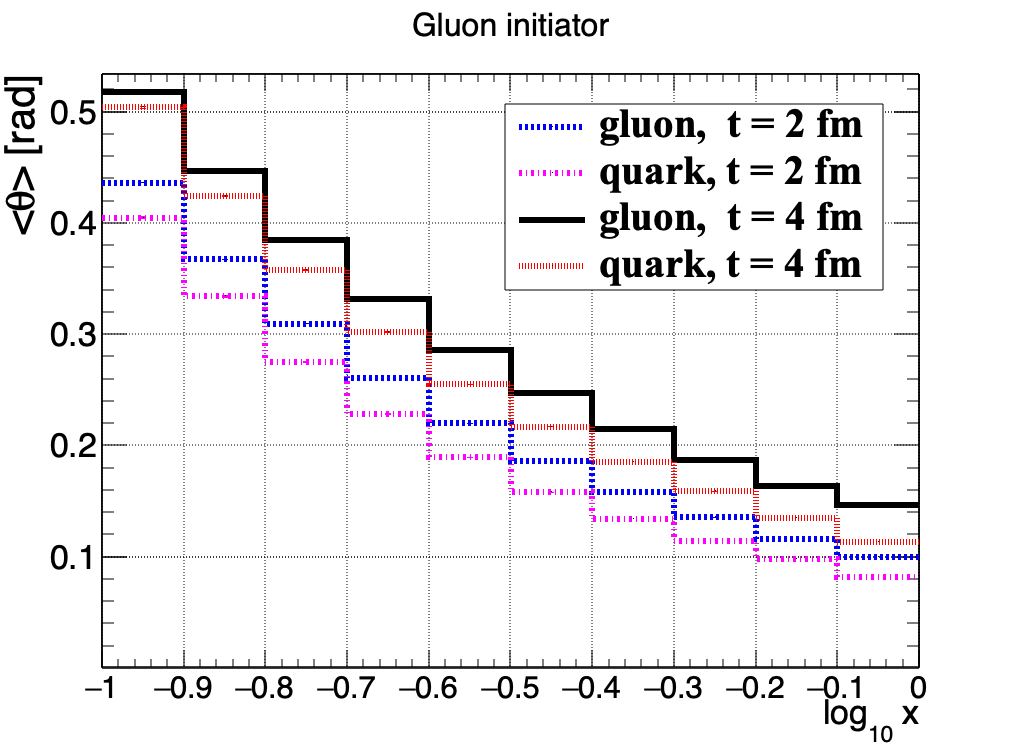}
    \includegraphics[width=0.35\linewidth]{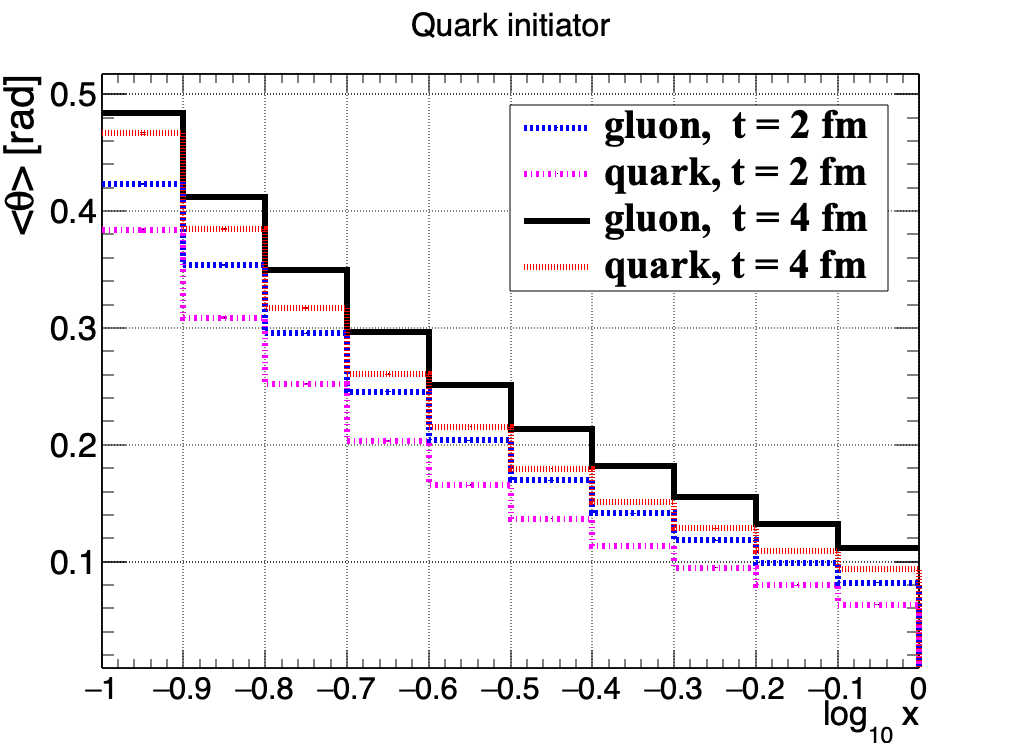}
    \caption{The average polar angle $\langle \theta \rangle$ versus $\log_{10}x$ 
    for the evolution equations (\ref{eq:BDIMsys1}) with 
    $w(\mathbf{l})\propto 1/[\mathbf{l}^2(m_D^2+\mathbf{l}^2)]$
    for the time-scales $t=0.1, 1, 2, 4\,$fm.}
    \label{fig:avtheta_sa}
\end{figure}
In order to draw further attention to small angles, close to the direction of the parent parton, we plot the average angle $\langle \theta \rangle$ as a function of the momentum fraction $x$ in Fig.~\ref{fig:avtheta_sa},
where $\theta$ (following the light-cone kinematics) is defined as
\begin{equation}
    \theta = \arccos\left(\frac{p_z}{E_p}\right), \qquad p_z = xE - \frac{\mathbf{l}^2}{4xE}\,, \quad E_p = xE + \frac{\mathbf{l}^2}{4xE}\,.
    \label{eq:theta}
\end{equation}
This sheds light on the internal structure of jets.
We observe that gluons in the cascade typically occupy larger angles than the quarks. The average angle for gluons also grows as we go to smaller $x$, which is caused by broadening. This is in line with earlier observations. We also note that the average angles are very similar for gluon-initiated and quark-initiated cascades from early times.

\begin{figure}[!t]
    \centering
    \includegraphics[width=0.49\linewidth]{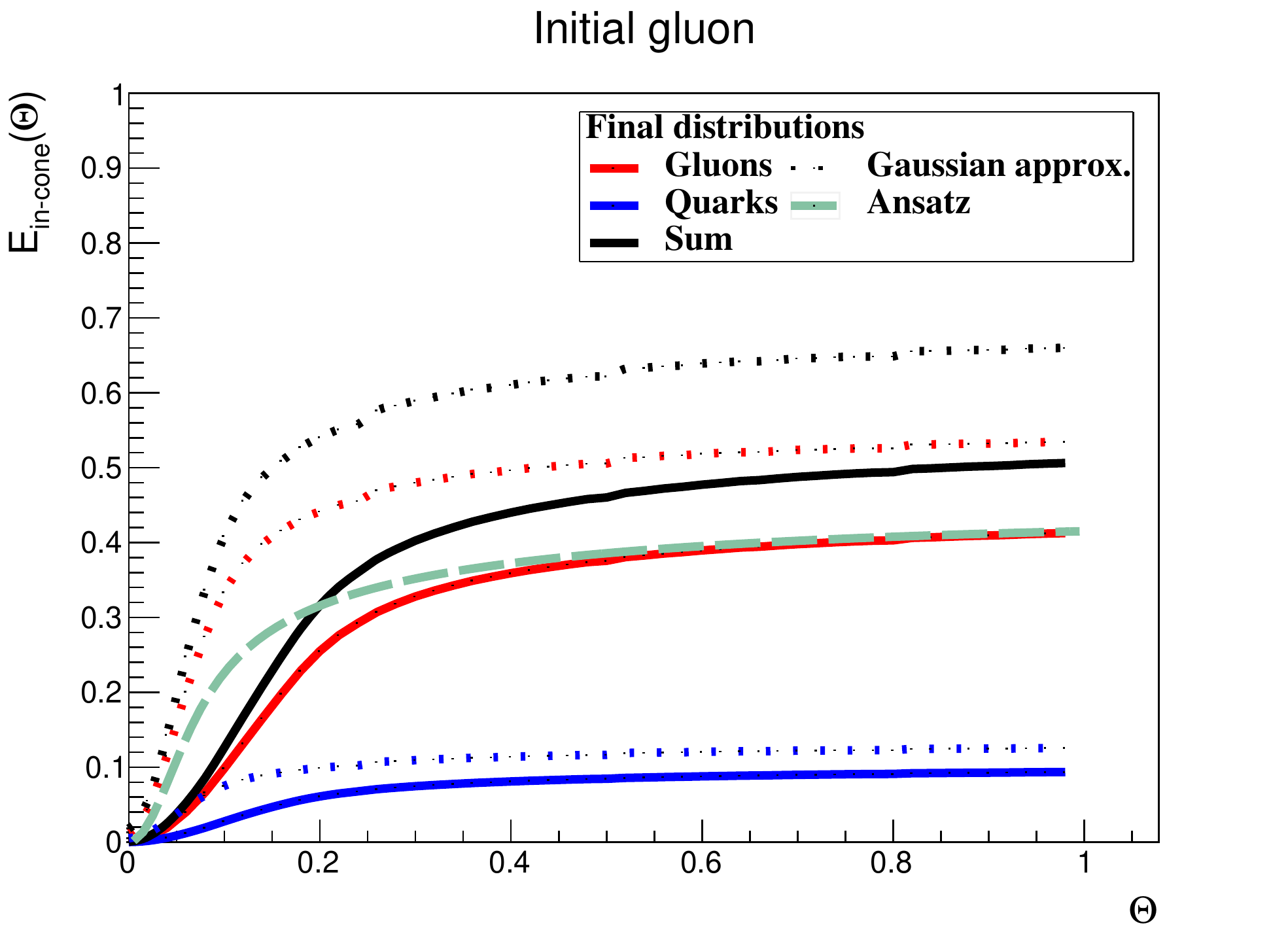}
    \includegraphics[width=0.49\linewidth]{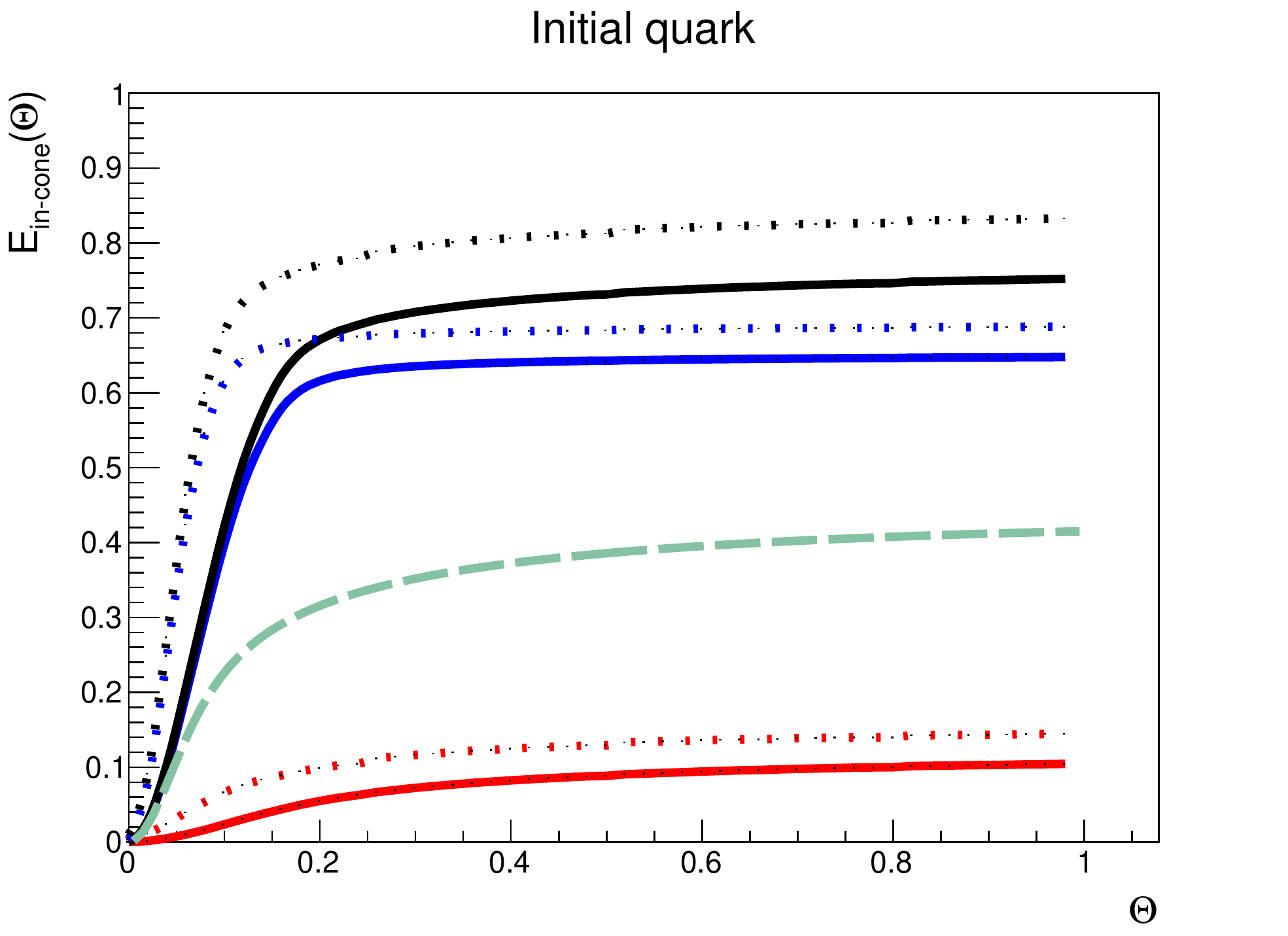}
    \caption{Energy in cone for an initial gluon jet (left) and for an initial quark jet (right) after 4fm in the case of $k_T$-dependent branching and $w(\mathbf{l})\propto 1/[\mathbf{l}^2(m_D^2+\mathbf{l}^2)]$ (solid lines), Gaussian approximation in Eq.~\eqref{eq:DG-2} (dotted lines) and analytical ansatz in Eq.~\eqref{eq:DA-2}(green dashed line). The curves for the analytical Ansatz are plotted only in the case of pure gluon jets.}
    \label{fig:Econe}
\end{figure}

The next characteristic feature that we study is the energy contained inside a cone-angle $\Theta$ around an initial parton defined as 
\begin{equation}
    E_{\text{in-cone}}(\Theta) = \int_0^1 {\rm d} x \int^{x E \sin{\Theta}}_0 {\rm d}k_T\,   \Tilde{D}(x,k_T,t) \,.
\end{equation}
The observable measures the amount of energy that is contained in a cone. Clearly, we see in Fig.~\ref{fig:Econe} that the configuration that maximises this observable is the one where the type of the parton does not change. Furthermore, since the quark contains more in-cone energy, 
we conclude that it is more collimated.

We can compare this to an analytical Ansatz of how the distribution should look like:
\begin{equation}
\label{eq:DA-2}
    D_\text{A}(x,\mathbf{k},t) = D_0(x,t)\, P(\mathbf{k},t) \,,
\end{equation}
where we approximate the $x$ distribution with the analytical solution for a purely gluonic cascade \cite{Blaizot:2013hx},
\begin{equation}
    D_0(x,t) = \frac{t/t_\ast}{\sqrt{x(1-x)^3}}\;\mathrm{exp}\left(-\pi\, \frac{(t/t_\ast)^2}{1-x}\right) \,,
\label{eq:xdistr}
\end{equation}
where now $t_\ast = N_c^{-3/2}\,\bar t_\ast$, and the broadening distribution is
\begin{equation}
  P(\mathbf{k},t) = \frac{4 \pi}{\hat q t}\; \mathrm{exp}\left(-\frac{\mathbf{k}^2}{\hat q t}\right) \,.
  \label{eq:ktformula}
\end{equation}
This Ansatz should, in principle, only be compared to the gluon distribution resulting from a fragmenting gluon.

Besides this, we compare the results to 
\begin{equation}
\label{eq:DG-2}
    D_\text{G}(x,\mathbf{k},t) = D(x,t) \, P(\mathbf{k},t) \,,
\end{equation}
where $P(\mathbf{k},t)$ is given by Eq.~$(\ref{eq:ktformula})$, while the $x$ distribution is given by a numerical solution of Eq.~(\ref{eq:eeintkqg}), i.e.\ the evolution equations for the energy distribution.
All of the results for the initial gluon feature universal slow growth if the angle is large enough, while for the processes initiated by the quark we see that the energy saturates.

Furthermore, by comparing the fully analytical Ansatz $D_\text{A}(x,\k,t)$ with other results, we see that both agree at large angles, while at small angles, the analytical results overestimate the amount of energy in the cone.  
The fully analytical Ansatz $D_\text{A}(x,\k,t)$ predicts more collimated jets than the other distributions. This is because, as realised in \cite{Kutak:2018dim}, the analytical distribution given by the Gaussian function $P(\mathbf{k},t)$ is narrower than the one obtained from the exact numerical solution which can be viewed as superposition of many Gaussians with different widths.
The above plots allow us also to conclude that the $k_T$ dependence controls the angle at which the distribution starts to saturate. 

\begin{figure}[!htbp]
    \centering
    \includegraphics[width=0.35\linewidth]{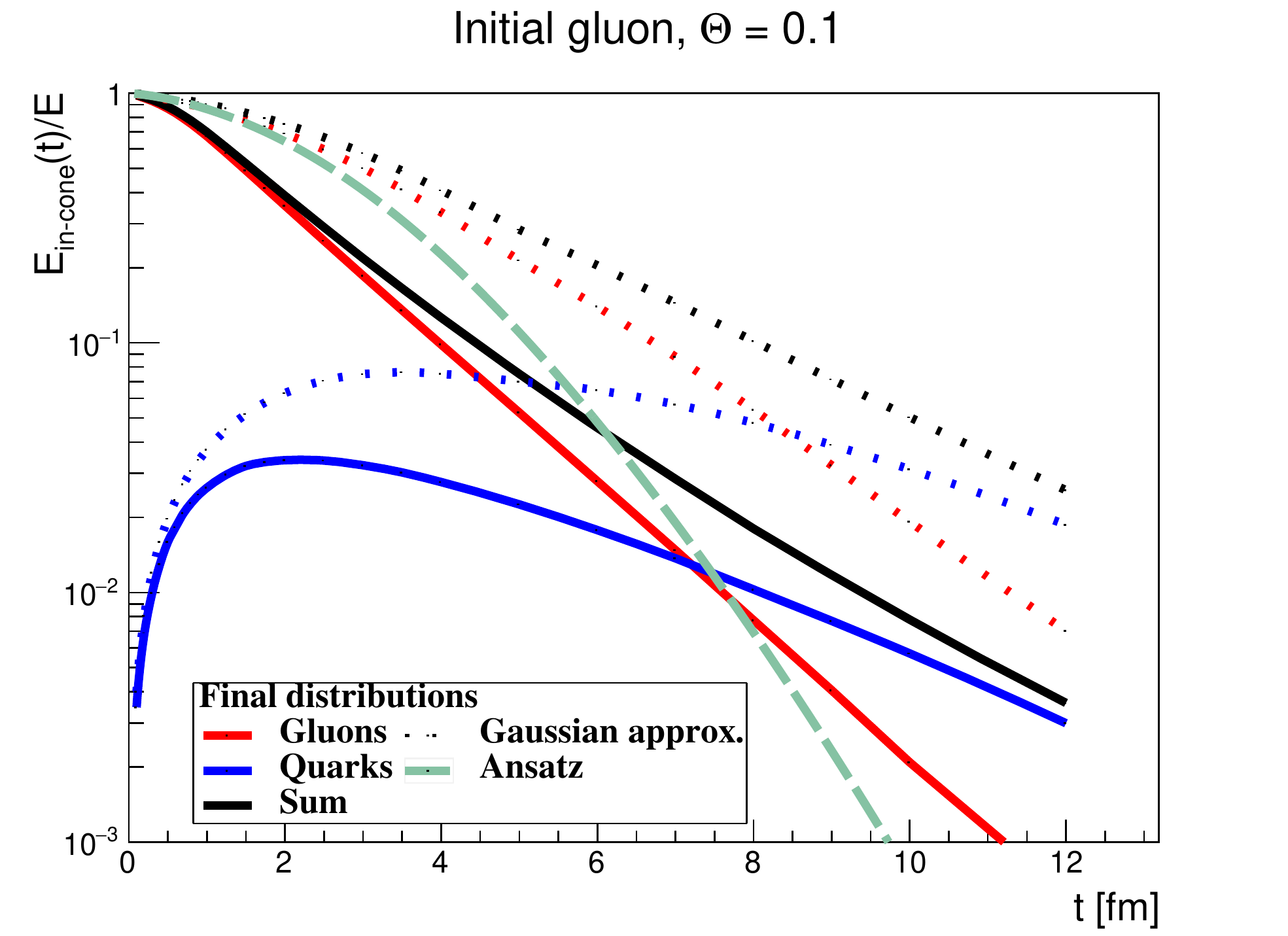}
    \includegraphics[width=0.35\linewidth]{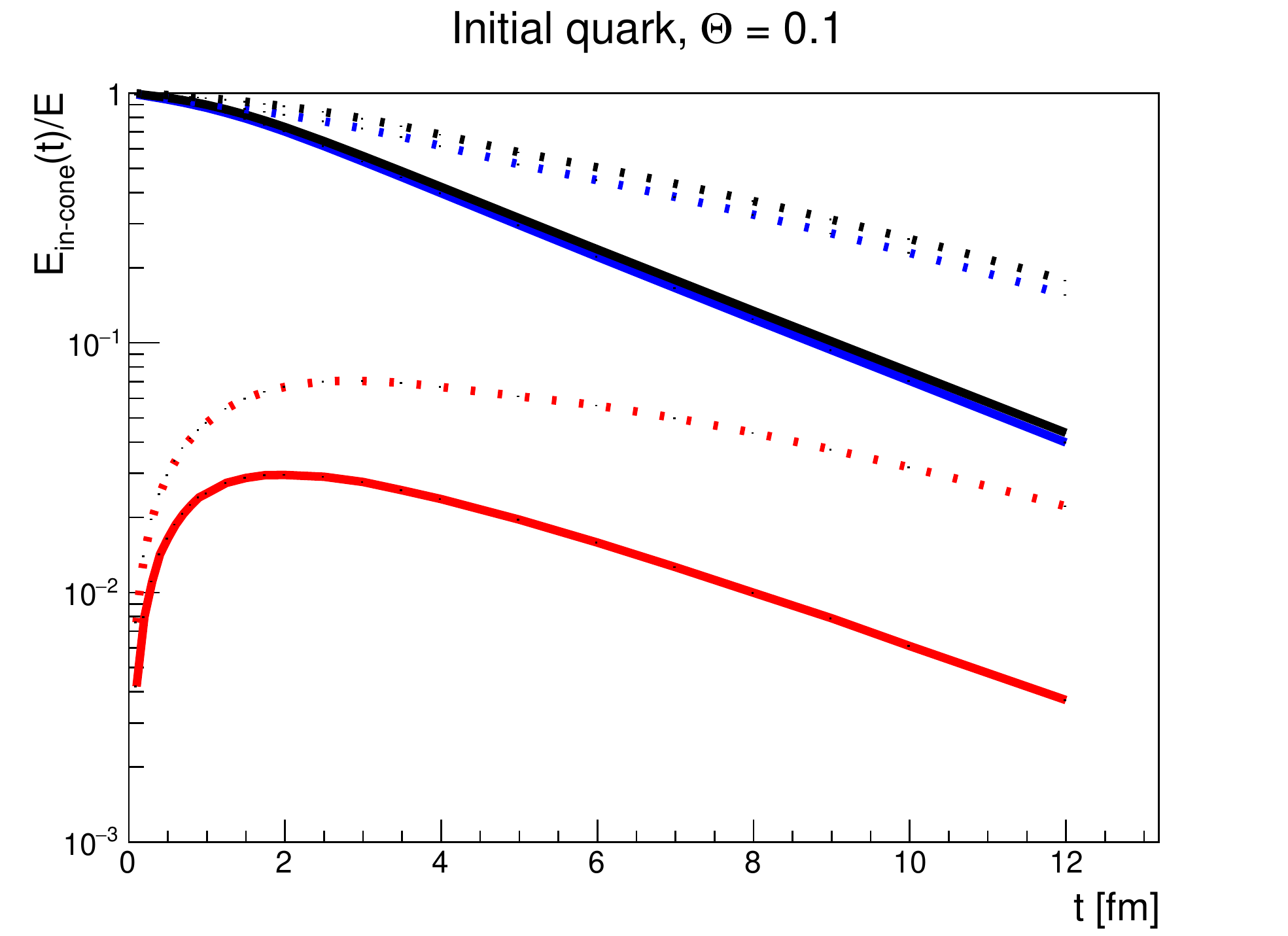}
    \includegraphics[width=0.35\linewidth]{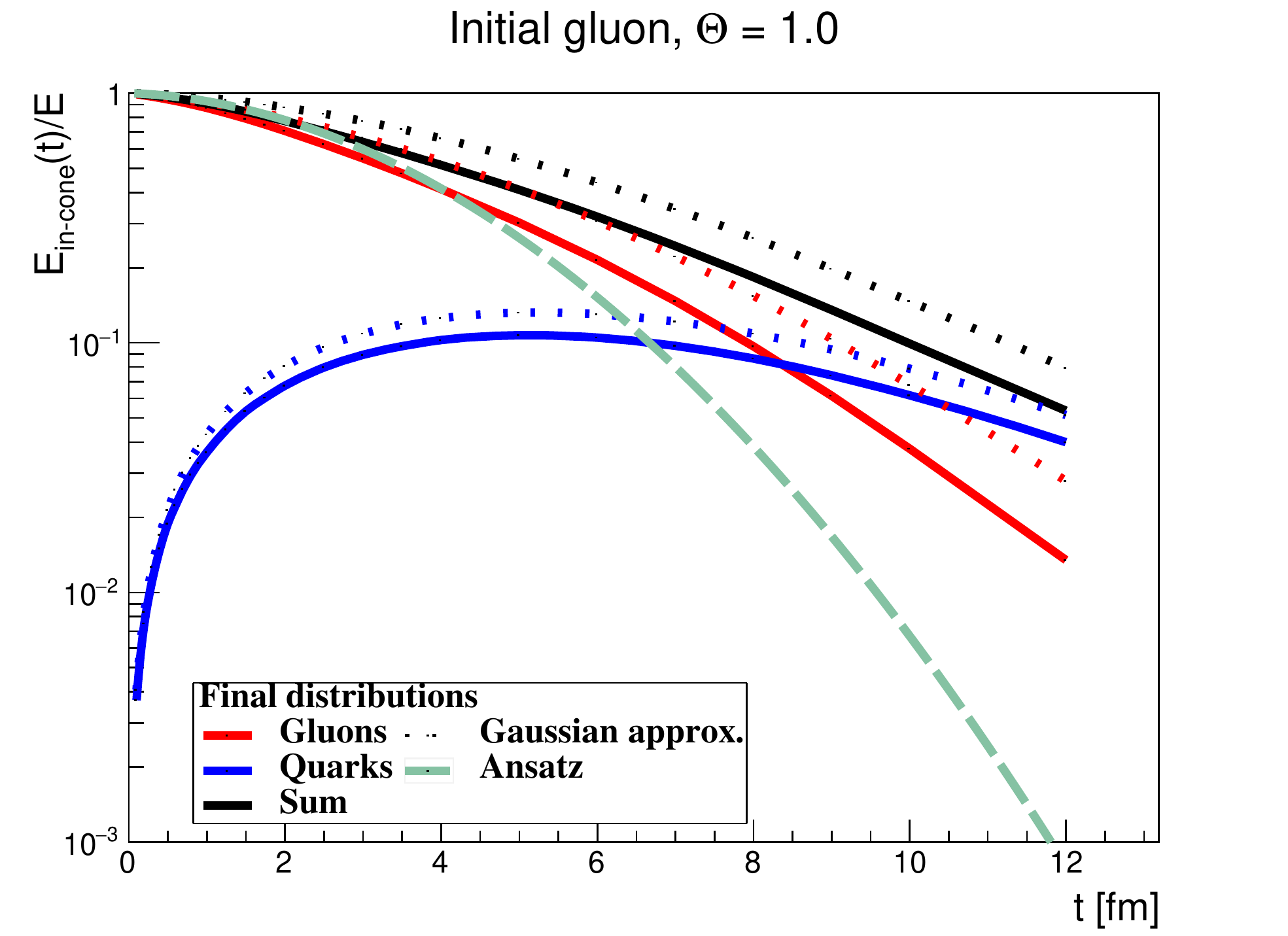}
    \includegraphics[width=0.35\linewidth]{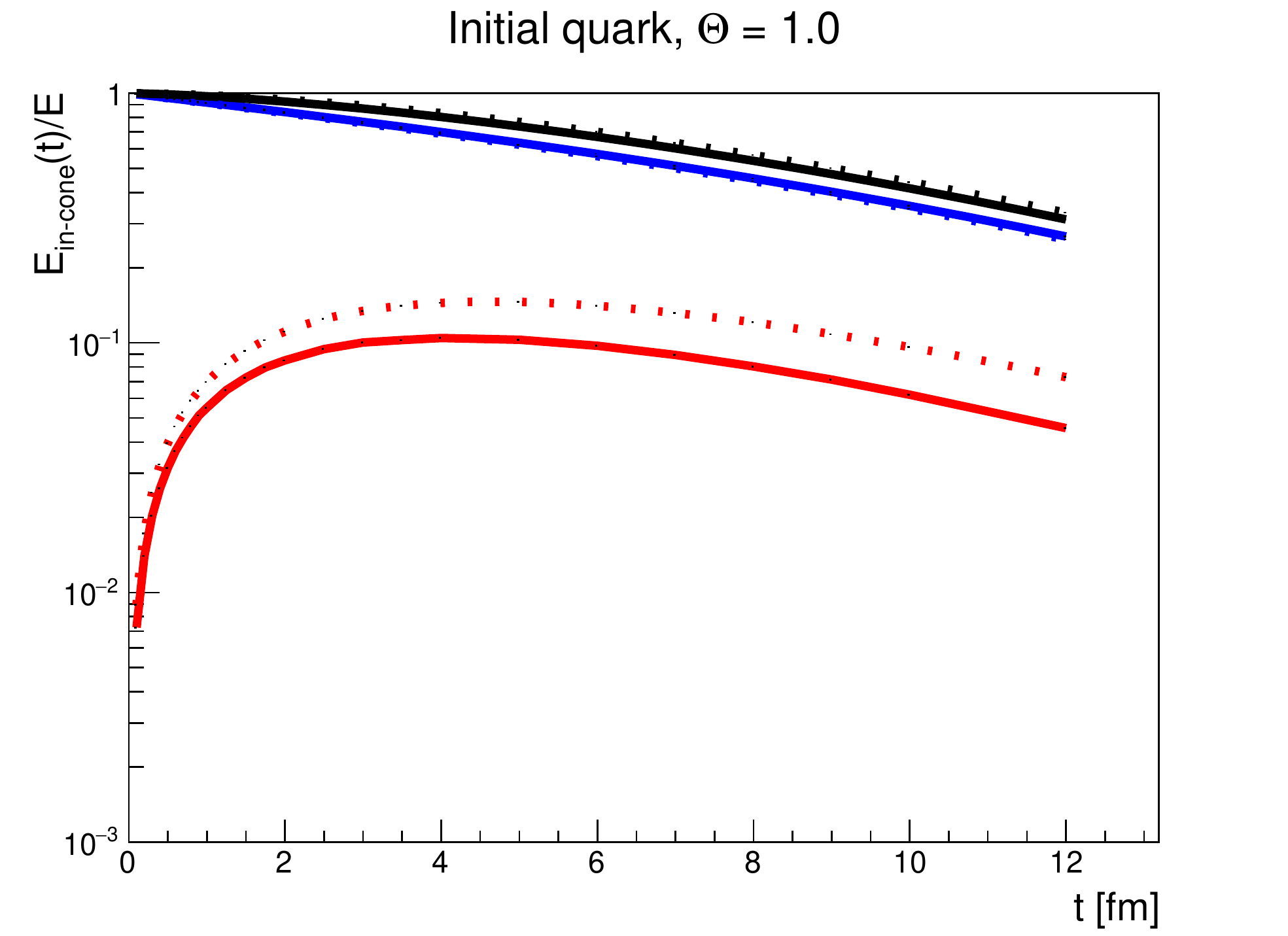}
    \caption{Time evolution of the jet energy in a fixed cone with $\theta=0.1$ (upper panels) and $\theta=1$ (lower panels) for the initial gluon jet (left panels) and the initial quark jet (right panels) in the case of the $k_T$-dependent branching and $w(\mathbf{l})\propto 1/[\mathbf{l}^2(m_D^2+\mathbf{l}^2)]$ (solid lines), the Gaussian approximation in Eq.~\eqref{eq:DG-2} (dotted lines) and the analytical Ansatz in Eq.~\eqref{eq:DA-2} (green dashed line); 
    the red lines correspond to the final gluon jets and the blue lines to the final quark jets. The curves for the analytical Ansatz are plotted only in the case of pure gluon jets.}
    \label{fig:Econet}
\end{figure}

The final feature that we discuss is the normalised amount of energy in a cone as a function of time. The results are presented in~Fig.~\ref{fig:Econet}. 
One can clearly see that for processes with quarks in the initial state, quarks dominate at any time-scales in both scenarios for the angles that we consider.
But what is more striking is that even when gluons are in the initial state, quarks start to dominate at late times.
One can also see that the analytical solution Eq.~(\ref{eq:DA-2}) overshoots the Monte Carlo solution of Eq.~(\ref{eq:BDIM1}) for the short time-scales while underestimates it for the long time-scales.

\section{Conclusions}

Our goal in this work was to study simultaneous evolution of quarks and gluons with transverse and longitudinal fragmentation functions in medium. In order to achieve that, first of all, we have introduced transverse-momentum-dependent splitting kernels that take into account broadening during branching. The derivation assumes that there is a local in time factorisation between longitudinal and transverse momentum dependence, valid when the transverse momentum is much smaller than the total energy of the collision. The obtained system of equations has been solved in full generality using Monte Carlo methods as well as by the Chebyshev-polynomial method applied to the equations for the energy distribution. 

Using the obtained solutions we have defined some characteristic features that allow us to demonstrate that
\begin{itemize}
\item as evolution in time progresses gluons broaden more than quarks, see Fig.~\ref{fig:kT_kzq_wq2};
\item at late times, for both quarks and gluons, there is universal distribution in $\langle k_T\rangle$ at low $x$; 
\item quarks are more collimated than gluons -- this we conclude from the calculation of energy contained in a cone (in the case of quarks more energy is contained in a cone);
\item quarks dominate at late times, see Fig.~\ref{fig:Econet}.
\end{itemize}
We should add here that our study breaks down at $x\sim 10^{-5}$ where one should account for thermalisation and possible recombination of gluons, solving an appropriate Boltzmann equation that also accounts for elastic rescattering. 

As an outlook, we mention here several improvements that are natural to consider for future work.
A natural step is to account for medium expansion. This was already implemented for a purely radiative cascade in \cite{Adhya:2019qse,Adhya:2021kws}, which made use of appropriate splitting kernels in an expanding background. Furthermore, it would be interesting to extend the formalism presented in this paper to include hard emissions, generated by rare, hard scatterings, as developed in \cite{Mehtar-Tani:2019tvy,Mehtar-Tani:2019ygg,Barata:2020sav,Barata:2020rdn,Barata:2021wuf} which also is well suited for expanding media, see also \cite{Feal:2018sml,Andres:2020vxs,Schlichting:2020lef,Schlichting:2021idr} for numerical approaches to evaluating the splitting kernel. Finally, we also notice that coherence effects are important to account for finite-size medium corrections \cite{Barata:2021byj}. These extensions are, however, beyond the scope of the current paper.

\section*{Acknowledgements}
\label{sec:Ack}
EB would like to thank Krzysztof Golec-Biernat for the useful discussion on the Chebyshev method for solving the integro-differential equations.

\appendix 

\section{Splitting functions}
\label{sec:splitting-func}
\renewcommand\exp{{\rm exp}}

Here we provide exact equations for the splitting functions we are using. 
The main object is the splitting function
\begin{equation}
    \mathcal{K}_{IJ}(z,y,\Q) = \frac{z}{\sqrt{y}} (1+\delta_{Ig}\delta_{Jg})\,
    \mathcal{P}_{IJ}(z)\, \mathcal{\tilde R}_{IJ}(\Q,z,y p_0^+) \,,
\end{equation}
where
\begin{equation}
    \mathcal{\tilde R}(\Q,z,p_0^+) = \frac{1}{\pi k_{\rm br}^2(z,p_0^+)}\, \sin \left[ \frac{\Q^2}{2 k_{\rm br}^2(z,p_0^+)} \right] \exp\left[-\frac{\Q^2}{2 k_{\rm br}^2(z,p_0^+)}\right] \,,
\end{equation}
and
\begin{equation} 
k_{\rm br}^2(z,p_0^+) = \sqrt{z(1-z) p_0^+ f_{IJ}(z)\hat{\bar q}} \,,
\end{equation}
where the functions $f_{IJ}(z)$ can be read off from Eq.~\eqref{eq:fij}.
Finally, the $z$-dependent in-medium splitting functions are defined as
\begin{equation}
    \mathcal{P}_{IJ}(z) = \frac{1}{2}P_{IJ}(z)\, \sqrt{\frac{f_{IJ}(z)}{z(1-z)}} \,,
\end{equation}
where we use the following expressions for the unregularised Altarelli--Parisi splitting functions:%
\footnote{In the expression for $P_{gg}$ we omit a factor of $2$, as used e.g.\ in Ref.~\cite{Ellis:1996mzs}, since it is accounted for as a multiplicative factor of the kernel $\Kc_{gg}$ in Eq.~(\ref{eq:BDIMsys1}).} 
\begin{align}
   P_{gg}(z) &= C_A\, \frac{\left[1-z(1-z)\right]^2}{z(1-z)} \,,\\
   P_{q_ig}(z) &= \frac{1}{N_F}\,P_{qg}(z) = T_R \left[z^2 + (1-z)^2 \right] \,,\\
   P_{gq_i}(z) &= P_{gq}(z) = C_F\, \frac{1+(1-z)^2}{z} \,,\\
   P_{q_iq_j}(z) &= \delta_{ij} P_{qq}(z), \qquad P_{qq}(z) = C_F\, \frac{1+z^2}{1-z} \,.
\end{align}

\section{Other forms of evolution equations}
\label{sec:evol-eq}

For completeness, we present here also the equations obtained from Eq.~(\ref{eq:BDIMsys1}) by neglecting the transverse momentum accumulated during branching, $k_{\rm br}^2 \approx 0$, and integrating over the momentum transfer $\mathbf{Q}$, we get
\begin{equation}
\begin{aligned}
\partialt D_g(x,\mathbf{k},\tau)
= & \intz  2\kgg(z)\left[\frac{1}{z^2}\sqrt{\frac{z}{x}} D_g\left(\frac{x}{z},\frac{\mathbf{k}}{z},\tau\right)-\frac{z}{\sqrt{x}}D_g(x,\mathbf{k},\tau)\right]
-\intz\, 2\kqg(z)\frac{z}{\sqrt{x}}D_g(x,\mathbf{k},\tau)
\\
& +\intz\,  \kgq(z) \frac{1}{z^2}\sqrt{\frac{z}{x}} \sum_{i} D_{q_i}\left(\frac{x}{z},\frac{\mathbf{k}}{z},\tau\right) + 
\bar t_\ast \int \frac{\rmd ^2\mathbf{l}}{(2\pi)^2} \,C_g(\mathbf{l})\, D(x,\mathbf{k}-\mathbf{l},\tau)\,,\\
\partialt \dq(x,\mathbf{k},\tau)
= & \intz\,\kqq(z)\left[ \frac{1}{z^2}\sqrt{\frac{z}{x}}\dq\left(\frac{x}{z},\frac{\mathbf{k}}{z},\tau\right)
-\frac{1}{\sqrt{x}}\dq\left(x,\mathbf{k},\tau\right)\right] \\
& + 
\frac{1}{N_F} \intz\, \kqg(z)  \frac{1}{z^2} \sqrt{\frac{z}{x}}\dg\left(\frac{x}{z},\frac{\mathbf{k}}{z},\tau\right)
+ \bar t_\ast \int \frac{\rmd ^2\mathbf{l}}{(2\pi)^2} \,C_q(\mathbf{l})\, D_q(x,\mathbf{k}-\mathbf{l},\tau)\,,
\end{aligned}
\label{eq:eeintQ}
\end{equation}
where $\tau = t/\bar t_\ast$ and the sum over $i$ runs over both quarks and antiquarks.
By integrating out completely the transverse momentum, we obtain 
\begin{equation}
\begin{aligned}
\partialt D_g(x,\tau)= 
& 
\intz\, 2\kgg(z)\left[\sqrt{\frac{z}{x}} D_g\left(\frac{x}{z},\tau\right)-\frac{z}{\sqrt{x}}D_g(x,\tau)\right]
-\intz\,2 \kqg(z)\frac{z}{\sqrt{x}}D_g(x,\tau)
\\
&
+\intz \, \kgq(z) \sqrt{\frac{z}{x}} \sum_{i} D_{q_i}\left(\frac{x}{z},\tau\right)\,,\\
\partialt \dq(x,\tau) = 
&
\intz\,\kqq(z)\left[\sqrt{\frac{z}{x}}\dq\left(\frac{x}{z},\tau\right)-\frac{1}{\sqrt{x}}\dq(x,\tau)\right]
+\frac{1}{N_F} \intz\,\kqg(z)\sqrt{\frac{z}{x}}\dg\left(\frac{x}{z},\tau\right).
\end{aligned}
\label{eq:eeintkqg}
\end{equation}
We can also rewrite the latter equations in the gluon/singlet/non-singlet basis \cite{Mehtar-Tani:2018zba}, where we find
\begin{equation}
\begin{aligned}
\partialt D_g(x,\tau)= 
&
\intz\,2 \kgg(z)\left[\sqrt{\frac{z}{x}}D_g\left(\frac{x}{z},\tau\right)
-\frac{z}{\sqrt{x}}D_g(x,\tau)\right] - \intz \, 2 \kqg(z)\frac{z}{\sqrt{x}}D_g(x,\tau)\\
 &
 +\intz\, \kgq(z)\sqrt{\frac{z}{x}}D_S\left(\frac{x}{z},\tau\right)\,,\\
\partialt D_S(x,\tau)=
&
\intz\, \kqq(z)\left[\sqrt{\frac{z}{x}}D_S\left(\frac{x}{z},\tau\right)-\frac{1}{\sqrt{x}}D_S(x,\tau)\right]+\intz\, 2 \kqg(z)\sqrt{\frac{z}{x}}D_g\left(\frac{x}{z},\tau\right)\,,\\
\partialt D_{NS}^{(i)}(x,\tau)= 
&
\intz \,\kqq(z)\left[\sqrt{\frac{z}{x}}D_{NS}^{(i)}\left(\frac{x}{z},\tau\right)-\frac{1}{\sqrt{x}}D_{NS}^{(i)}(x,\tau)\right]\,,
\end{aligned}
\label{eq:eeintk}
\end{equation}
with 
\begin{equation}
D_S=\sum_{i=1}^{N_F}\big[ D_{q_i}+D_{\bar{q}_i}\big] \,,\qquad D_{NS}^{(i)}=D_{q_i}-D_{\bar{q}_i}\,.
\label{eq:DSDNS}
\end{equation}
At this level, we see that it is quite natural to absorb an extra factor 2 into the gluon--gluon and gluon--quark splitting functions, respectively. Compared to the notation in \cite{Mehtar-Tani:2018zba}, we identify their splitting functions with ours as: $\mathcal{K}_{gg}(z) = 2\mathcal{P}_{gg}(z)$,  $\mathcal{K}_{qg}(z) = 2\mathcal{P}_{qg}(z)$, $\mathcal{K}_{gq}(z) = \mathcal{P}_{gq}(z)$ and $\mathcal{K}_{qq}(z) = \mathcal{P}_{qq}(z)$.

{\color{black}
\section{First-order perturbative estimate of the distribution functions}
\label{sec:evol-first}

It is very instructive to compare the result of the full medium evolution with a perturbative estimate which only accounts for one single emission or elastic scattering with the medium. In this Appendix, we summarise the results for such a first-order perturbative estimate based on formulas .

For the partonic gluon initiator,
\begin{equation}
    D^{(0)}_g(x,\k) = x \delta(1-x) \, \delta^{(2)}(\k) \,, \qquad \text{and}\quad D^{(0)}_q(x,\k) = 0 \,,
\end{equation}
we find the first-order perturbative estimate to be,
\begin{align}
    D_g^{(1)}(x,\k,\delta t) &= D_g^{(0)}(x,\k)\left\{1 - \delta t \left[\alpha_s \int_0^1 \rmd z\, \left(\Kc_{gg}(z,p_0^+) + \Kc_{qg}(z,p_0^+) \right) + \int \frac{\rmd^2 \q}{(2\pi)^2} \, w_g(\q)\right] \right\} \nn
    & + \frac{\delta t}{(2\pi)^2} \left[2 \alpha_s \,x \Kc_{gg}(\k,x,p_0^+) + x\delta(1-x) w_g(\k) \right] \,,\\
    D_q^{(1)}(x,\k,\delta t) &= \frac{\delta t}{(2\pi)^2}\, \frac{\alpha_s}{N_F} x
\Kc_{qq}(\k,x,p_0^+) \,,
\end{align}
where $\Kc_{ij}(x,p_0^+) = \int \frac{\rmd^2\k}{(2\pi)^2}\, \Kc_{ij}(\k,x,p_0^+)$. The term in curly brackets next to the initial condition $D_g^{(0)}(x,\k)$ is responsible for probability conservation. Now, integrating out the transverse momentum $\k$ or the momentum fraction $x$, respectively, we get the following distributions:
\begin{equation}
    D(x, t) \equiv \int \rmd^2\k \, D(x,\k,t) \,, \qquad D(\k,t) = \int_0^1 \rmd x \, D(x,\k,t) \,,
\end{equation}
which respects $\int_0^1 \rmd x\, D(x, t) = 1$ and $\int \rmd^2\k \, D(\k, t) =1$.
The expressions for these take the form,
\begin{align}
    D_g^{(1)}(x,\delta t) &=  \delta t\,2 \alpha_s\, x \Kc_{gg}(x,p_0^+) \,,\\
    D_q^{(1)}(x,\delta t) &=  \delta t \, \frac{\alpha_s}{N_F} x\Kc_{qg}(x,p_0^+) \,,
\end{align}
and, finally,
\begin{align}
    D_g^{(1)}(\k,\delta t) &= \frac{\delta t}{(2\pi)^2}\, \int_0^1 \rmd x\,2 \alpha_s\, x \Kc_{gg}(\k,x,p_0^+) + \frac{\delta t}{(2\pi)^2} w_g(\k) \,,\\
    D_q^{(1)}(\k,\delta t) &= \frac{\delta t}{(2\pi)^2}\,\frac{\alpha_s}{N_F} \int_0^1 \rmd x \,x \Kc_{qg}(\k,x,p_0^+) \,.
\end{align}
For readability, we have dropped terms that are proportional to the initial condition. They are however crucial in order to restore the normalisation of the distributions. Finally, in order to make contact with Eq.~\eqref{eq:Dtilde}, we note that $D(\k,\delta t) = D(\k^2,\delta t)$, so that
\begin{equation}
    \label{eq:tilde-dkt-def}
    \tilde D^{(1)}_i(k_T,\delta t) = 2\pi\, k_T\, D^{(1)}_i(k_T,\delta t) \,,
\end{equation}
where $k_T \equiv |\k|$.

\begin{figure}[!htbp]
    \centering
    \includegraphics[scale=0.4]{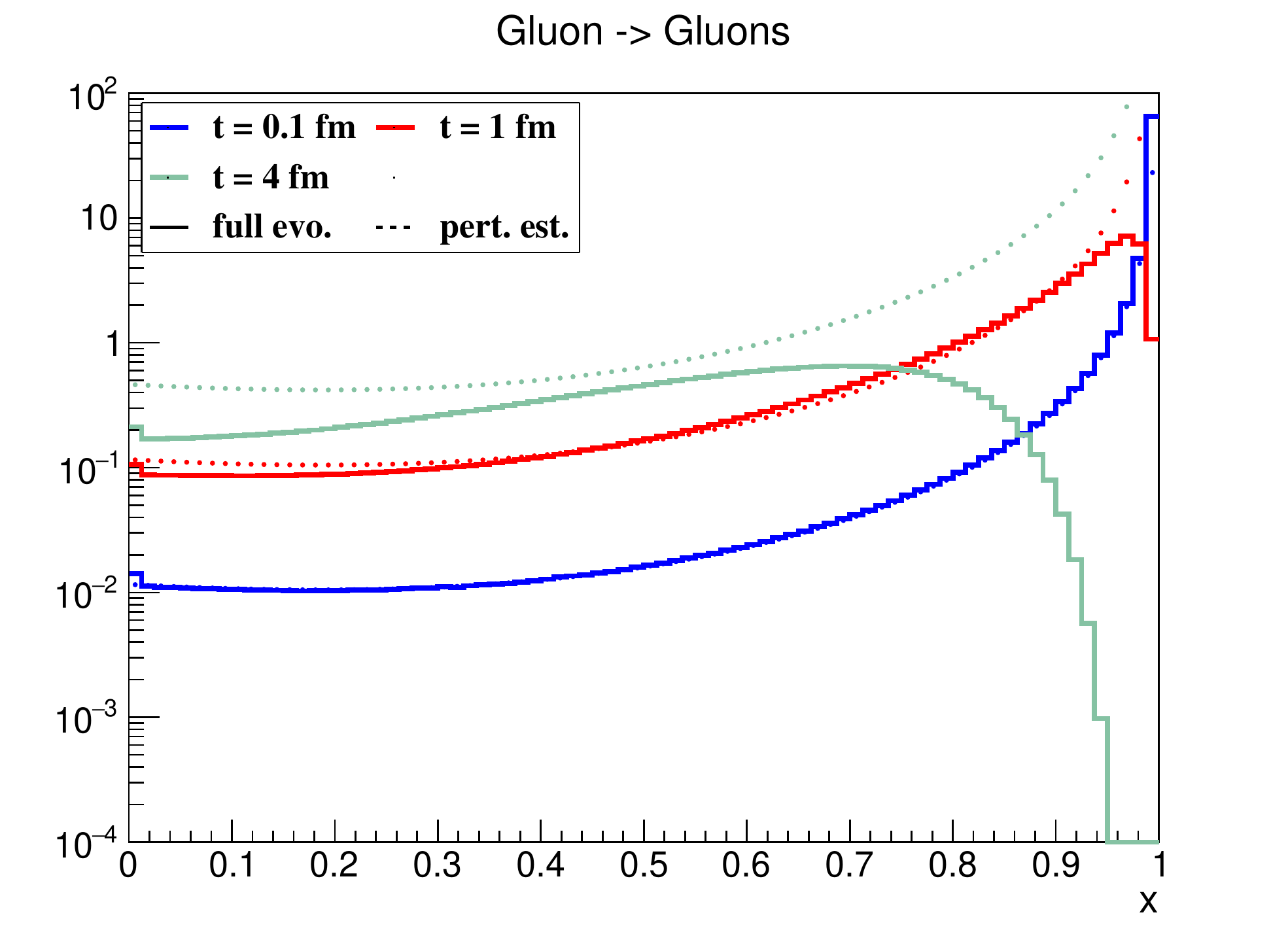}
    \includegraphics[scale=0.4]{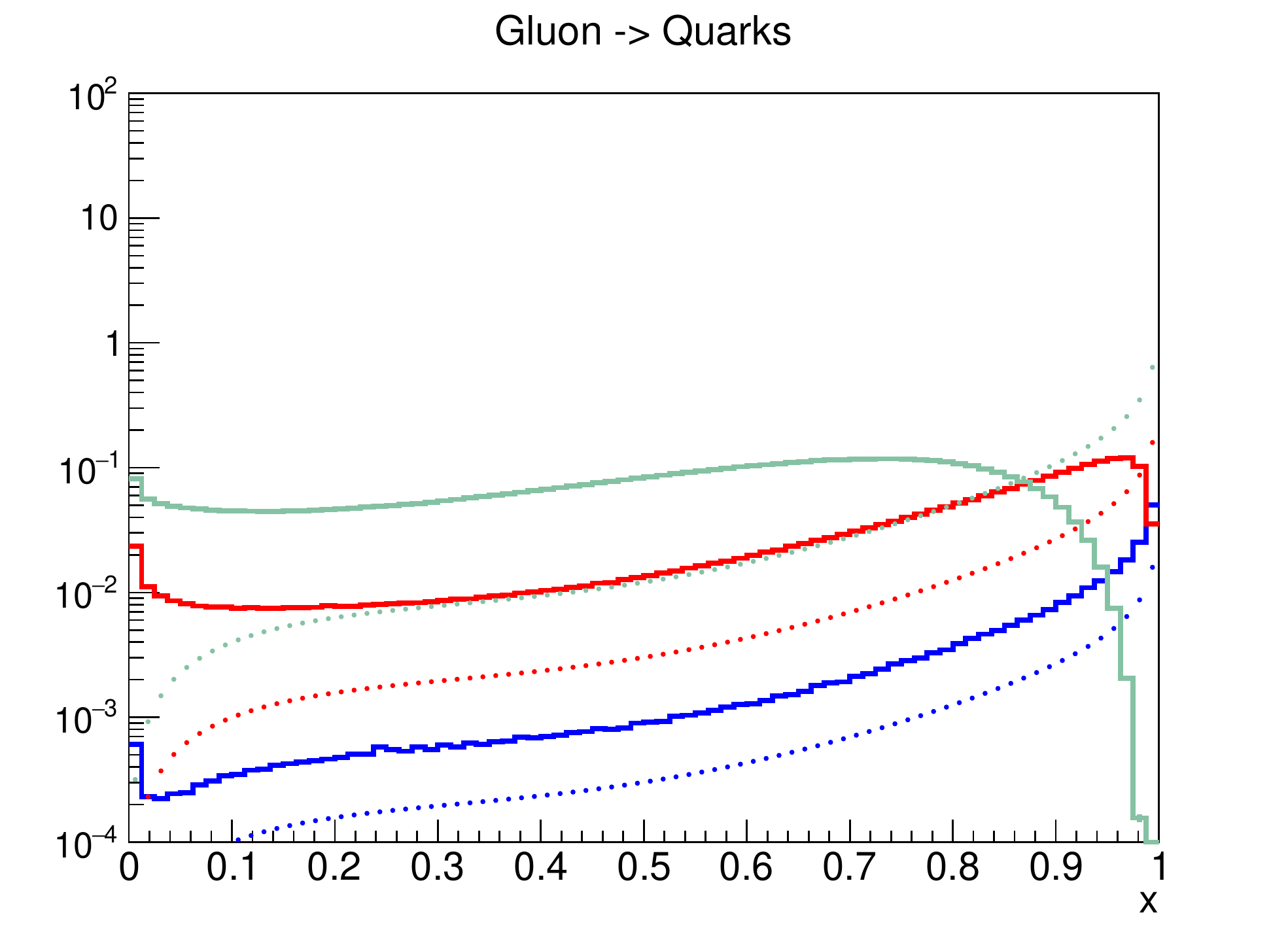}
    \caption{The $\sqrt{x}D(x,t)$ distributions 
    at the time-scales $t=0.1, 1, 4\,$fm:
    cascades initiated by gluon. The solid lines correspond to the full medium evolution and the dotted line to the perturbative estimate.}
\end{figure}

\begin{figure}[!htbp]
    \centering
    \includegraphics[scale=0.4]{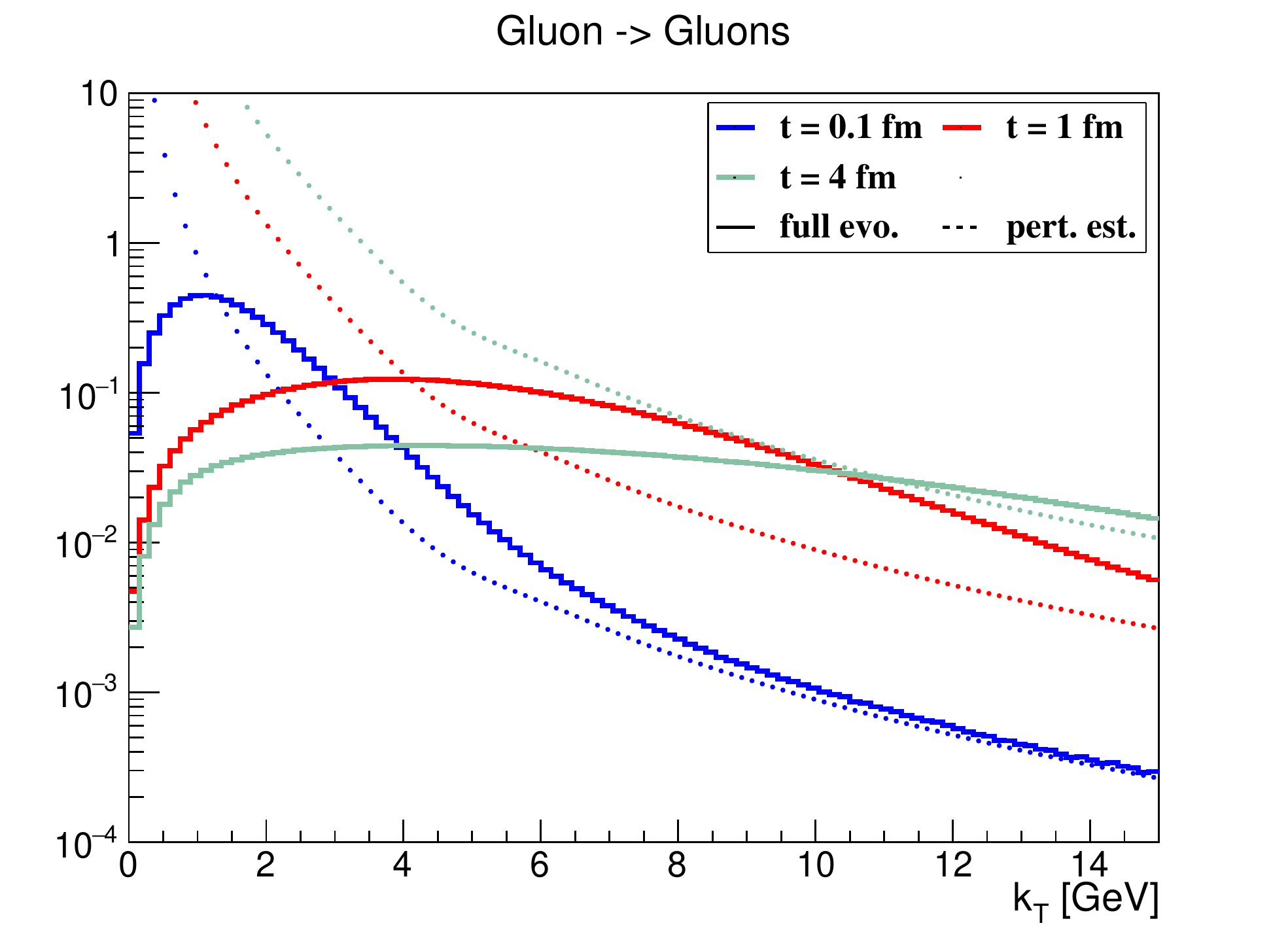}
    \includegraphics[scale=0.4]{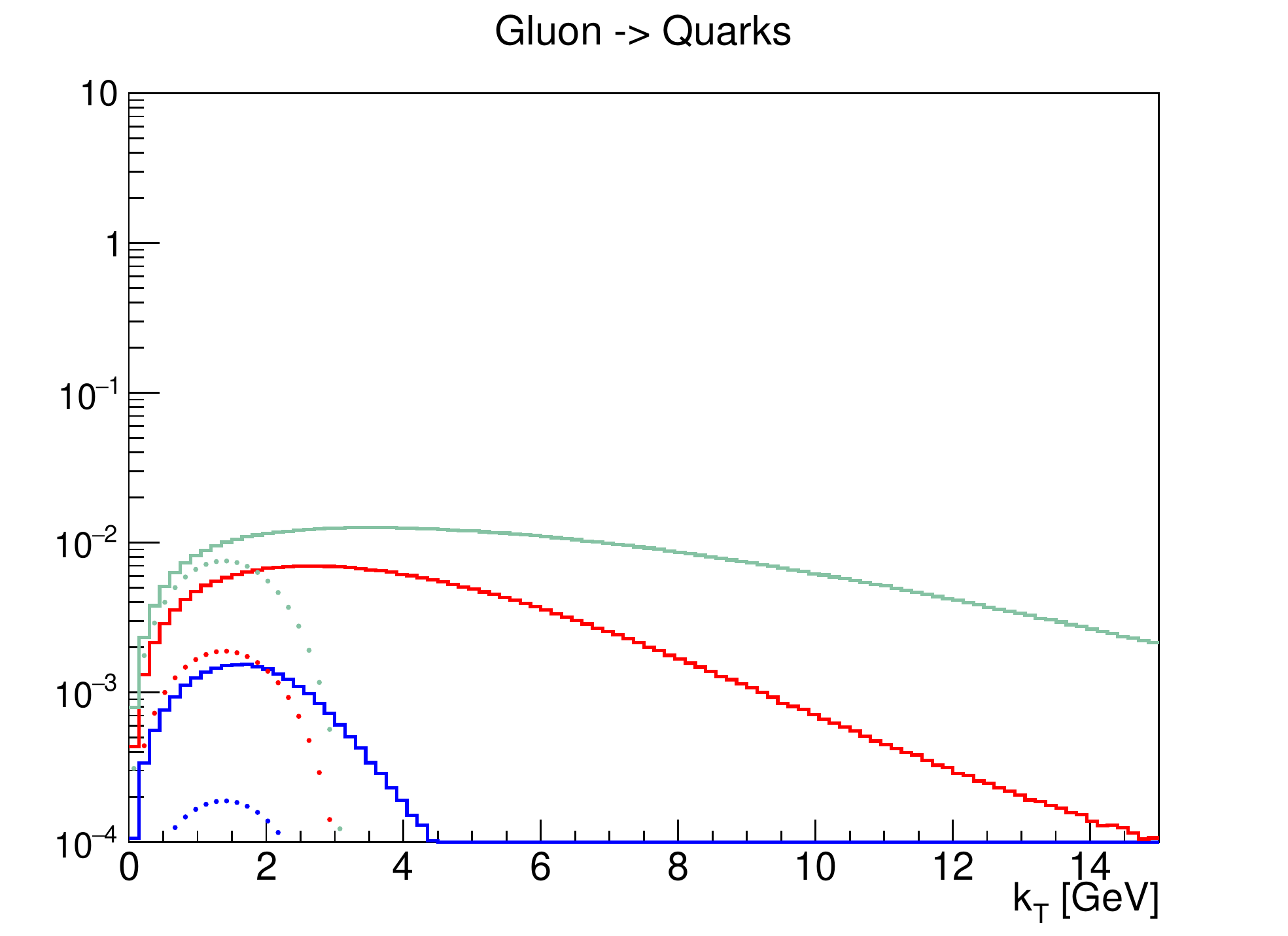}
    \caption{The $\tilde{D}(k_T,t)$ distributions 
    at the time-scales $t=0.1, 1, 4\,$fm:
    cascades initiated by gluon. The solid lines correspond to the full medium evolution and the dotted line to the perturbative estimate.}
\end{figure}

For the quark initiator,
\begin{equation}
    D^{(0)}_g(x,\k) = 0 \,, \qquad \text{and}\quad D^{(0)}_q(x,\k) = x \delta(1-x) \, \delta^{(2)}(\k) \,,
\end{equation}
we obtain,
\begin{align}
    D_g^{(1)}(x,\k,\delta t) &= \frac{\delta t}{(2\pi)^2}\, \alpha_s\, N_F\, x \Kc_{gq}(\k,x,p_0^+) \,,\\
    D_q^{(1)}(x,\k,\delta t) &= D_q^{(0)}(x,\k)\left\{1 - \delta t \left[\alpha_s \int_0^1 \rmd z\, \Kc_{qq}(z,p_0^+) + \int \frac{\rmd^2 \q}{(2\pi)^2} \, w_q(\q)\right] \right\} \nn
    & + \frac{\delta t}{(2\pi)^2} x\delta(1-x) w_q(\k) \,.
\end{align}
The integrated distributions now read, omitting the terms proportional to the initial condition,
\begin{align}
    D_g^{(1)}(x,\delta t) &=  \delta t\, \alpha_s\, N_F\, x \Kc_{gq}(x,p_0^+) \,,\\
    D_q^{(1)}(x,\delta t) &=  \delta t \, \alpha_s\, x\Kc_{qq}(x,p_0^+) \,,
\end{align}
and, finally,
\begin{align}
    D_g^{(1)}(\k,\delta t) &= \frac{\delta t}{(2\pi)^2}\, \int_0^1 \rmd x\, \alpha_s\, N_F\, x \Kc_{gq}(\k,x,p_0^+)\,,\\
    D_q^{(1)}(\k,\delta t) &= \frac{\delta t}{(2\pi)^2}\, \alpha_s\, \int_0^1 \rmd x \,x \Kc_{qq}(\k,x,p_0^+) \,.
\end{align}
}

\section{Numerical-solution methods}
\label{sec:numerics}

\subsection{Monte Carlo algorithms}

Below we briefly sketch appropriate algorithms implemented in two independent Monte Carlo programs, 
\mincas\ and \tmdice. 
One essential difference between these approaches is that \mincas\ provides Monte Carlo solutions for the fragmentation functions that follow Eqs.~(\ref{eq:BDIMsys1}), while \tmdice\ solves the equivalent equations
for the multiplicity distributions, i.e.\ with the replacement: 
 $D_I(x,\mathbf{k},t) \rightarrow F_I (x,\mathbf{k},t)= D_I(x,\mathbf{k},t)/x$.

First, let us rewrite Eq.~(\ref{eq:qgkQitsol}) in a probabilistic form which is suited for 
the MCMC algorithm. To this end we define 
two probability distribution functions (p.d.f.s):
\begin{equation}
\varrho_{J_{i-1}}(\tau_i) = \Psi_{J_{i-1}}(x_{i-1}) \,e^{-\Psi_{J_{i-1}}(x_{i-1}) \,,
(\tau_i - \tau_{i-1})}\,\theta(\tau_i-\tau_{i-1})
\label{eq:qgProbtaui}
\end{equation}
for generation of the random variable $\tau_i$, and
\begin{equation}
 \Xi_{J_iJ_{i-1}}(z_i,\mathbf{Q}_i,\mathbf{l}_i) = \frac{\EuScript{G}_{J_iJ_{i-1}}(z_i,x_{i-1},\mathbf{Q}_i,\mathbf{l}_i)}{\Psi_{J_iJ_{i-1}}(x_{i-1})}\,,
 \qquad
 \Psi_{J_iJ_{i-1}}(x_{i-1}) = \Phi_{J_iJ_{i-1}}(x_{i-1}) + W_{J_i},
 \label{eq:qgprobzQl}
\end{equation}
for generation of the random variables $(z_i,\mathbf{Q}_i,\mathbf{l}_i)$,
and the conditional probability for generating a new parton flavour $J_i$, given the flavour $J_{i-1}$:
\begin{equation}
 p_{J_i} \equiv p(J_i|J_{i-1}) = \frac{\Psi_{J_i J_{i-1}}(x_{i-1}) \,
 e^{-\Psi_{J_{i-1}}(x_{i-1}) (\tau_i - \tau_{i-1})} }{\varrho_{J_{i-1}}(\tau_i)}\,.
\label{eq:qgProbJi}
\end{equation}
One may easily check that they are properly normalised to $1$.

Then, Eq.~(\ref{eq:qgkQitsol}) can be written as
\begin{equation}
 \begin{aligned}
D_I (x,\tau) 
 = & \; d(\tau_0) \int_0^1 \rmd x_0 \int \rmd ^2 \mathbf{k}_0 \, \chi(x_0,\mathbf{k}_0) \sum_{J_0} p_{J_0} 
 \Bigg\{  \int_{\tau}^{+\infty} \rmd \tau_1 \,\varrho_{J_0}(\tau_1) \, 
\delta_{J_0 I}\,\delta(x-x_0)\,\delta(\mathbf{k}-\mathbf{k}_0)
\\&
+ \, \sum_{n=1}^{\infty} \,\sum_{J_1,\ldots J_n}\, \prod_{i=1}^n \left[ \int_0^{\tau} \rmd \tau_i \, \varrho_{J_{i-1}}(\tau_i)\, p_{J_i} 
\int_0^1 \rmd z_i \int \rmd ^2\mathbf{Q}_i \int \rmd ^2\mathbf{l}_i \, \Xi_{J_i J_{i-1}}(z_i,\mathbf{Q}_i,\mathbf{l}_i) \right]
\\&
\hspace{22mm}
\int_{\tau}^{+\infty} \rmd \tau_{n+1} \,\varrho_{J_n}(\tau_{n+1})\, \delta_{J_n I}\,\delta(x-x_n)\,\delta(\mathbf{k}-\mathbf{k}_n)  \Bigg\} \,,
\end{aligned}
\label{eq:qgxkIterProb}
\end{equation}
where the initial functions are
\begin{equation}
 \begin{aligned}
d(\tau_0) &= \int_0^1 \rmd x_0 \int \rmd ^2 \mathbf{k}_0 \sum_J D_J(x_0, \mathbf{k}_0,\tau_0)\,,
\\
\chi(x_0,\mathbf{k}_0) &= \frac{\sum_J D_J(x_0, \mathbf{k}_0,\tau_0)}{d(\tau_0)}\,,
\\
p_{J_0} &= \frac{D_J(x_0, \mathbf{k}_0,\tau_0)}{\sum_J D_J(x_0, \mathbf{k}_0,\tau_0)}\,.
\end{aligned}
\label{eq:qgxkdtau0}
\end{equation}
The formula of Eq.~(\ref{eq:qgxkIterProb}) is a basis of the following MCMC algorithm:
\begin{enumerate}
\item
Set the initial values of $(x_0,\mathbf{k}_0)$ and $J_0$ or generate them according to the probabilities $\chi(x_0,\mathbf{k}_0)$ 
and $p_{J_0}$, respectively.
\item
Having generated the random-walk leap $i-1\, (i=1,2,\ldots)$, generate $\tau_i$ according to the p.d.f.\ $\varrho_{J_{i-1}}(\tau_i)$: 
if  $\tau_i > \tau$ set $x=x_{i-1},\, \mathbf{k} =\mathbf{k}_{i-1},\,I=J_{i-1}$ and {\bf stop}. 
\item
Otherwise, i.e.\ $\tau_i \leq \tau$:
\begin{itemize}
\item[(a)] 
generate the parton flavour $J_i$ according the probability $p_{J_i}$, 
\item[(b)] 
generate the variables $(z_i,\mathbf{Q}_i,\mathbf{l}_i)$ according to the p.d.f.\ $\Xi_{J_iJ_{i-1}}(z_i,\mathbf{Q}_i,\mathbf{l}_i)$,
 \item[(c)] 
set $x_i = z_i\, x_{i-1}$ and $\mathbf{k}_i = z_i\mathbf{k}_{i-1} + \mathbf{Q}_i + \mathbf{l}_i$,
\item[(d)] 
increment $i \rightarrow i+1$ and go to step 2.
\end{itemize}
\end{enumerate}
For the Monte Carlo solution of 
the evolution equations for the multiplicity distributions, as implemented in \tmdice,
the algorithm is analogous, while the probability distributions change: 
instead from $\Xi_{J_iJ_{i-1}}(z_i,\mathbf{Q}_i,\mathbf{l}_i)$, the variables 
in step 3(b) of the algorithm are generated from the distribution
\begin{equation}
    \tilde{\Xi}_{J_iJ_{i-1}}(z_i,\mathbf{Q}_i,\mathbf{l}_i) = \frac{\tilde{\EuScript{G}}_{J_iJ_{i-1}}(z_i,x_{i-1},\mathbf{Q}_i,\mathbf{l}_i)}{\Psi_{J_iJ_{i-1}}(x_{i-1})}\,,
\end{equation}
where 
\begin{equation}
\tilde{\EuScript{G}}_{IJ}(z,y,\mathbf{Q},\mathbf{l}) =
\tilde{\Kc}_{IJ}(z,y,\mathbf{Q})\theta(1 - \epsilon - z)\delta(\mathbf{l}) +  
\bar t_\ast \frac{w_I(\mathbf{l})}{(2\pi)^2} \,\theta(|\mathbf{l}| -l_{\mathrm{min}})\,\delta(1-z) \delta(\mathbf{Q})\,,
\label{eq:qgesG_mult}
\end{equation}
where $\mathcal{\tilde K}_{IJ}(z,y,\Q) = \Kc_{IJ}(z,y,\Q)/z$.
Please note that the functions $\Phi_{J_iJ_{i-1}}$, $\Psi_{J_iJ_{i-1}}$, and $W_{J_i}$ are the same 
in both the Monte Carlo solution methods, \tmdice\, and \mincas.

The Monte Carlo weight corresponding to the $n$-leap trajectory $\gamma_n$ is
\begin{equation}
w_I^{\gamma_n}(x,\mathbf{k},\tau) = d(\tau_0)\, \delta_{{J_n} I}\, \delta(x-x_n)\,\delta(\mathbf{k}-\mathbf{k}_n) \,.
\label{eq:qgwgtxk}
\end{equation}
One can prove that the expectation value of such a weight corresponds to the distribution
function $D_I (x,\mathbf{k},\tau)$:
\begin{equation}
E\left[w_I^{\gamma}(x,\mathbf{k},\tau)\right] = D_I (x,\mathbf{k},\tau) \,.
\label{eq:qgEwgtxk}
\end{equation}
In Monte Carlo computations, based on the law of large numbers (LLN), 
the above expectation value is approximated by
the arithmetic mean of the Monte Carlo weights for the given $(x,\mathbf{k},\tau)$.
In practice, one makes histograms of desired variables with
the event weight given by Eq.~(\ref{eq:qgwgtxk}).
In the above MCMC algorithm, it is assumed that all the necessary random variable can be
sampled from the respective probability distribution functions. In practice, however, this is
not straightforward, as the corresponding evolution kernels used to construct these p.d.f.s are
complicated functions, in particular they cannot be integrated analytically. Therefore, one
cannot apply basic Monte Carlo techniques, such as analytical inverse transform method,
to generate the respective random variables.

One possible solution is to perform numerical integration and use numerical inverse transform method
for random variables generation. This, however, can be slow in terms of CPU and prone to numerical
instabilities, therefore it has to be done with great care. These methods have been implemented
in the Monte Carlo program \tmdice\ for the case of the the evolution of the multiplicity distributions. 

The other way to deal with this problem is to replace the exact p.d.f.s in Eq.~(\ref{eq:qgxkIterProb}) 
with their approximations, i.e.\ some simpler functions that on the one hand are as close as possible to 
the original functions, but on the other hand can be integrated analytically and used in the analytical 
inverse transform method for the random variables generation. In such a case, these approximate p.d.f.s
are used to generate all the necessary random variables, and all the simplifications are compensated by
appropriate Monte Carlo weights being the ratios of the exact to approximate p.d.f.s.
Here, two Monte Carlo algorithms are possible: (1) the weighted-event (or variable-weight) 
algorithm in which each generated event is associated with the corresponding (variable) 
Monte Carlo weight
and (2) the veto algorithm \cite{Sjostrand:2006za} which uses a dedicated acceptance-complement method
to generate unweighted (i.e.\ weight $=1$) events. These two algorithms have been implemented
in the Monte Carlo event generator \mincas. They have comparable efficiency in generation of $(x,\mathbf{k})$
distributions in terms of the CPU time. It has been checked that both these algorithms give the same
numerical results, which constitutes an important internal test of their implementation in \mincas.

More details on the above Monte Carlo algorithms as well as the programs \mincas\ and \tmdice\ 
will be given in separate publications.

\subsection{Chebyshev method}
\label{sec:ChebM}

As a test of the obtained Monte Carlo solutions, we have used semi-analytical method based on the Chebyshev polynomials to solve equations~(\ref{eq:eeintk}). The idea is to expand the solution with the Chebyshev polynomials:
\begin{equation} \label{eq:Cheb.exp}
D_p(x,\tau) \simeq \frac{2}{N}\sum_{i=0}^{N-1}{}'\sum_{j=0}^{N-1}
D_p(x_j,\tau)T_i(y_j)T_i(y(x)), \quad p=g,NS,S,
\end{equation}

where
\begin{itemize}
\item $T_i$ are the Chebyshev polynomials of the first kind,
\item $\sum{}'$ means that the first term in the sum is divided by 2,
\item $\{y_i\}_{i\in\llbracket 1;N \rrbracket}$ are the nodes (or zeros) of $T_N$: 
$y_i = \cos{\frac{\pi}{N}(i+\frac{1}{2})}, \ i\in\llbracket 1;N \rrbracket$,
\item in the following, we use the notation:  $T_{ij} \equiv T_i(y_j)$, 
$\forall\, i,j\in\llbracket 1;N \rrbracket$,
\item $y:[0,1]\to[-1,1]$ is an arbitrary bijection,
\item $\{x_i\}_{i\in\llbracket 1;N \rrbracket}$ are the arguments of $\{y_i\}$ by $y$: 
$\forall\, i\in\llbracket 1;N \rrbracket, \ x_i = y^{-1}(y_i)$.
\end{itemize}
The idea is then to solve the equations in the nodes of the Chebyshev polynomials in $\tau = t/\bar t_\ast$,
\begin{equation}
\begin{split}
\frac{\partial D_g(x_k,\tau)}{\partial \tau} \simeq &
\frac{2}{N}\sum_{i=0}^{N-1}{}'\sum_{j=0}^{N-1}D_g(x_j,\tau)T_{ij}\Bigg[
-\int_0^{1}\rmd z\,\delta_{jk}\mathcal{K}_{qg}(z)\frac{z}{\sqrt{x_k}}
-\int_0^{x_k}\rmd z\,\delta_{jk}\mathcal{K}_{gg}(z)\frac{z}{\sqrt{x_k}}
\\ & \hspace{2.5cm} + \int_{x_k}^1 \rmd z\, \mathcal{K}_{gg}(z)\left(
\sqrt{\frac{z}{x_k}}T_i\left(y\left(\frac{x_k}{z}\right)\right)
-\delta_{jk}\frac{z}{\sqrt{x_k}}\right) \Bigg] \\
&+\frac{2}{N}\sum_{i=0}^{N-1}{}'\sum_{j=0}^{N-1}D_S(x_j,\tau)T_{ij}
\int_{x_k}^1 \rmd z\mathcal{K}_{gq}(z)\sqrt{\frac{z}{x_k}}T_i\left(y\left(\frac{x_k}{z}\right)\right), \\
& \\
\frac{\partial D_S(x_k,\tau)}{\partial \tau} \simeq &
\frac{2}{N}\sum_{i=0}^{N-1}{}'\sum_{j=0}^{N-1}D_S(x_j,\tau)T_{ij}\Bigg[
-\int_0^{x_k}\rmd z\,\delta_{jk}\mathcal{K}_{qq}(z)\frac{z}{\sqrt{x_k}}
\\ & \hspace{2.5cm} + \int_{x_k}^1\, \rmd z \mathcal{K}_{qq}(z)\left(
\sqrt{\frac{z}{x_k}}T_i\left(y\left(\frac{x_k}{z}\right)\right)
-\delta_{jk}\frac{z}{\sqrt{x_k}}\right) \Bigg] \\
&+\frac{2}{N}\sum_{i=0}^{N-1}{}'\sum_{j=0}^{N-1}D_g(x_j,\tau)T_{ij}
\int_{x_k}^1 \rmd z\,\mathcal{K}_{qg}(z)\sqrt{\frac{z}{x_k}}T_i\left(y\left(\frac{x_k}{z}\right)\right), \\
& \\
\frac{\partial D_{NS}^{(l)}(x_k,\tau)}{\partial \tau} \simeq &
\frac{2}{N}\sum_{i=0}^{N-1}{}'\sum_{j=0}^{N-1}D_{NS}^{(l)}(x_j,\tau)T_{ij}\Bigg[
-\int_0^{x_k}\rmd z\,\delta_{jk}\mathcal{K}_{qq}(z)\frac{z}{\sqrt{x_k}}
\\ & \hspace{2.5cm} + \int_{x_k}^1 \rmd z\, \mathcal{K}_{qq}(z)\left(
\sqrt{\frac{z}{x_k}}T_i\left(y\left(\frac{x_k}{z}\right)\right)
-\delta_{jk}\frac{z}{\sqrt{x_k}}\right) \Bigg].
\end{split}
\end{equation}

The integrals  from 0 to 1 in (\ref{eq:eeintk}) are split in integral from 0 to $x_k$ and from $x_k$ to 1 and evaluated with another Chebyshev decomposition (with adequate bijections).
Actually, we use the logarithmic bijections (implying a low-$x$ cut-off) 
which are more suitable here.
Finally, a simple Euler method is used to solve the equation in time, with a narrow Gaussian function as a starting distribution:
\begin{equation}
D_0(x;\varepsilon) = \sqrt{\frac{2}{\pi}}\,\frac{1}{\varepsilon}\,\mathrm{exp}\left[-\frac{(x-1)^2}{2\varepsilon^2}\right]\,,
\label{eq:D0Gauss}
\end{equation}
where $\varepsilon=10^{-2}$,
as the Chebyshev polynomials are unable to reproduce the Dirac $\delta$-function. 
The inability to approach peaked functions translates into oscillations 
of the solution in the low evolution times
(particularly visible on the quark distributions for the quark-initiated cascade).

Unfortunately, this method has not shown suitable for the $k_T$-dependent evolution equations (\ref{eq:BDIMsys1}). 

\subsection{Numerical cross-checks}

Here, we present some results of numerical cross-check of the presented above
methods for solving of the evolution equations (\ref{eq:BDIMsys1}) and (\ref{eq:eeintk}).
They have been obtained for the input parameters given in Section~\ref{sec:NumRes}. 


\begin{figure}[!htbp]
    \centering
    \includegraphics[scale=0.43]{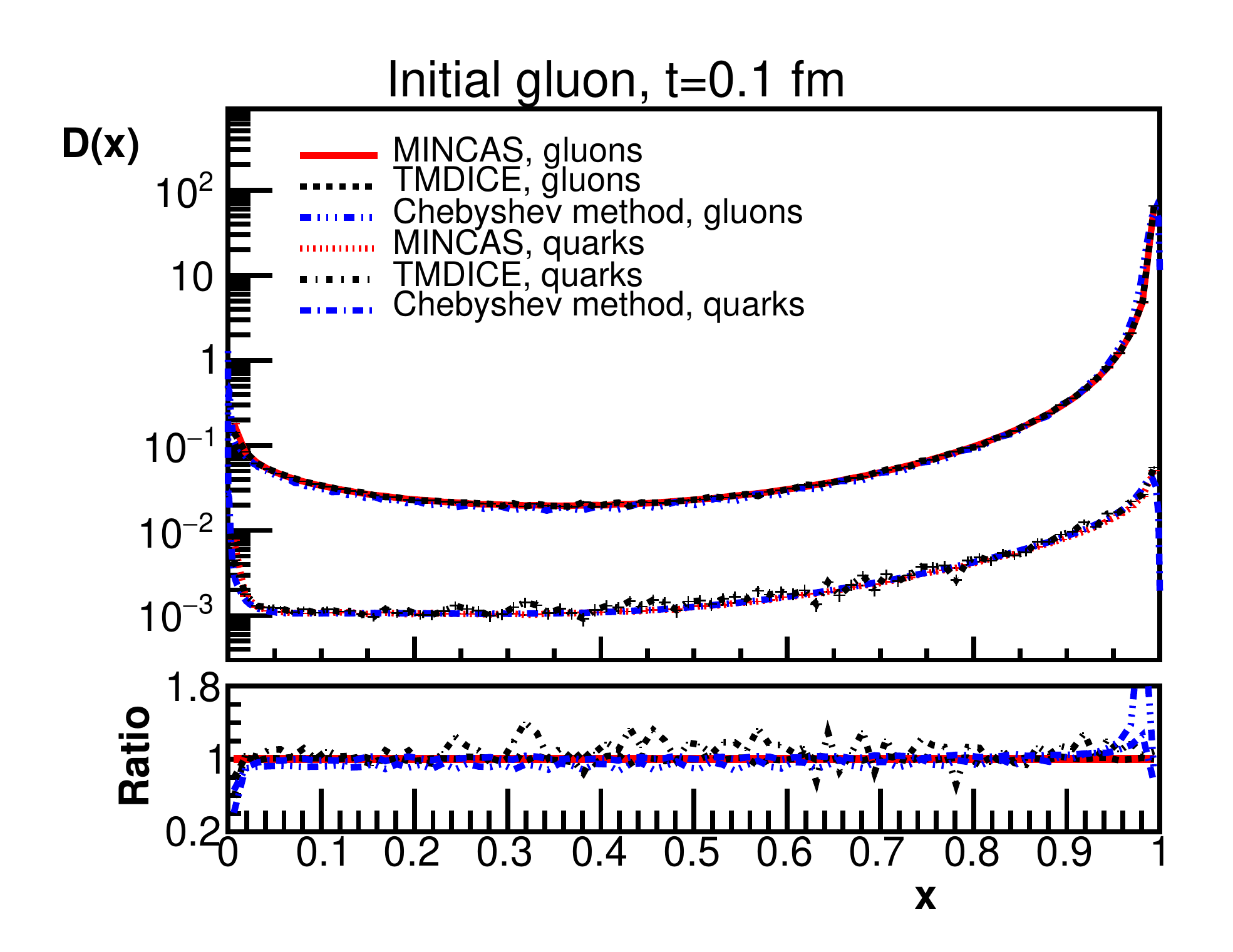}
    \includegraphics[scale=0.43]{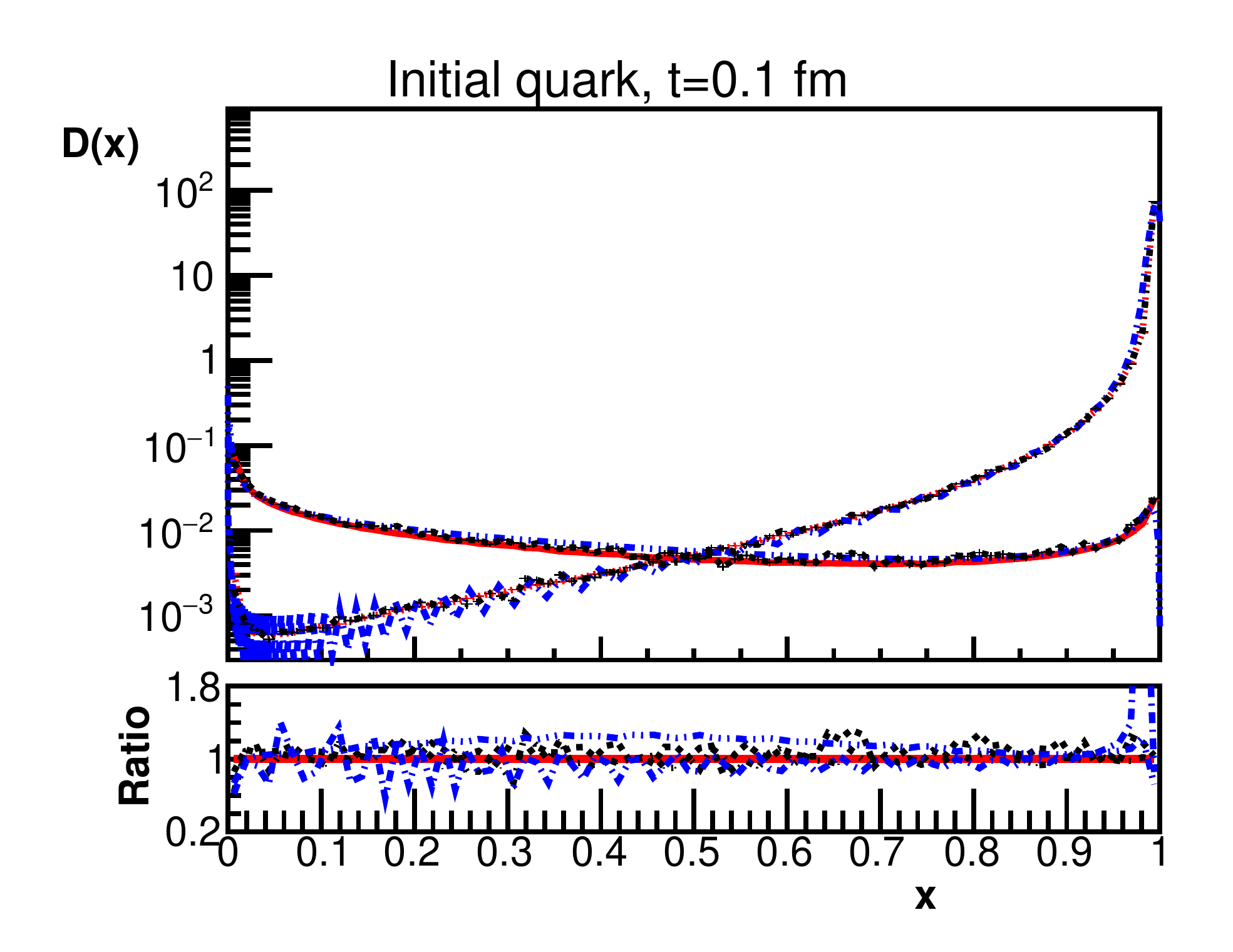}
    \includegraphics[scale=0.43]{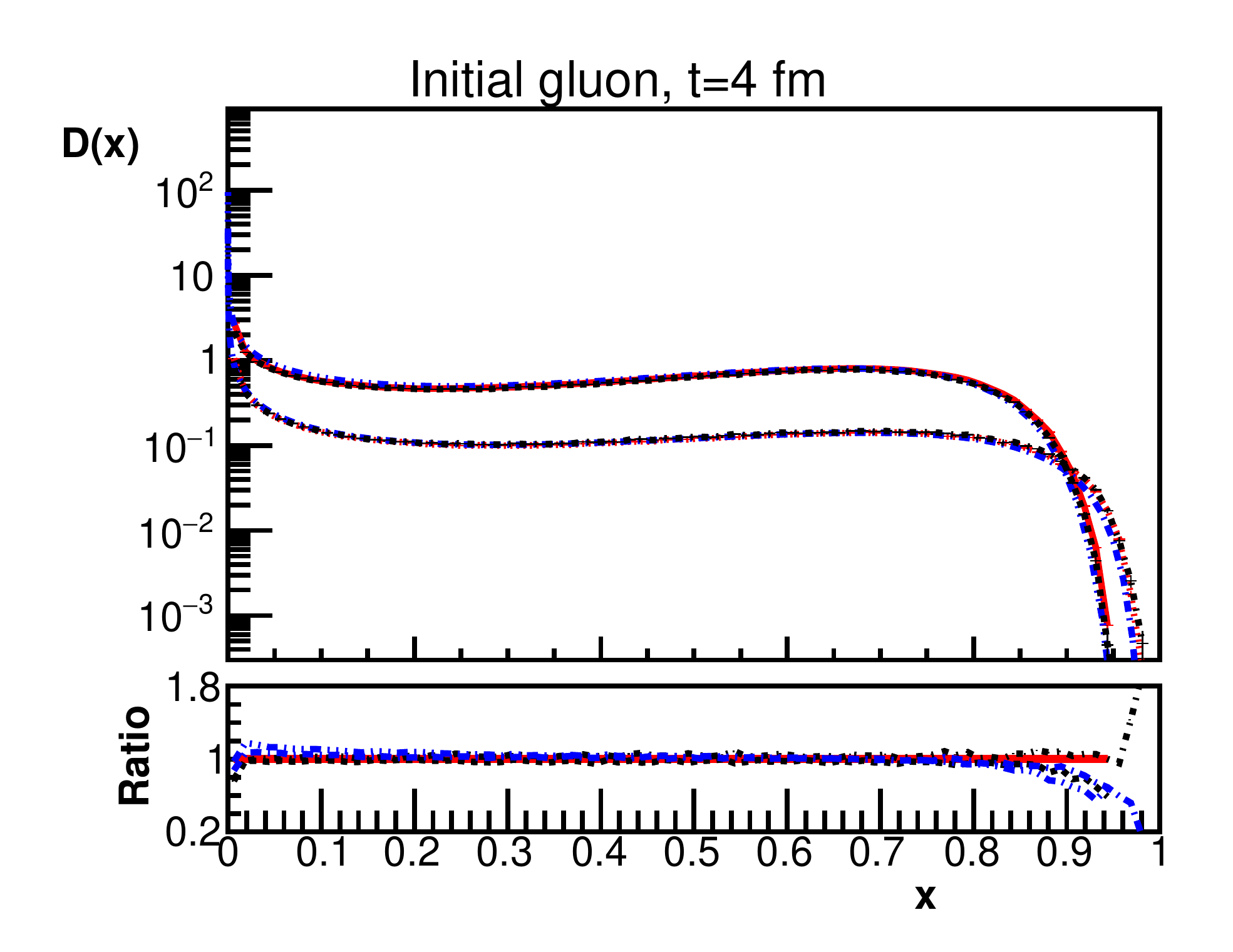}
    \includegraphics[scale=0.43]{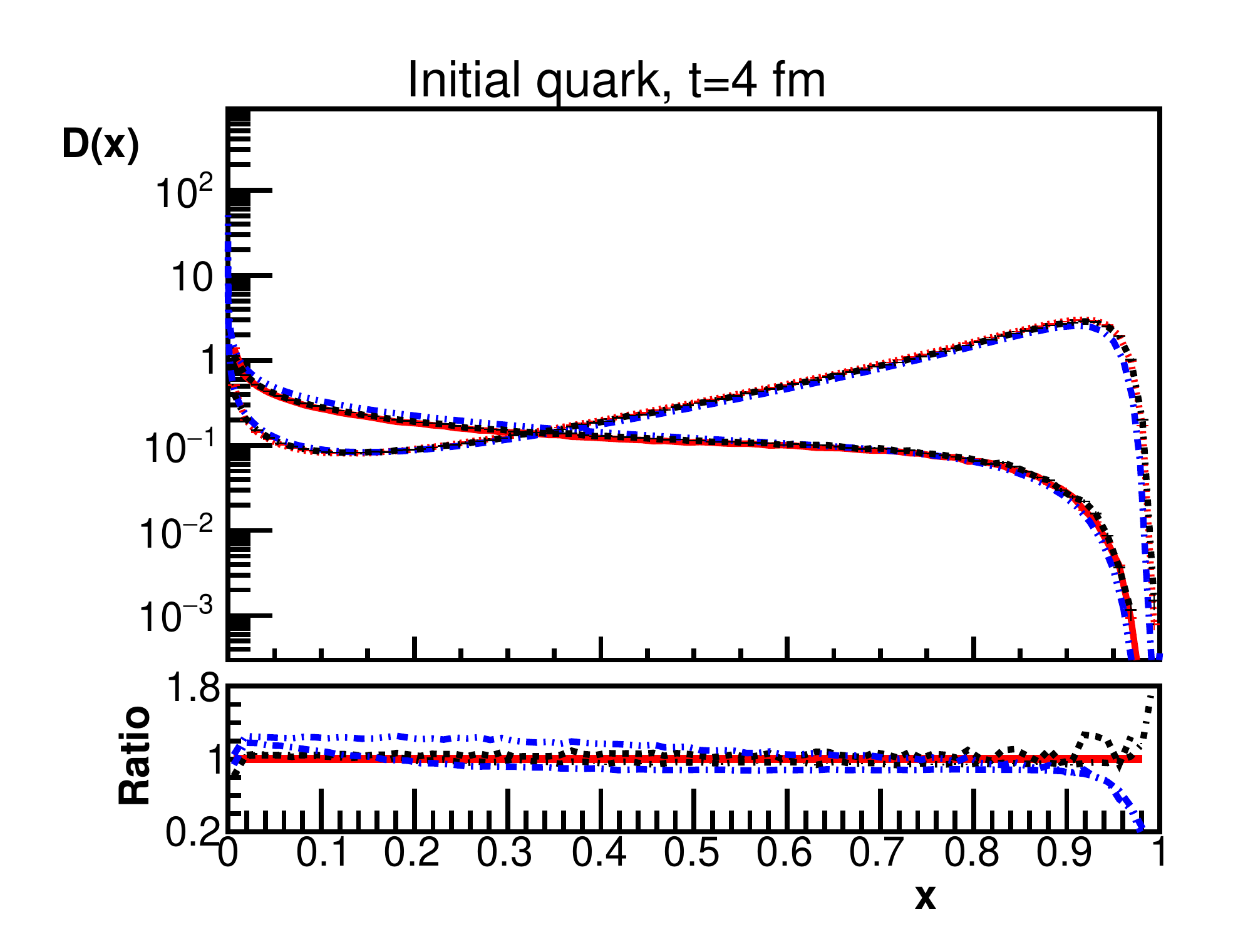}
    \caption{Comparisons of the $x$ distributions from the Monte Carlo programs \mincas\ and \tmdice\ 
    for the evolution equations (\ref{eq:BDIMsys1}) with the results from the Chebyshev method for the evolution equations (\ref{eq:eeintk}) 
    at the time-scales $t=0.1$ and $4\,$fm:
    cascades initiated by gluons (left) and quarks (right).
       The bottom panels show the ratios to the corresponding results from \mincas.
    }
    \label{fig:x_kzq_wq2_v2}
\end{figure}

We start from the $D(x,t)$ distributions for which we can compare all three numerical-solution methods.
In Fig.~\ref{fig:x_kzq_wq2_v2} we present comparisons for the results from the Monte Carlo programs
\mincas\ and \tmdice\ for the evolution equations (\ref{eq:BDIMsys1})
with the results from the Chebyshev method for the evolution equations 
(\ref{eq:eeintk}) at the time-scales $t=0.1$ and $4\,$fm.
The results from \mincas\ and \tmdice\ are obtained for the full $(x,\mathbf{k})$ evolution
but they are integrated by Monte Carlo method over $\mathbf{k}$. 
One can see a very good agreement of the three methods for both the final gluon and quark distributions
in both the gluon and quark initiated cascades.
In the case of the Monte Carlo programs, this also shows that they reproduce correctly
the $D(x,t)$ distribution from the full evolution for the $D(x,\mathbf{k},t)$ distribution
given by Eq.~(\ref{eq:BDIMsys1}).
A similar agreement has been found for other values of the evolution time values.
The Chebyshev method reveals some oscillatory behaviour for $t=0.1\,$fm, 
i.e.\ for short evolution time, but works well for higher time-scales, $t>1\,$fm.

\begin{figure}[!htbp]
    \centering
    \includegraphics[scale=0.43]{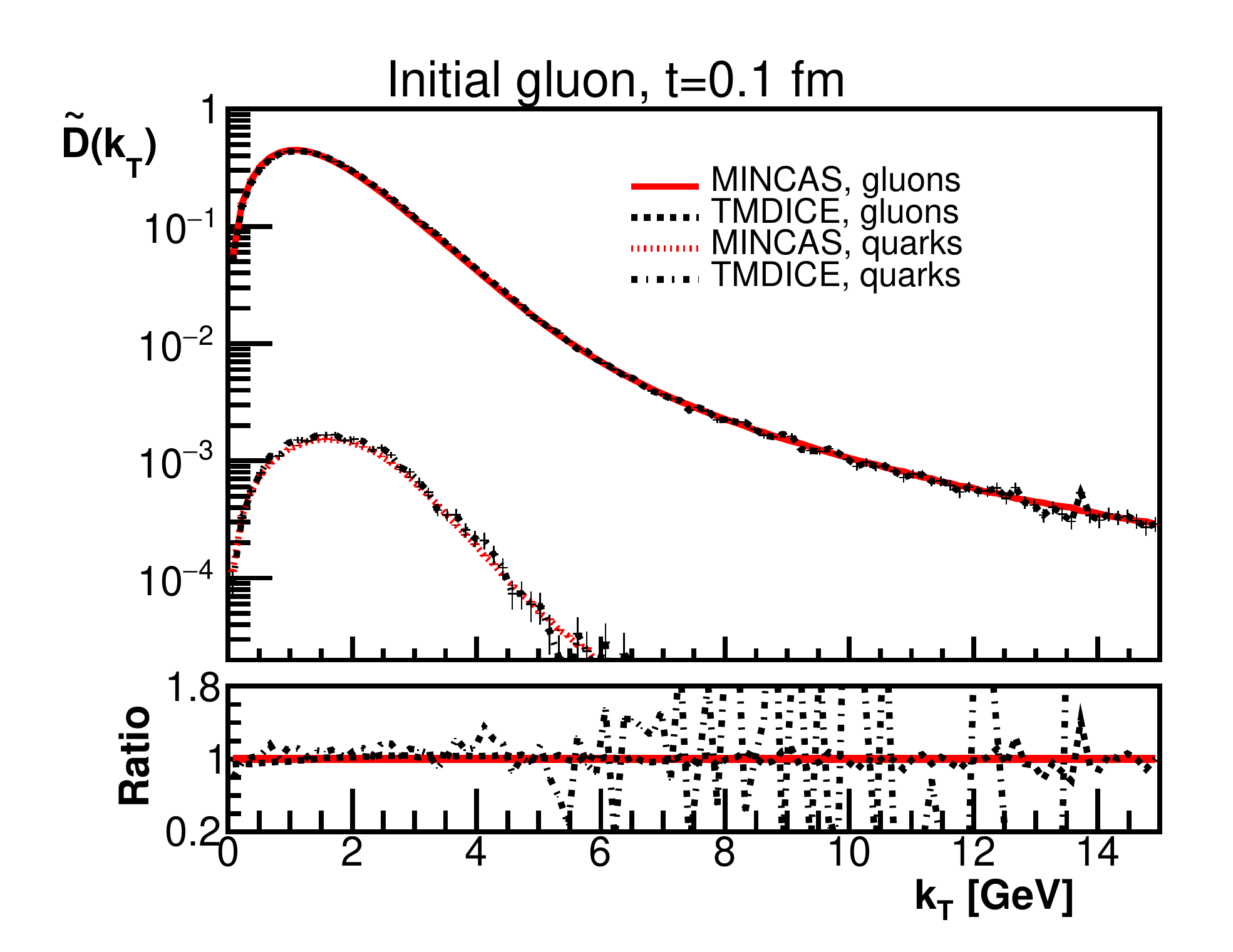}
    \includegraphics[scale=0.43]{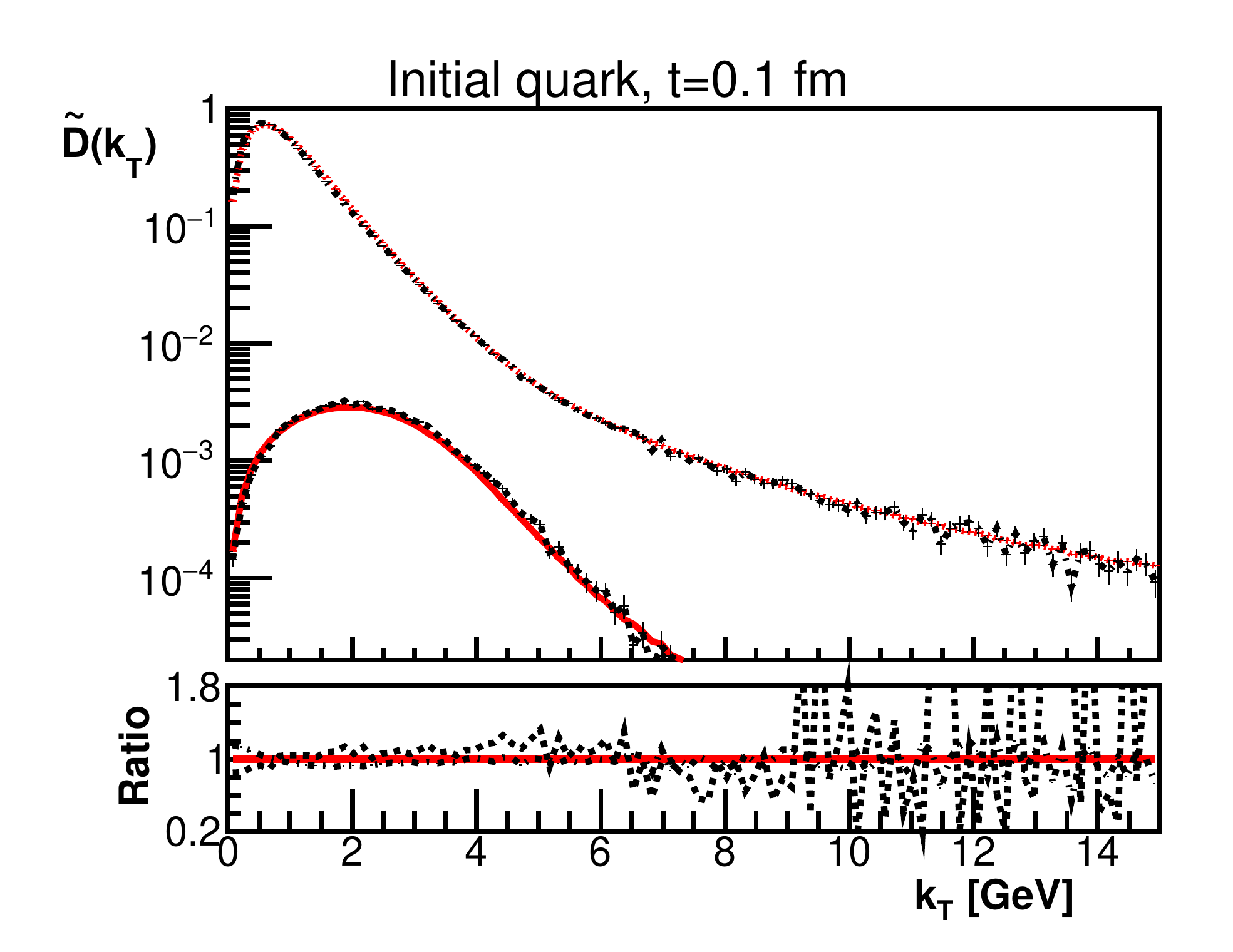}
    \includegraphics[scale=0.43]{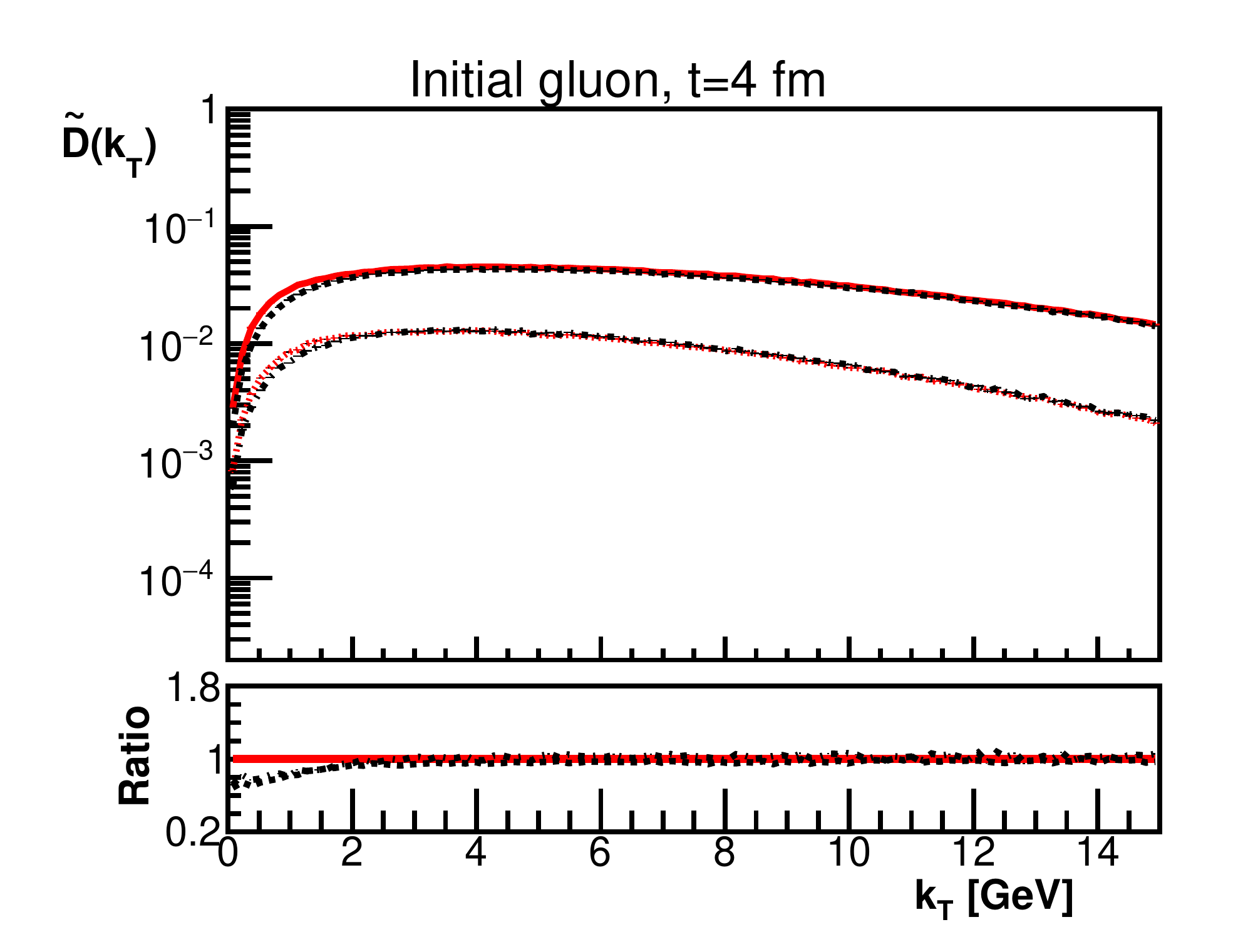}
    \includegraphics[scale=0.43]{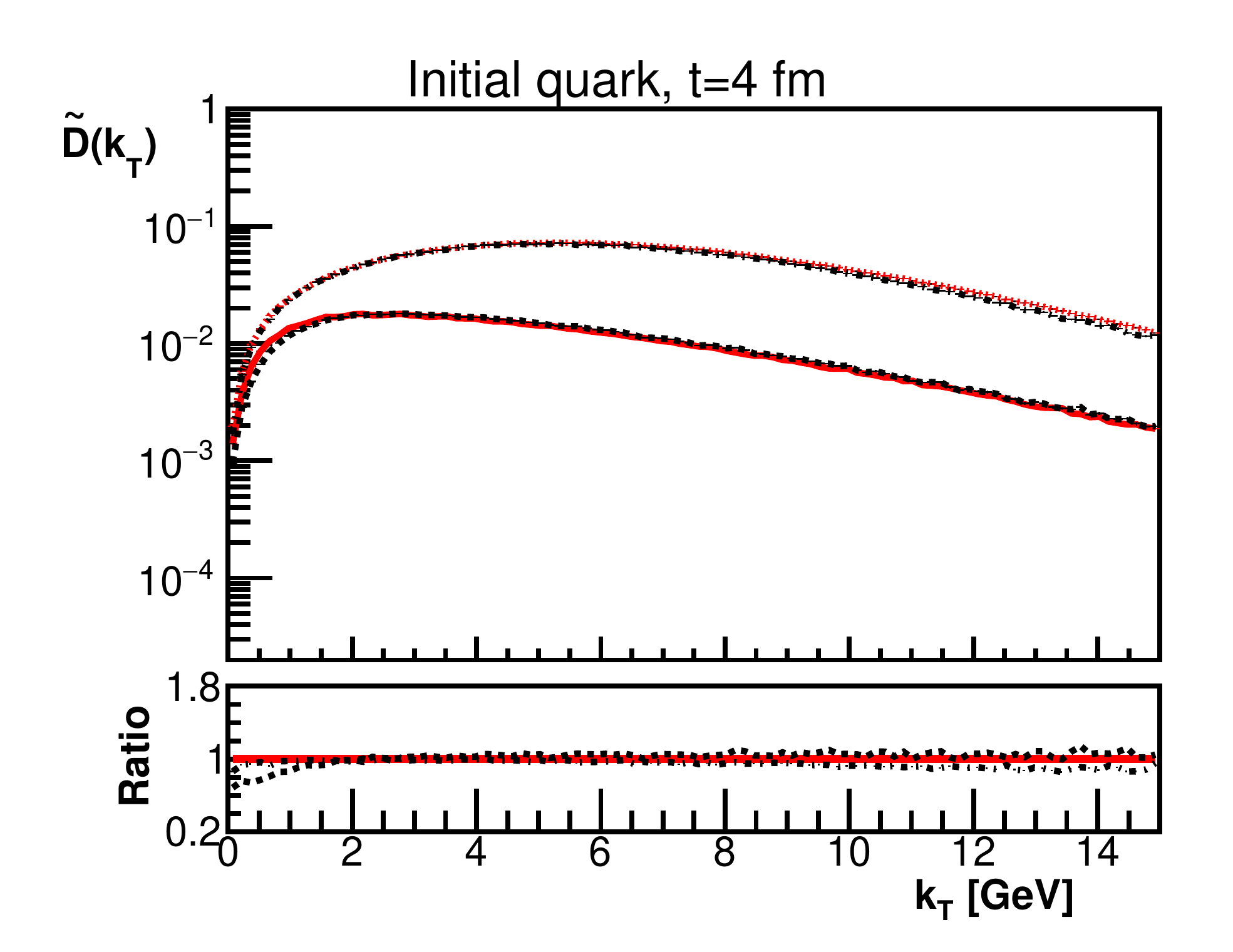}
    \caption{Comparisons of the $k_T$ distributions from \mincas\ and \tmdice\ 
    for the evolution equations (\ref{eq:BDIMsys1}) with 
    $w(\mathbf{l})\propto 1/[\mathbf{l}^2(m_D^2+\mathbf{l}^2)]$ 
    at the time-scales $t=0.1$ and $4\,$fm:
    cascades initiated by gluons (left) and quarks (right).
    The bottom panels show the ratios to the corresponding results from \mincas.
    }
    \label{fig:kt_kzq_wq2}
\end{figure}

In Fig.~\ref{fig:kt_kzq_wq2}, we present comparisons of 
the $\Tilde{D}(k_T,t)$ distributions for the evolution equations (\ref{eq:BDIMsys1})
obtained from two Monte Carlo programs:
\mincas\ and \tmdice, for the cascades initiated by gluons (left) and quarks (right)
at evolution time-scales $t=0.1$ and $4\,$fm. As one can see, a perfect agreement
between the two independent Monte Carlo algorithms and programs is found. 
We have checked that such an agreement holds also for other evolution time-scales,
other splitting kernels and other medium-scattering functions.
Unfortunately, for the equations including the $k_T$-dependence, 
the Chebyshev method has turned out to be unfeasible.

\bibliography{refs}{}
\bibliographystyle{utphys_spires_tit}
\end{document}